\documentclass[12pt]{article}
\usepackage[a4paper]{geometry}

\usepackage{etoolbox}

\usepackage[utf8]{inputenc}
\usepackage{lipsum}

\usepackage{amsmath,amssymb}
\usepackage{color}
\usepackage{mathtools}
\usepackage[ruled]{algorithm2e}

\def\keywords{\vspace{.5em}
{\textit{Key words}:\,\relax\footnotesize%
}}

\usepackage{graphicx}
\usepackage[caption = false]{subfig} 
\usepackage{enumerate}
\usepackage{comment}
\usepackage{natbib, float}

\newtheorem{theorem}{\it  Theorem}
\newtheorem{definition}{\it Definition}

\newcommand{\RR}{\mathbb R}

\newcommand{\Prob}{{\rm Pr}}

\newcommand{\A}{\mathcal A}
\newcommand{\calR}{\mathcal R}

\newcommand{\D}[1]{\mathrm{d}{#1}}

 \def\Deq{{\ {\buildrel {\rm D}\over =}\ }}
 
\newcommand{\R}{r}
\newcommand{\Ar}{\A_\R}
\newcommand{\Set}{\mathcal{F}}

\newcommand{\simplex}{\mathcal{S}}
\newcommand{\simplexSub}{\mathcal{W}}
\newcommand{\radialSet}{\mathcal{U}}
\newcommand{\exceedanceSets}{\mathcal{A}}
\newcommand{\indicatorFun}{1}
\newcommand{\degree}{^{\circ}}

\def\bRev#1{{\color{black} #1}}%
\usepackage[normalem]{ulem}

\title{Functional Peaks-over-threshold Analysis}

\author{Rapha\"el de Fondeville \thanks{raphael.de-fondeville@epfl.ch} 
\\
{\small\textit{Ecole Polytechnique F\'ed\'erale de Lausanne, Station 14,}} \\ {\small\textit{1015 Lausanne, Switzerland}} \\
 Anthony C. Davison \thanks{anthony.davison@epfl.ch}
 \\ 
{\small\textit{Ecole Polytechnique F\'ed\'erale de Lausanne, EPFL-FSB-MATH-STAT, Station 8,}}\\ {\small\textit{1015 Lausanne, Switzerland}} }
\date{}

\begin{document}
\maketitle
\begin{abstract}
Peaks-over-threshold analysis using the generalized Pareto distribution  is widely applied in modelling tails of univariate random variables, but much information may be lost when complex extreme events are studied using univariate results.  In this paper, we extend peaks-over-threshold analysis to extremes of functional data.  Threshold exceedances defined using a functional $\R$ are modelled by the generalized $\R$-Pareto process, a functional generalization of the generalized Pareto distribution that covers the three classical regimes for the decay of tail probabilities, and that is the only possible continuous limit for $\R$-exceedances of a properly rescaled process.
We give construction rules, simulation algorithms and inference procedures for generalized $\R$-Pareto processes, discuss model validation, and apply the new methodology to extreme European windstorms and heavy spatial rainfall.
\end{abstract}


\section{Introduction}\label{sec: intro}
Extreme value theory provides a mathematical framework for the description and modelling of tails of statistical distributions that can be used to extrapolate beyond observed events.  This theory has been studied extensively  \citep{Fisher1928,gnedenko43,Pickands:1975,Embrechts1997,Beirlant2006,Coles.Tawn:1991,
Heffernan.Tawn:2004} and is widely used in applications  \citep{Hosking1987,Coles2001,Buishand1989}.  However many complex phenomena must be  bowdlerised to be  modelled using univariate or even multivariate methods, so richer approaches to the analysis of  high-dimensional data have been explored over the past decade.

\begin{figure}
\begin{center}
\begin{tabular}{cc}
\multicolumn{2}{c}{
\includegraphics[width=0.4\textwidth]{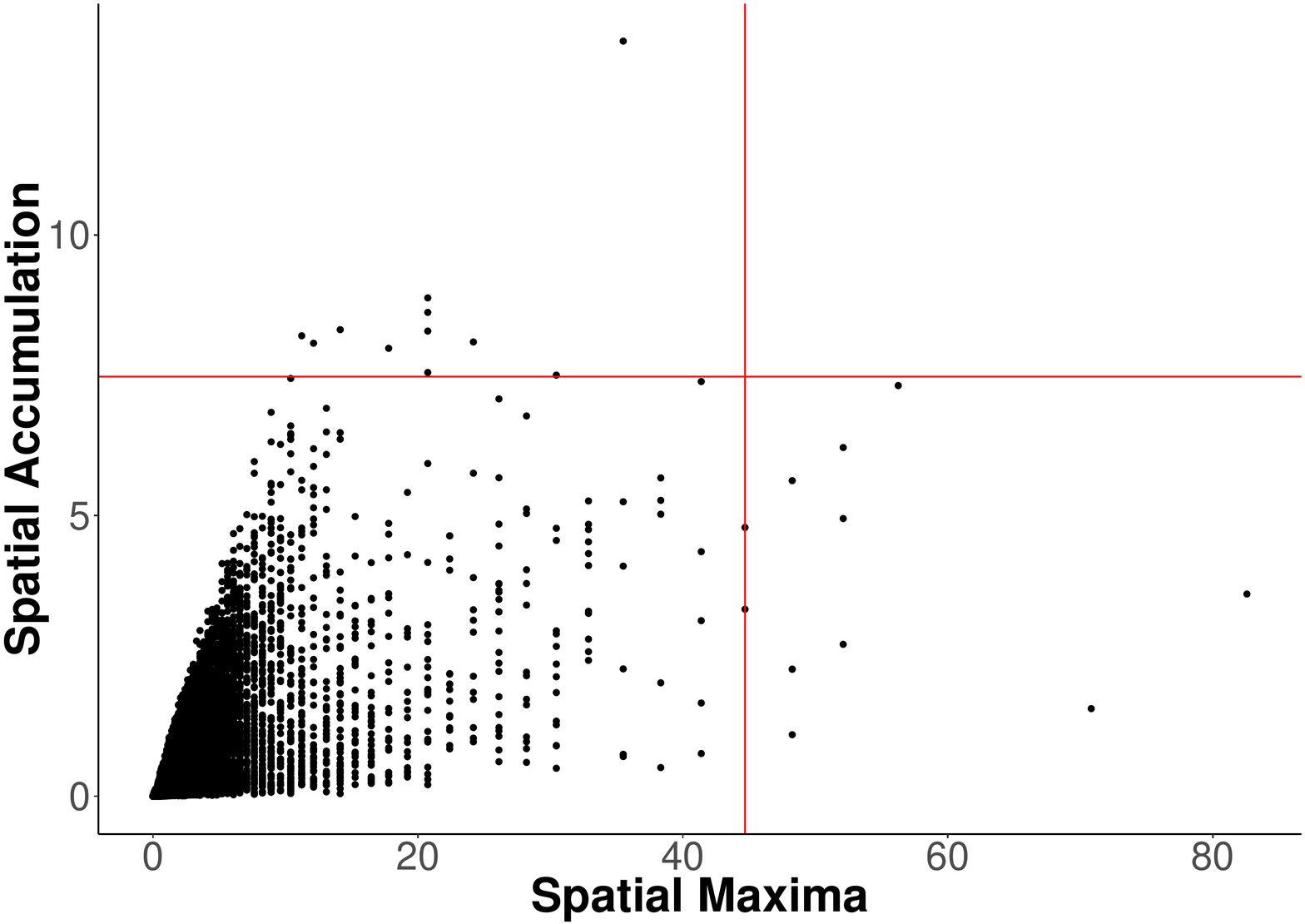}}  
\\ 
\includegraphics[width=0.45\textwidth]{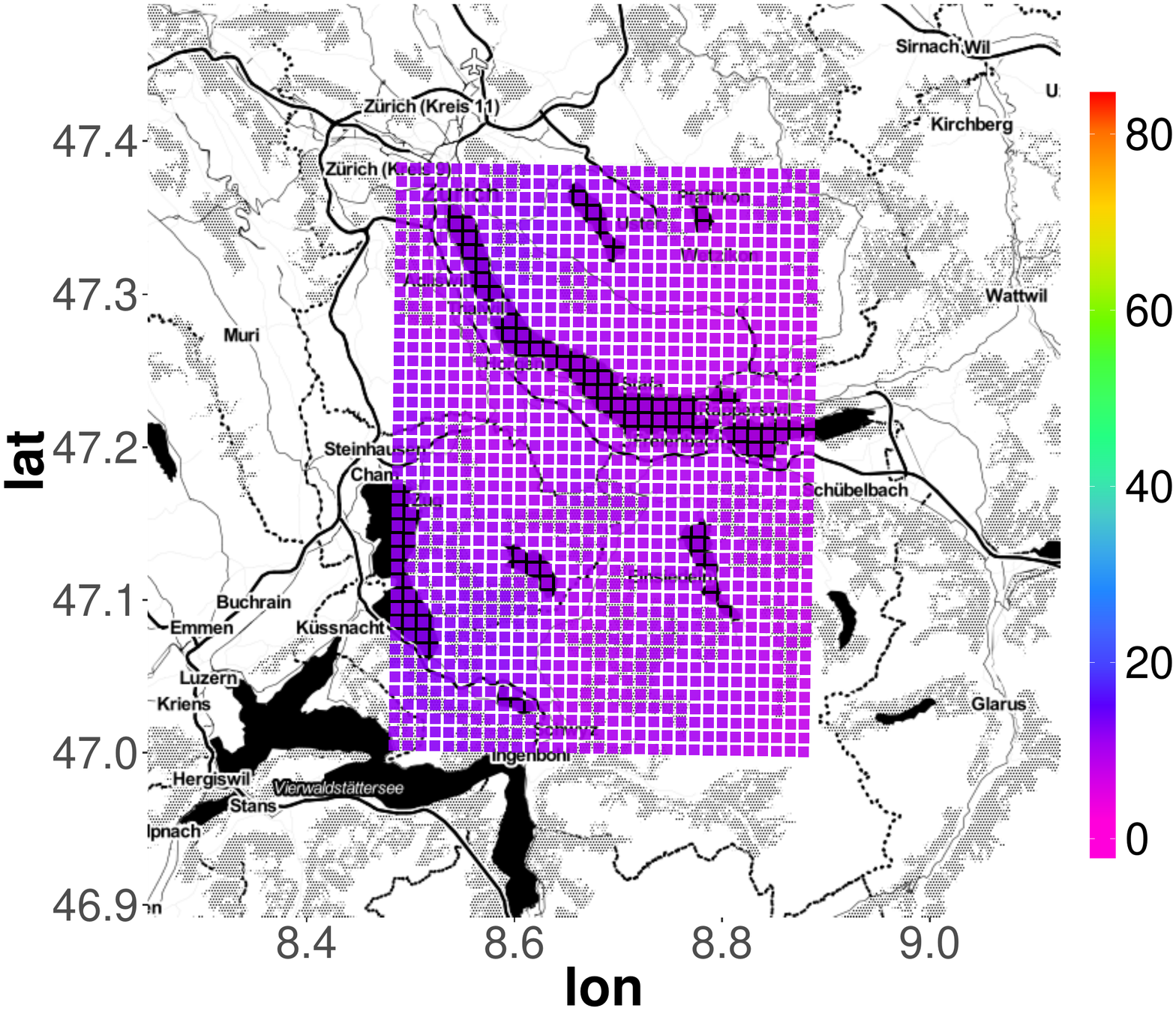} & \includegraphics[width=0.45\textwidth]{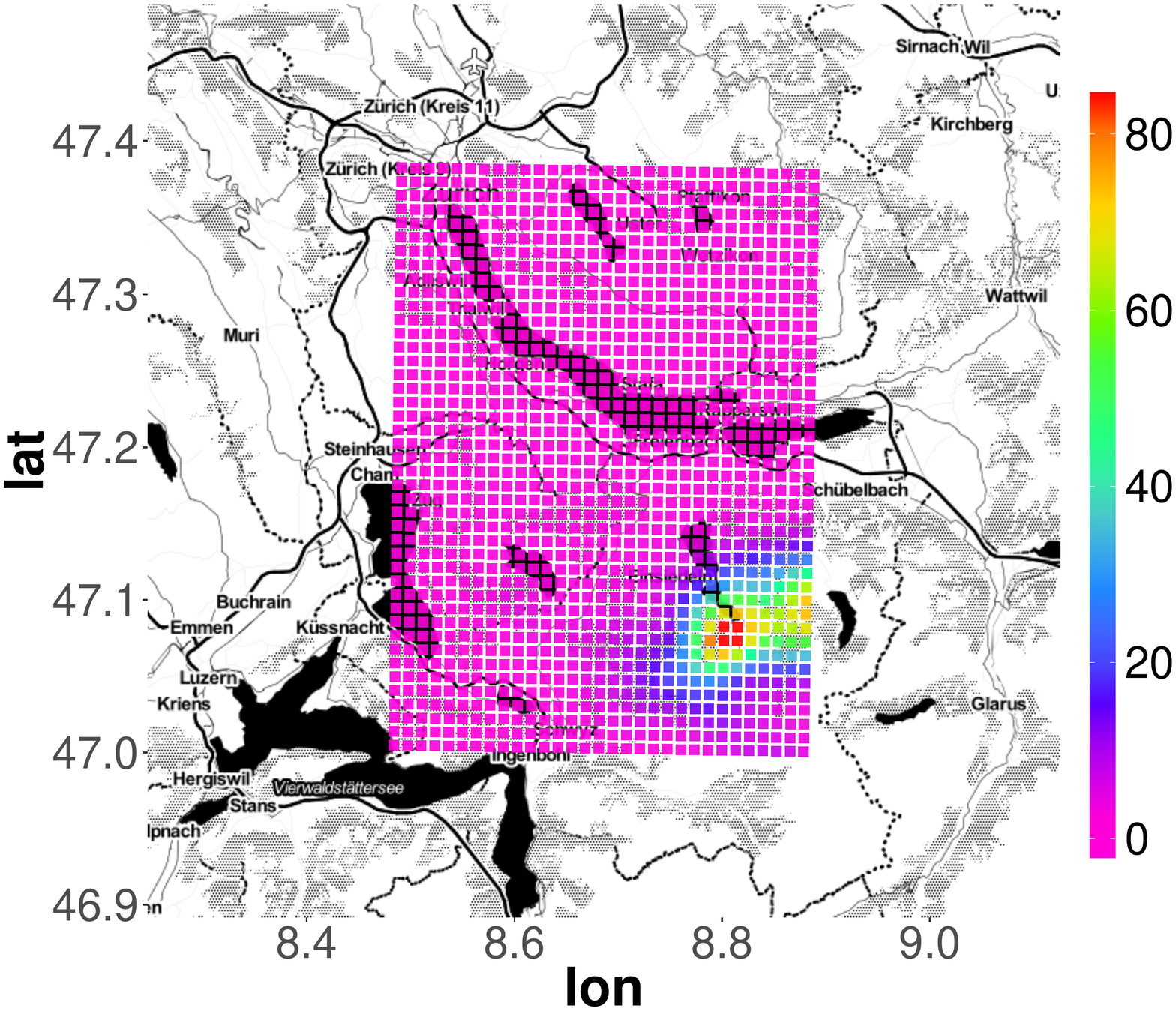}  \\
\end{tabular}
\end{center}
\caption{Extreme hourly rainfall events in the  Zurich region, 2013--2018, computed using radar rainfall measurements $X(s)$ (mm) on a grid $S$.  Top:  spatial averages $|S|^{-1}\int_S X(s)\, \D{s}$ and spatial maxima $\max_{s \in S} X(s)$, with red thresholds demarcating the largest $11$ events of each type.
Bottom line: events corresponding to the largest spatial average (left) and the largest spatial maximum (right).}
\label{fig: rainfall observations}
\end{figure}

Max-stable processes \citep[Section~9.2]{DeHaan1984,DeHaan2006} provide a functional extension of the classical extreme value distributions and have successfully been used to model maxima,  but are difficult to fit in high dimensions \citep{Huser2013a}.  Moreover they conflate individual extremal events and hence discard information, making it difficult to detect phenomena such as mixtures of tail behaviours.  For example, in some regions rainfall events are either convective, and hence locally very intense, or cyclonic, with larger spatial accumulations of water but lower local intensities.   Although driven by different weather patterns, both may lead to flooding, and, as suggested by Figure~\ref{fig: rainfall observations}, the marginal distributions of their tails and their spatio-temporal structures may differ greatly. Even though large-scale events may also be damaging, the use of maxima tends to drive modelling to focus on small-scale intense events that yield most maxima.



In the one-dimensional case the analysis of threshold exceedances is often preferred to that of block maxima.  The approach originated in hydrology under the name of `peaks over threshold' (POT) or `partial duration series' analysis \citep{Todorovic.Zelenhasic:1970,Todorovic.Rousselle:1971,NERC:1975}, its goal being to include all large individual events  and thus access more information than can be extracted from block, typically annual, maxima.  This approach is particularly valuable when the data are limited and there is an appreciable seasonal component.  A probabilistic basis for threshold modelling was provided by \citet{Balkema.deHaan:1974}, \citet{Pickands:1975} and \citet{Leadbetter:1991} and statistical aspects were developed by \citet{Davison:1984}, \citet{Smith:1984} and \citet{Davison1990}.   The basic idea is to fit the generalized Pareto distribution to the exceedances of a variable such as river flow or pollution level over a threshold.  A large literature has built on this early work and the method and its many variants have been applied in numerous other contexts.

In some applications it is helpful to reduce multivariate data to scalar `structure variables' \citep{Coles1994} that can be analysed using POT or other univariate methods, but this approach gives no insight into the combinations of variables  yielding a rare event.  Different structure variables may have different tail behaviours  due, for example, to the presence of several underlying physical processes.  Functional peaks-over-threshold analysis modifies this approach to give different perspectives on the dependence structure and provides a theoretical foundation for the detection of mixtures of tail behaviours through definitions of functional extremes tailored to particular types of events, as illustrated in Figure~\ref{fig: rainfall observations}. 

{Existing functional peaks-over-threshold procedures rely on particular types of exceedances \citep{Ferreira2014} or are limited to settings where the data must have unbounded support and share the same polynomial-type tail decay \citep{Dombry2013}.
Observations can be transformed to have a common marginal distribution, such as the unit Fr\'echet \citep[Section~ 5]{Coles.Tawn:1991} or unit Pareto \citep{Kluppelberg2008}, and exceedances may be defined on this transformed scale, but, as many extreme phenomena are most naturally characterized on the scale of the original data, the use of transformations can require the user to trade off interpretability against mathematical rectitude; for example, \citet{Fondeville2017} attempt to characterize different types of rainfall after data  transformation. In univariate extreme value theory, the generalized Pareto distribution provides a single framework for modelling the original data in any of the classical Weibull, Gumbel or Fr\'echet regimes.
The present paper provides a similar unified formulation for functional peaks-over-threshold analysis under the assumptions that the process has the same rate of tail decay over its domain and that its limiting tail distribution presents some level of dependence.
The tail decay restriction is needed to define the exceedances directly on the original process, as otherwise the region or location with the heaviest tail would dominate the limit distribution, leading to unrealistic models.
We extend the work of \citet{Dombry2013} by introducing the generalized $\R$-Pareto process, allowing more flexible definitions of rare events and generalized Pareto margins for tails.
The generalized $\R$-Pareto process is the only continuous limit of exceedances of a properly rescaled process.  For some definitions of exceedances, it allows the Monte Carlo simulation of events with a fixed intensity, i.e., events for which the level of risk has a prescribed return level.}
These results rely on a specific type of convergence that  excludes independence of the limiting tail distribution; although our results can be generalized to deal with this, we leave this for future work.

Section~\ref{sec: theory} reviews classical univariate results and introduces functional peaks-over-threshold analysis.
We derive convergence results for the three tail decay regimes, define and characterize the generalized $\R$-Pareto process, present simulation algorithms, and discuss the scope of our models.
Section~\ref{sec: model form} introduces a general model for functional exceedances. 
In Section~\ref{sec: stat inference} we discuss statistical inference and  in Section~\ref{sec: model validation} we describe methods for model validation.
In Section~\ref{sec: windstorms} we use our ideas to develop a stochastic weather generator for windstorms over Europe, and Section~\ref{sec: rainfall} illustrates the importance of risk definition when studying potential flooding in the city of Zurich.
Technical details and proofs of the main results are relegated to Appendices.

\section{Modelling threshold exceedances}\label{sec: theory}
\subsection{Univariate exceedances}\label{sec: univariate results}

If a scalar random variable $X$ has distribution function $F$ and there exist sequences of constants $a_n > 0$ and $b_n$ such that
\begin{equation}\label{eq: mda}
n\, \Prob\left(\dfrac{X - b_n}{a_n}> x \right) \rightarrow -\log G(x), \quad n \rightarrow \infty,
\end{equation}
where $G$ is a non-degenerate distribution function, then $X$ is said to belong to the max-domain of attraction of $G$ \citep[p.~12]{Resnick1987}.
For a large enough threshold $u<\inf\{x:F(x)=1\}$ and $x > 0$, the tail behaviour of $X$ can be described using a generalized Pareto distribution,  
\begin{equation}\label{eq: gpd}
\Prob\left( X> x+u \mid X>u\right) \approx H_{\xi,\sigma}(x)  = \left\{
\begin{array}{ll}
\left(1+\xi x/\sigma\right)_+^{-1/ \xi}, & \xi \neq 0, \\
\exp\left(- x/\sigma\right),  & \xi = 0,
\end{array}
\right. 
\end{equation}
where {$\sigma = \sigma(u)> 0$} and, here and below, $a_+ = \max(a, 0)$ for real $a$. 
The shape parameter $\xi$ is also called the tail index.  If $\xi$  is negative then $X-u$ lies in the interval $[0,-\sigma/\xi]$, and otherwise $X-u$ can take any positive value. 
The random variable $X$ is said to belong to the Weibull, Gumbel or Fr\'echet domains of attraction if $\xi$ is respectively negative, zero or positive.  The max-domain of attraction conditions are broadly but not universally satisfied \citep[e.g.,][pp.~59, 62, 72]{Beirlant2006}.

\citet{Davison1990} use equation~\eqref{eq: gpd} as the basis of the approximation 
\begin{equation}
\label{eq:GPD}
F(x) \approx 1 - \zeta_u H_{\xi,\sigma}(x-u), \quad x > u,
\end{equation}
where $\zeta_u$ denotes the probability that $X$ exceeds the threshold $u$. This offers a general, flexible and unified model for distribution tails and is widely used to estimate probabilities of rare events. 

In its simplest form equation~\eqref{eq:GPD} applies to independent and identically distributed variables, but it is also used more broadly.  
The modelling of exceedances has been extended to multivariate settings \citep{Rootzen2006,Rootzen2018, Rootzen2017} and to continuous processes \citep{Ferreira2014, Dombry2013}.

\subsection{Functional exceedances}

Let $S$ be a compact subset of $\RR^D$, let $\Set$ denote the space of real-valued continuous functions on $S$ equipped with norm $\|\cdot\|$, and let $\Set_+ $ denote the subset of $\Set$ containing only non-negative functions that are not everywhere zero; this excludes the zero function and hence avoids the appearance of degenerate probability measures when taking limits. 

Exceedances for a random function $X=\{X(s): s\in S\}$ can be defined using risk functionals and $\R$-exceedances.  A risk functional $\R$ is defined as a continuous mapping from $\Set$ into $\RR$ and an $\R$-exceedance is defined to be an event of the form $\{\R(X) \geq u\}$ for some $u \geqslant 0$, i.e., an event for which the scalar $\R(X)$ exceeds a threshold $u$. This definition was introduced by \citet{Dombry2013} for homogeneous `cost functionals' on $\Set_+$, i.e., functionals for which there exists $\kappa > 0$ such that $\R(ay) = a^\kappa \R(y)$ when $y \in \Set_+$ and  $a> 0$.  The term `radial aggregation function' was earlier used by \citet{Opitz2013}, but in our view the term `risk functional' better reflects how $\R(X)$ measures the severity of $X$ in terms of the risk summarised by $\R$.

\citet{Ferreira2014} studied threshold exceedances  for continuous processes using the functional $\R(X) = \sup_{s \in S}X(s)$, but this treats as extreme all events with an exceedance at at least one point in $S$ only. \citet{Tawn1996} had earlier modelled areal rainfall via large values of $\int_{S} X(s)\, \D{s}$, and other functionals such as $\int_{S} X^2(s)\, \D{s}$ for a proxy of the energy inside a climatic system \citep{Powell2007},
$X(s_0)$ for risks impacting a specific location $s_0$, and so forth, may arise in  applications.
Likewise one might project climate data onto scalar signals of particular weather patterns and examine the behaviour of their $\R$-exceedances.  The motivation behind the present paper is to define risk functionals tailored to specific types of physical processes, and this may yield different models based on different functionals. If a single model that merges different notions of risk is required, consistency between definitions can be enforced in our framework by studying 
\begin{equation}\label{mix-risk.eq}
\R(X) = \max\left\{{\R_1(X)}-{u_1}, \dots, {\R_M(X)}-{u_M}\right\},
\end{equation}
where $\R_1, \dots , \R_M$ are functionals of interest and $u_1,\dots, u_M $ the corresponding thresholds.

Below we generalize $\R$-exceedances under minimal assumptions on the risk functional and derive limit distributions for the three tail decay regimes.

\subsection{Functional $\R$-exceedances}\label{sec: generalized functional pot}

\subsubsection{Notation, assumptions and convergence}

\begin{figure}
    \centering
    \begin{tabular}{ccc}
    $\xi > 0$ & $\xi = 0$ & $\xi < 0$ \\
    \includegraphics[width=0.3\textwidth]{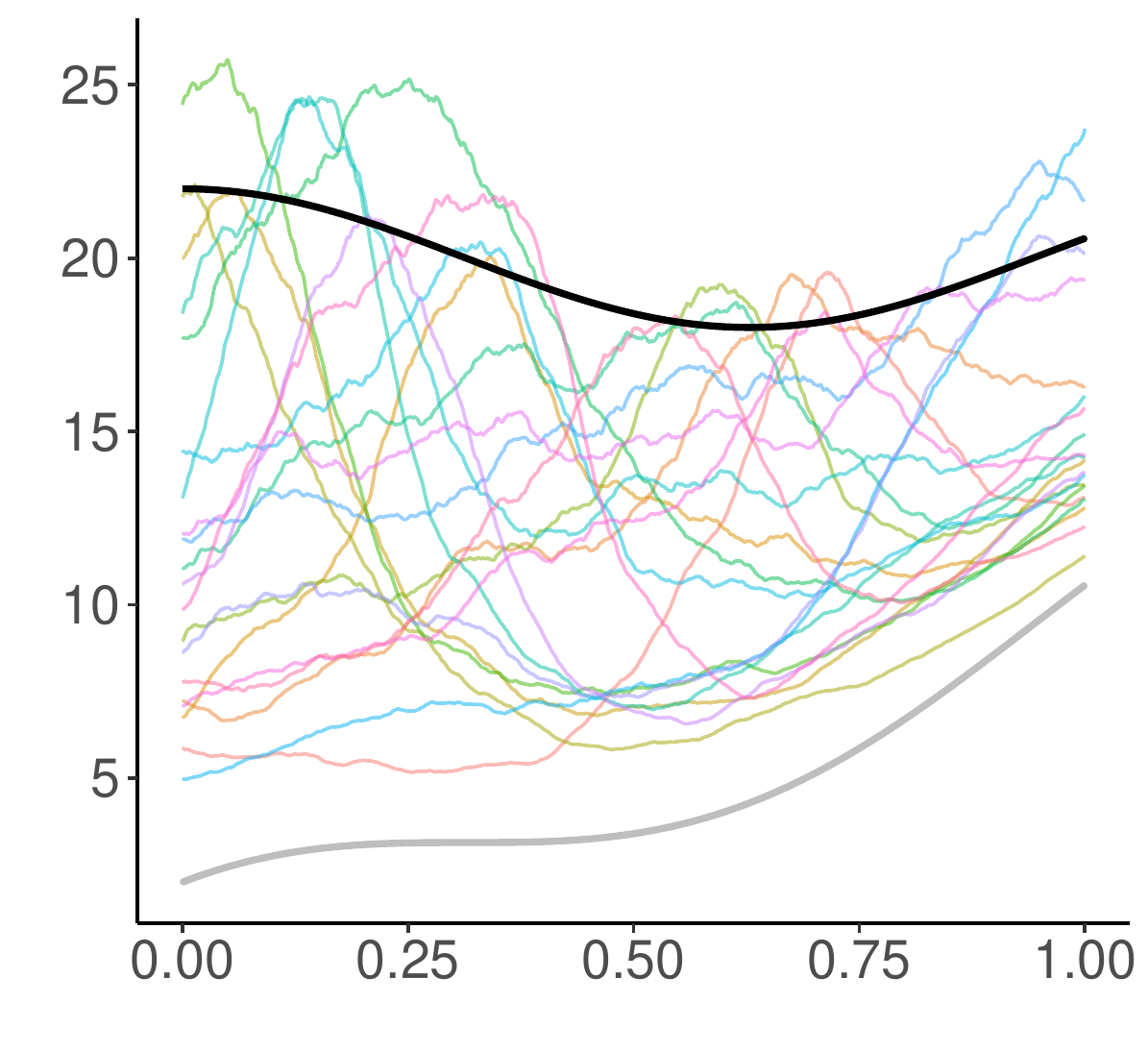} & \includegraphics[width=0.3\textwidth]{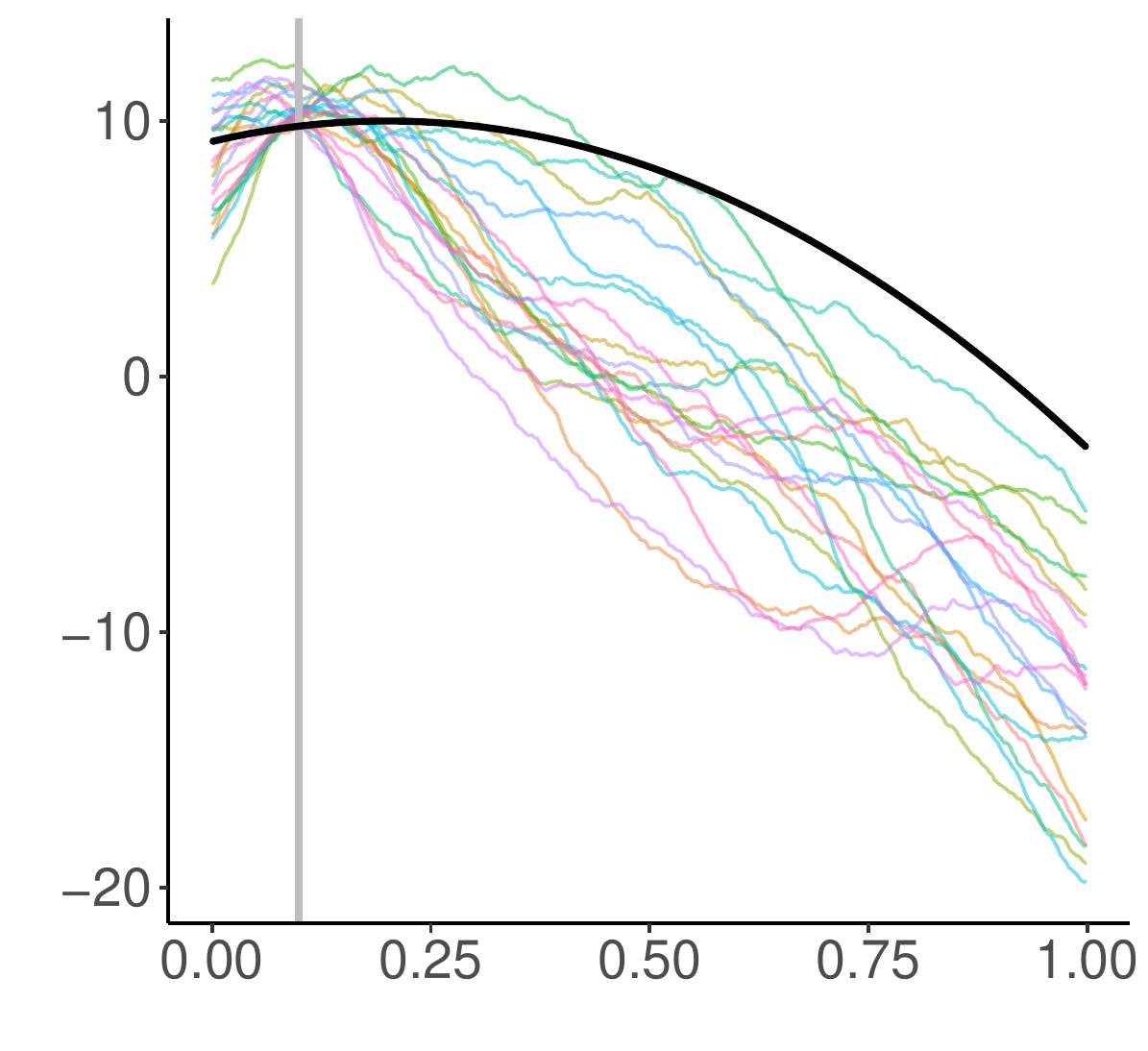} & \includegraphics[width=0.3\textwidth]{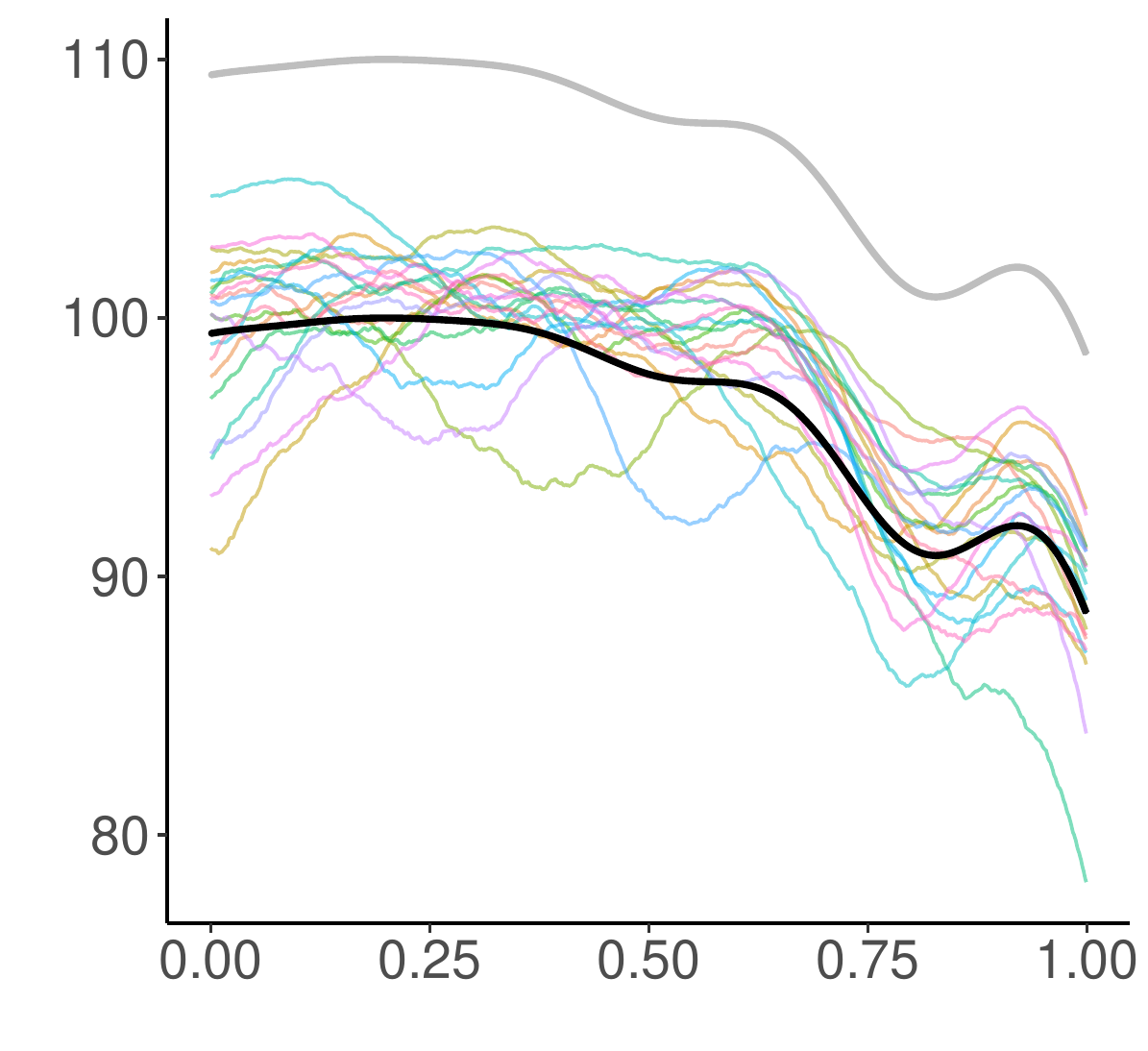} \\
    $\R(x) = \sup_{s \in [0,1]} x(s)$ & $\R(x) =  x(0.1)$ & $\R(x) = \int_{s \in [0,1]} x(s)\D{s}$ 
    \end{tabular}
    \caption{Realisations of generalized $\R$-Pareto processes on $S=[0,1]$, with location function $b$ (heavy black line) and lower/upper bound {$b-\xi ^{-1}a$} of $\Set^{\xi,a,b}$ (heavy grey line). Left: realisations $x$ with shape parameter $\xi>0$ and risk functional $\R(x) = \sup_{s \in S} x(s)$, with the parts of the $x$s below the threshold function $b$ shown as dotted lines. Middle: realisations with $\xi = 0$ and risk functional $\R(x) = x(0.1)$ representing evaluation at $s=0.1$ (vertical line). Right: realisations $x$ with $\xi<0$ and risk functional $\R(x) = \int_{s \in S}x(s)\D{s}$.} 
    \label{fig: example Pareto}
\end{figure}

Let $\xi $ be a scalar shape parameter, and let $a\equiv a(s) > 0$ and $b\equiv b(s)$ be continuous functions defined for $s\in S$, with $\equiv$ denoting equivalent notations. For given $\xi$, $a$ and $b$ define the set
\begin{equation}\label{F.eq}
\Set^{\xi,a,b} = 
\left\{\begin{array}{ll}  \Set_+ - (b-\xi ^{-1}a ), & \xi > 0, \\
\Set, & \xi = 0, \\
(b-\xi ^{-1}a )- \Set_+ , & \xi < 0,
\end{array}\right.
\end{equation}
i.e., when $\xi\neq 0$, the positive quadrant in $\Set$, shifted by $b-a/\xi$, and also reflected when $\xi<0$.  Figure~\ref{fig: example Pareto} illustrates~\eqref{F.eq} in the three possible tail regimes: $\Set^{\xi,a,b}$ is bounded below by $b - \xi^{-1}a$ when $\xi>0$,  is unbounded when $\xi=0$, and  is bounded above by $b - \xi^{-1}a$ when $\xi<0$.

In this section, $X$ denotes a stochastic process with sample paths in $\Set$ for which there exist a real number $\xi$ and  sequences $\{a_n\}_{n = 1}^\infty > 0$ and $\{b_n\}_{n = 1}^\infty$ of continuous functions {such that} the conditions for the univariate approximation~\eqref{eq: gpd} are satisfied  for each $s \in S$, i.e., 
\begin{equation}\label{eq: pot}
\lim_{n\to\infty} n\Pr\left\{\dfrac{X(s) - b_n(s)}{a_n(s)} > x\right\}= \left\{\begin{array}{lll}
  \left(1 +  \xi x\right)_+^{-1/\xi}, & \xi \neq 0,\\
 \exp\left(-x\right),& \xi = 0.
\end{array}\right. 
\end{equation}
In a functional setting it is natural to extend this by assuming the existence of a boundedly finite non-zero measure $\Lambda$ on the space of non-negative and non-zero functions over $S$ such that
\begin{equation}\label{eq:gen rv}
\left.\begin{array}{ll}
\lim_{n\to\infty}n\Pr\left[\left\{1 + \xi\left(\dfrac{X - b_n }{a_n} \right)\right\}_+^{1/\xi} \in \cdot\right] , & \xi \neq 0, \\
\lim_{n\to\infty}n\Pr\left\{\exp \left(\dfrac{X - b_n}{a_n} \right)\in \cdot\right\} , & \xi = 0,\\
\end{array}\right\} =  \Lambda(\cdot), 
\end{equation}
where $\{a\}_+ = \max\{a(s),0\}$ is a function of $s$.

Assumption~\eqref{eq:gen rv} is weak in general, and any functional model using approximation~\eqref{eq:GPD} should be linked to some limiting measure $\Lambda$.
Here we assume that $\Lambda$ is non-zero only on the set of continuous functions $\Set_+$, which rules out some types of extremal dependence; see Section~\ref{sec: limitations}.
In this case, Equation~\eqref{eq:gen rv} involves a specific type of convergence {described in the Appendices} and defines a general form of functional regular variation \citep{Hult2005} introduced by \citet{Ferreira2014};  we write  $X \in {\rm GRV}\left(\xi, a_n, b_n, \Lambda\right)$, where GRV stands for generalized regular variation.  The limiting measure $\Lambda$ is homogeneous of order $-1$: $\Lambda( t\A) = t^{-1}\Lambda(\A)$ for any positive scalar $t > 0$ and Borel set $\A\subset \Set_+$ \citep[Theorem 3.1]{Lindskog2014}.

Equation~\eqref{eq:gen rv} requires that $\xi$ is constant over $S$.  As we wish to compute the risk directly from $X$, useful limiting results are obtained only if the shape parameter is constant---if $\xi$ varies then either those locations with the highest values of $\xi$ or those with the highest upper bound determine the asymptotic tail behaviour and the limiting dependence cannot be modelled.
For environmental applications, $\xi$ can be considered as stemming from the physical process, such as convective rainfall, that is characterised by the functional $\R$.  Imposing constant $\xi$ could be avoided by transforming the data to have a common rate of tail decay throughout $S$, for example by studying the limiting $\R$-exceedances of the rescaled processes $\{1 + \xi (X-b_n)/a_n\}^{1/\xi}_{\color{red} +}$ with $\xi$ a function that varies smoothly over $S$, as in \citet{Ferreira2014}, but typically this entails losing the physical interpretation of the risk in terms of the original data.

We also suppose that there exists a sequence of real numbers $a_n'$ and a continuous strictly positive function $A$ on $S$ such that
\begin{equation}\label{eq: asym cond}
\lim_{n \rightarrow \infty} \sup_{s \in S} \left| \dfrac{a_n(s)}{a_n'} - A(s)\right| = 0,
\end{equation}
so $a_n(s) \approx a_n'A(s)$ for large $n$.  A similar assumption was used in \citet{Ferreira2012a} and \citet{Engelke} and seems reasonable in many applications. 
For instance, assuming that the marginal distributions belong to a location-scale family $F[\{x(s) - B(s)\} / A(s)]$ that describes the behaviour of the underlying physical process characterized by the risk functional~$\R$ implies both a common limiting shape parameter $\xi$ and that we can choose $a_n(s) =a_n'A(s)$ and $b_n(s) = b_n'A(s) + B(s)$ with real sequences $a_n'>0$ and $b_n'$.  We also assume that the support of $X$ is bounded below for $\xi>0$ and above for $\xi<0$.

A risk functional $\R: \Set \rightarrow \mathbb{R}$ is said to be valid for the process $X \in {\rm GRV}\left(\xi, a_n, b_n, \Lambda\right)$ if the \bRev{set of $\R$-exceedances of the limiting process has positive finite measure.
The properties needed for $\R$ to be valid depend on whether the risk is defined directly in terms of the original process $X$ or using some transformation such as $(X-b_n)/a_n$, with possibly non-constant scale and location functions $a_n$ and $b_n$.   Expressions~\eqref{eq:gen rv} and~\eqref{eq: asym cond} imply that any risk functional $\R$ applied to $(X-b_n)/a_n$ will be valid if it satisfies the condition}
\begin{equation}\label{eq: cond r}
  \begin{array}{llll} 
  \text{if } \xi > 0, \text{ then } \R(-A{\xi}^{-1}) < 0 
  , \\
\text{if }  \xi \leqslant 0, \text{ then } \R(x) \rightarrow -\infty \text{ as } x \rightarrow -\infty;
  \end{array}
\end{equation}
\bRev{the function $A$ is uniformly equal to $1$ with this rescaling.} If we further assume that $\R$ is monotonic and that there exists a positive finite real scalar $\alpha$ such that
\begin{equation}\label{eq:marginal conditions 2}
\lim_{n \rightarrow \infty} \frac{\R(a_n)}{a_n'} = \alpha, 
\end{equation}
then one can replace $a_n$ by the scalar $\R(a_n)$, i.e., apply $\R$ to $(X -b_n)/\R(a_n)$.  
In this case \bRev{$A$ may be non-constant and we can rescale $a_n'$ and $A$ so that $\alpha=1$; then~$a_n'$ can be replaced by $\R(a_n)$ in~\eqref{eq: asym cond}. We shall do this below.} Expression~\eqref{eq:marginal conditions 2} applies for instance to $1$-homogeneous functionals, for which $\R$ can be applied directly to $X - b_n$. For linear functionals, non-degenerate limits will arise when $\R$ is computed  directly on the original process.

We focus on $\R$-exceedances of  $(X - b_n) / \R(a_n)$ as it is most technical case; linear functionals are treated in Appendix~\ref{app: general risk}. Generalizations to $\R$-exceedances of $(X - b_n)/a_n$ are obtained by replacing $\R(a_n)$ by $a_n$ and removing $A$.  With the maxima taken pointwise and $x\in\Set$, we let 
\begin{equation}\label{floor.eq}
\lfloor x \rfloor = 
\begin{cases}
\max(x, -A\xi^{-1} ),& \xi>0,\\
x, & \xi\leq 0.
\end{cases}
\end{equation}
  Our first main result is the following.

%

\begin{theorem}\label{th: generalized convergence}
Let $X$ be a stochastic process whose sample paths are continuous functions on $S$.  If $u\geq 0$, the risk functional $\R$ is valid for $X \in {\rm GRV}\left(\xi, a_n, b_n, \Lambda\right)$ and~\eqref{eq: asym cond}--\eqref{eq:marginal conditions 2} hold, then
\begin{equation}\label{eq: convergence non-linear}
\Prob\left.\left[\left\lfloor\dfrac{X - b_n}{\R(a_n)}\right\rfloor \in \; \cdot \; \right| \R\left\{\dfrac{X - b_n}{\R(a_n)}\right\} \geqslant u \right] \rightarrow \Prob( P \in \; \cdot \; ), \quad n \rightarrow \infty,
\end{equation}
where  $P$ is a generalized $\R$-Pareto process with tail index $\xi$, scale {function} $A$, location function zero and measure $\Lambda$. 

\end{theorem}

Theorem \ref{th: generalized convergence} states that generalized $\R$-Pareto processes, which we discuss in Section~\ref{sec: gen Pareto}, appear as limits for any properly rescaled stochastic process $X$ that is regularly varying in the sense of~\eqref{eq:gen rv}, conditional on $\R$-exceedances of $(X - b_n) / \R(a_n)$.
Whether or not these exceedances are themselves large will depend on the choice of scaling functions $a_n$ and $b_n$. For linear risk functionals, which satisfy $\R(x + y) = \R(x) + \R(y)$ for any $x,y \in \Set$, the {conditioning event} in~\eqref{eq: convergence non-linear} simplifies to $\R(X) \geqslant u_n$ with $u_n = u\R(a_n) + \R(b_n)$, i.e., generalized $\R$-Pareto processes appear as the limit tail {processes} of increasingly large $\R$-exceedances of $X$.

The linear transformation $x \mapsto (x  - b_n) / \R(a_n)$ required in Theorem \ref{th: generalized convergence} before characterizing the {risk} is both simpler and closer to the original data than classical marginal transforms \citep{Kluppelberg2008}, as it does not modify the tail decay regime.
For homogeneous functionals and $\xi >0$, we can choose $b_n = 0$; then Theorem~\ref{th: generalized convergence} retrieves the work of \citet{Dombry2013},  which describes the limiting distribution of $X$ for increasingly high thresholding of $\R(X)$.

\subsubsection{Generalized $\R$-Pareto processes}
\label{sec: gen Pareto}

We now describe generalized $\R$-Pareto processes, give their properties, describe simulation algorithms and link them to max-stable processes.
For a given shape parameter $\xi \in \RR$ and positive function $a\equiv a(s)$, let $A = a/\R(a)$ and define the set of positive functions
\begin{equation}\label{eq: reference sets}
\Ar = \left\{ \begin{array}{ll}
\left\{ y \in \Set_+ : \R\left(A \dfrac{y^\xi- 1}{\xi}\right) \geqslant 0\right\}, & \xi \neq 0, \\
\left\{ y \in \Set_+ : \R\left(A\log y\right) \geqslant 0\right\}, & \xi = 0,
\end{array}\right.
\end{equation}
which contains possible sample paths for $P$ in~\eqref{eq: convergence non-linear} after transformation to a scale with the same marginal tail distribution.

\begin{definition}\label{def: generalized pareto process}
Let $a >0$ and $b$ be continuous functions on $S$, let $\R: \Set \rightarrow \mathbb{R}$ be a valid risk functional and let $\Lambda$ be a $(-1)$-homogeneous measure on $\Set_+$.
The generalized $\R$-Pareto process $P$ associated to the measure $\Lambda$ and tail index $\xi \in \mathbb{R}$ is the stochastic process {taking values in} $\{x \in \Set^{\xi,a,b} : \R\{(x - b)/\R(a)\} \geqslant 0\}$ and defined as
\begin{equation}
P = \left\{
\begin{array}{ll}
a(Y_\R^\xi - 1)/\xi + b, & \xi \neq 0, \\
a \log Y_\R + b, &  \xi = 0,
\end{array}
\right.
\end{equation}
where $Y_\R$ is the stochastic process on $\Ar$ with probability measure $\Lambda(\cdot) / \Lambda(\Ar)$.
\end{definition}

Generalized $\R$-Pareto processes are thus closely related to the stochastic processes $Y_\R$.  A standard approach to dependence modelling, the use of copulas,  requires that all the components of a random vector be transformed to follow uniform distributions.  Similarly,   marginal properties and dependence are typically handled separately in extreme-value modelling, with the marginal variables standardized to a distribution such as the unit Pareto.  Here we  use $Y_\R$, whose margins lie in the Fr\'echet domain of attraction with tail index $\xi=1$, as the process of reference. Other standardizations are possible, using for instance a Gumbel domain of attraction \citep[e.g.,][]{Rootzen2017}, but we focus on the Fr\'echet case to keep the exposition concise.

Following \citet{Dombry2013} and \citet{Fondeville2017}, the polar decomposition
\begin{equation}\label{eq: polar Y}
Y_\R \bRev{\Deq} RW \;\left|\; \R[A\xi^{-1}\{(RW)^\xi - 1\}]\right. \geq 0,
\end{equation}
where $R$ and $W$ are independent, the scalar $R$ is unit Pareto and $W$ is a stochastic process with state space $S$ and taking values in $\simplex= \{y \in \Set_+: \|y\|_{1} = 1\} $ with probability measure
\begin{equation}\label{eq: angular Y}
\sigma_0(\cdot)  = \dfrac{\Lambda\left\{y \in \Set_+: y / \|y\|_1 \in \cdot \;, \bRev{\|y\|_1 \bRev{\geq 1}} \;\right\}}{\Lambda\left\{y \in \Set_+: \|y\|_1 \bRev{\geq 1}\right\}},
\end{equation}
where $\|\cdot\|_1$ denotes the $1$-norm on $\Set_+$. 
The equality~\eqref{eq: polar Y} \bRev{holds in distribution, denoted by $\Deq$}, and is convenient because  it allows the Monte Carlo simulation of $W$ at a large number of locations for many common models \citep{Thibaud2013,Dombry2015}.

One desirable feature of generalized $\R$-Pareto processes is that for each $s_0 \in S$,  $P(s_0)$ has a generalized Pareto distribution after suitable conditioning: if for a threshold  $u_0 \geqslant 0$ sufficiently high that for any $t > 1$,
$$
x \in \A_{u_0} 
\Rightarrow tx \in \A_{u_0},
$$
where $\A_{u_0} = \left\{ x \in \Set^{\xi,a,b} : x(s_0) \geq u_0, \R \left\{(x - b) / \R(a)\right\} \geqslant 0\right\}$, then
\begin{equation}\label{eq: gpp marginals}
\Pr\left\{ P(s_0) > x_0  \mid P(s_0) > u_0  \right\} = \left\{1 + \xi  (x_0-u_0)/\sigma(s_0)\right\}^{-1/\xi}, \quad x_0>u_0,
\end{equation}
where $\sigma(s_0) = \R(a)A(s_0) + \xi \{u_0 - b(s_0)\}$. 
 Unfortunately there is no simple general expression for the distribution of $\R\{(P - b)/\R(a)\}$, but if necessary it can be estimated using Monte Carlo methods.  
If the risk functional is linear, generalized $\R$-Pareto processes also admit a pseudo-polar decomposition and the distribution of the risk $ \R(P)$ above $\R(b)$ is  generalized Pareto, with shape and scale parameters $\xi$ and $\R(a)$ (Appendix~\ref{app: gen r pareto marginals}). 
In univariate extreme-value theory the marginal assumptions of equation~\eqref{eq: pot} are equivalent to convergence of rescaled block maxima toward the generalized extreme value (GEV) distribution. There is a similar relation between generalized $\R$-Pareto processes and the functional extensions of GEV variables known as max-stable processes; see Appendix~\ref{sec: ms process}.

\subsubsection{Simulation}\label{sec: sims}
 The pseudo-polar decomposition~\eqref{eq: polar Y} is key to the construction of generalized $\R$-Pareto processes and to their simulation.  Simple algorithms to draw samples from $Y_\R$ are available for risk functionals such as $\R_1(x) = \|x\|_1$ or $\R_2(x) = \sup_{s \in S} x(s)$; see \citet{Asadi2015}, for example.
 We generalize the principle of \citet[Section~2.3]{Fondeville2017} to develop an accept-reject algorithm for the generalized $\R$-Pareto process when $\xi \neq 0$; modification for $\xi = 0$ is straightforward.
If we can find a threshold $u > 0$ such that
\begin{equation}\label{eq: sim cond 1}
\Ar \subset \left\{y \in \Set_+ : \|y\|_{1} \geqslant u \right\},
\end{equation}
then Algorithm~\ref{algo1} enables simulation of $P$ when an algorithm for $Y_\R$ with $\R(\cdot) = {\|\cdot\|_1}$ is available. In the algorithm, every unit Pareto variable is independent of every other and has distribution function $1-1/v$ for $v\geq 1$. Its efficiency is determined by the capacity to find the largest possible $u$, $u_{\sup}$, say, such that~\eqref{eq: sim cond 1} is satisfied, and its acceptance rate is the ratio of the measures of the two sets in~\eqref{eq: sim cond 1}. Simulated generalized $\R$-Pareto processes on $[0,1]$ are displayed in Figure~\ref{fig: example Pareto} for three different risk functionals: for $\xi>0$, exceedances are defined as positive values of $\sup_{s \in [0,1]} x(s) - b(s)$;  for $\xi=0$, they are functions that are large at $s_0 = 0.1$; and for negative tail index exceedances are functions with exceptionally high integrals over $[0,1]$. When the risk functional is linear, an alternative algorithm in Appendix~\ref{app: sim linear} allows simulation of such processes with a pre-determined risk $\R(P)$.

\begin{algorithm}
\SetAlgoLined
\textbf{Inputs}: scaling functions $a$, $b$, $A$, threshold function $u$, and scalar shape parameter $\xi$.

 Set $Y_\R = 0$;
 
  \While{$\R[A\xi^{-1}\{(Y_{\R})^{\xi} - 1\} ] < 0$}{
  generate a unit Pareto random variable  $R$\;
  generate $W$ on $\simplex$ with probability measure $\sigma_0$ in~\eqref{eq: angular Y}\;
  set $Y_{\R} = uRW$;
 }
 Set $P = a\xi^{-1}\{(Y_{\R})^{\xi} - 1\} + b$;
 \caption{Simulation of generalized $\R$-Pareto process, $P$}
 \label{algo1}
\end{algorithm}

\subsection{Limitations on the asymptotic dependence regime}\label{sec: limitations}

The derivations above presuppose  the existence of a limiting measure $\Lambda$ in~\eqref{eq:gen rv} with non-zero mass on the space of continuous functions. This assumption precludes asymptotic independence \citep{Trust2017} throughout $S$, but mixed regimes, in which asymptotic independence replaces asymptotic dependence at distances greater than some finite radius, are possible. The methodology could be extended to asymptotic independence throughout $S$ by assuming positivity of $\Lambda$ on more general functional spaces.  For instance, asymptotic independence would be covered by considering measures that place positive mass on the set of functions that are zero everywhere except at a specific location.  The study of such functional spaces would require more general notions of convergence than in \citet{Hult2005} and has not been undertaken, so far as we know.

The positivity of $\Lambda$ on $\Set_+$ implies homogeneity of order $-1$, i.e., dependence at `low' levels of intensity is extrapolated further into the tail.
In practice this implies that the average size of a region on which the threshold is locally exceeded is independent of the intensity of the event.  Decreased dependence at high intensities has, however, been observed in numerous environmental phenomena, for which the asymptotic models described in this paper may over-estimate dependence for high intensities.  
Sub-asymptotic models with decreasing dependence have recently been investigated \citep{Huser2019}, but they correspond to processes that are asymptotically independent throughout $S$ and thus may under-estimate dependence at extreme levels especially for close-by locations.
In general, the choice of asymptotic dependence regime should be determined by the investigator's tolerance of risk. 
Asymptotically dependent sub-asymptotic models fitting into the above framework could provide more realistic alternatives than asymptotic models, but do not yet exist, so far as we are aware.  The current methodology is for now the only functional approach for risk-averse policy makers:
simulations of generalized $\R$-Pareto processes provide scenarios whose extent may be over-estimated but that can be fed into impact models to assess the potential for damage to  infrastructure.

\cite{Wadsworth2019} propose an approach that encompasses both asymptotic independence and asymptotic dependence, can be applied in high dimensions and involves conditioning on the process being extreme at any one of a number of locations, but is based on \citet{Heffernan.Tawn:2004} and thus does not construct an overall statistical model for the data. 

\section{Functional peaks-over-threshold modelling}\label{sec: model form}

We now describe a general approach to modelling $\R$-exceedances over a high threshold.
Theorem~\ref{th: generalized convergence} suggests that in principle the choice of risk functional should not impact the model parameters, but in practice it affects what events are considered extreme, especially in the presence of a mixture in the tail behaviour, as illustrated by Figure~\ref{fig: rainfall observations}. 
The choice of \bRev{the} risk functional allows the user to focus on one component of a possible mixture by incorporating domain-specific expertise, while improving sub-asymptotic behaviour by fitting the model using only those observations most relevant to the chosen type of extreme event.

Suppose we have a {valid risk functional} $\R$ whose exceedances occur for a single physical process, such as cyclonic rainfall, and that for such events it is reasonable to use a uniform tail index $\xi$.
More specifically, let $X \in {\rm GRV}\left(\xi, a_n, b_n, \Lambda\right)$ and suppose that the marginal distributions of $X(s)$ form a location-scale family with continuous positive scale function $A(s)$, continuous real location function $B(s)$, 
and distribution function  $F$ satisfying equation~\eqref{eq: mda} with real-valued sequences $a_n'>0$ and $b_n'$.  If so, the normalizing functions $a_n(s)$ and $b_n(s)$ for $X(s)$ satisfy
\begin{equation}\label{eq: loc scale fam}
a_n(s) = A(s)a_n', \qquad b_n(s) = B(s) + A(s)b_n', \quad s \in S,
\end{equation}
 yielding the asymptotic decomposition implied by~\eqref{eq: asym cond}.

We impose a parametric structure on the extremal dependence of $X$ and on the marginal scale and location functions $A$ and $B$, which are assumed to belong to parametric families of functions $A_{\theta_A}$ and $B_{\theta_B}$.  The limiting measure $\Lambda_{\theta_W}$ is supposed to be parametrised by the distribution of $W$, which depends on parameters $\theta_W$.

The dependence properties of the limiting generalized $\R$-Pareto process are determined by the angular process $W$, which takes values in $\simplex= \{y \in \Set_+: \|y\|_{1} = 1\} $.  
To characterize and compare angular process models, we need a measure of dependence, but classical measures such as the covariance function or the semi-variogram 
$$
\gamma(h) = \dfrac{1}{2} \text{var}\{X(s')- X(s)\}
$$
rely on the existence of  moments and may be undefined in our setting.
A more suitable dependence measure is  \citep{Fondeville2017}
\begin{equation}\label{eq: cond prob}
\pi_\R(s',s) = \lim_{q \rightarrow 1}\Pr\left[X(s') > u_q(s') \mid \{X(s) > u_q(s)\} \cap \{\R(X) \geqslant u\} \right], \quad s,s'\in S,
\end{equation}
where $u_q(s)$ denotes the $q$ quantile of $X(s)$ and $u\geq 0$.
Equation~\eqref{eq: cond prob} summarizes the pairwise extremal dependence between $X(s)$ and $X(s')$; it extends the extremogram \citep{Davis2009} to $\R$-exceedances and generalizes the extremal dependence coefficient $\chi$ \citep{Trust2017} to processes. Expression~\eqref{eq: cond prob} matches the  extremogram for high enough $q$, i.e., if  the risks $\R(X)$ for all those $X$ for which $X(s) > u_q(s)$ and $X(s') > u_q(s')$ also exceed $u$, then the additional condition $\R(X) \geqslant u$ in~\eqref{eq: cond prob} has no theoretical impact, though in practice it allows one to disentangle tail mixtures and thus to identify any differences in tail dependence regimes.
Although other dependence measures exist \citep{Smith1990, Cooley2006}, we prefer $\pi_\R$ for its interpretability.

\begin{figure}[!t]
\begin{center}
\begin{tabular}{cc}
 \includegraphics[scale = 0.25]{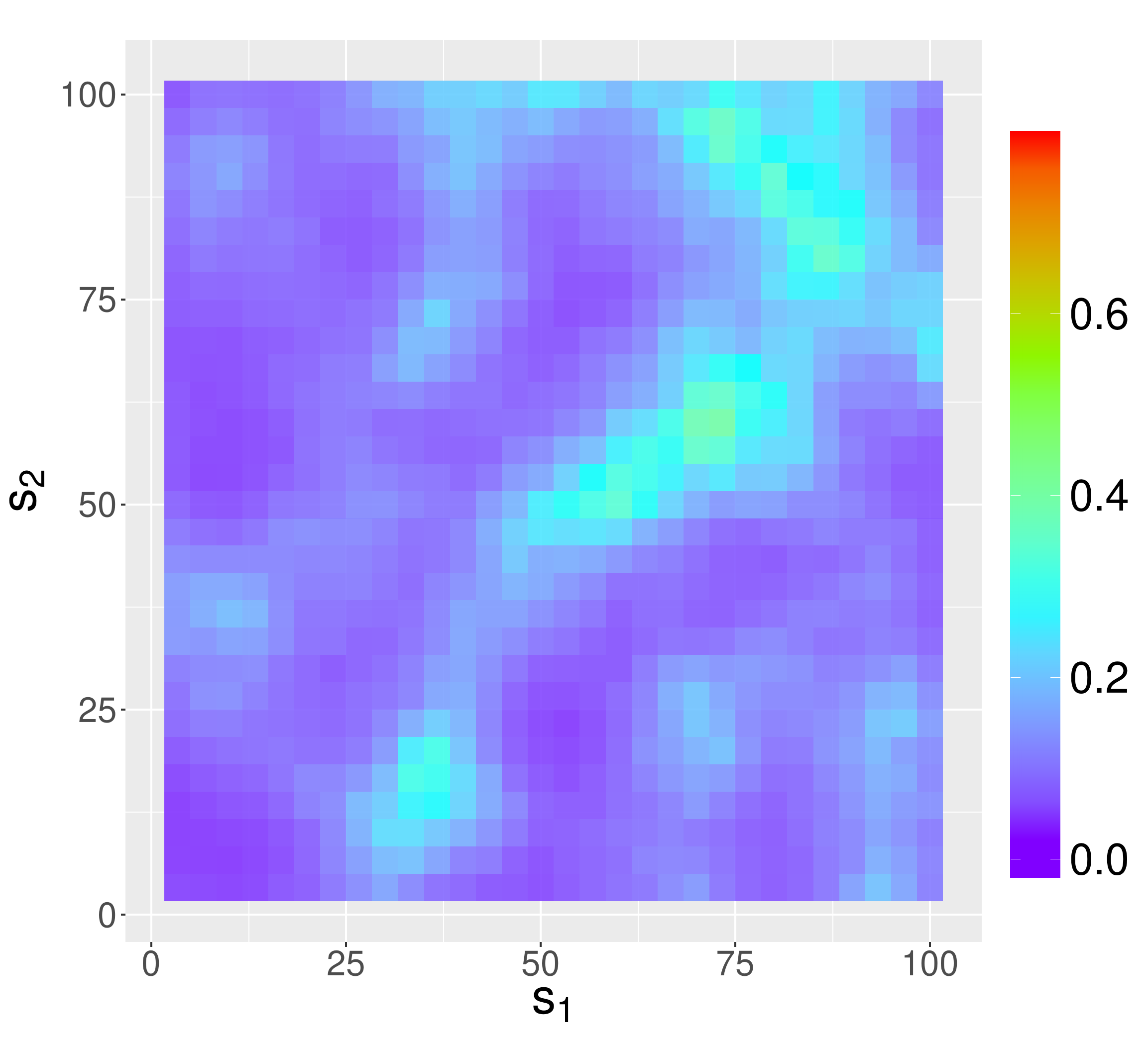} & \includegraphics[scale = 0.25]{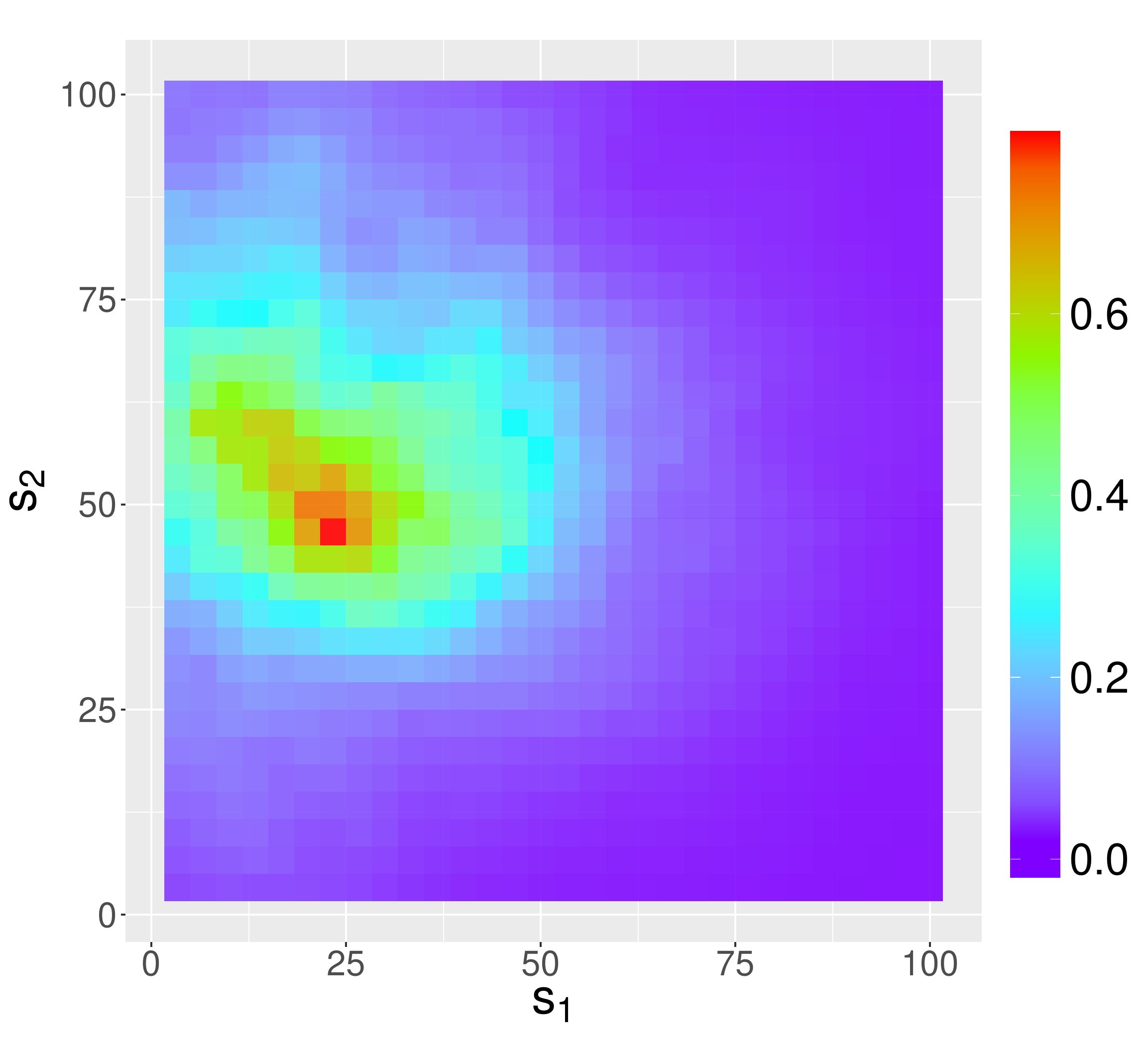} \\
 $\gamma(h) = 0.5 \left[ 1 - \exp\{- ( \|h\| / 15)^{1.8}\} \right]$ & $\gamma(h) = ( \|h\| / 30)^{1.8} $
\end{tabular}
\end{center}
\caption{Simulated generalized $\R$-Pareto processes with $\R(X) = \int_S X(s) ds = 100$ for two semi-variogram functions $\gamma(h)$. Left: bounded power-exponential semi-variogram function. Right: unbounded power variogram.}
\label{fig: properties vario}
\end{figure}

The literature on max-stable processes suggests several parametric models for $W$. 
The Gaussian extreme value process  \citep{Smith1990} relies on deterministic Gaussian kernels randomly shifted in space and is attractive for its computational tractability and relative simplicity, but it yields unrealistic random fields.  Under the Brown--Resnick~(\citeyear{Brown1997})  model the angular process  $W$ is a log-Gaussian random function whose underlying Gaussian process has stationary increments and semi-variogram $\gamma$. In this case,~\eqref{eq: cond prob} reduces to
$$
2\left( 1 - \Phi\left[\left\{{\gamma(h)}/{2}\right\}^{1/2}\right]\right),
$$
where $h=s-s'$ and $\Phi$ denotes the standard normal cumulative distribution function. 
The Brown--Resnick model is particularly attractive because many semi-variogram functions available in the spatial statistics literature furnish models for extremal dependence. 
The behaviour of $\gamma$ as $h\to 0$  determines the smoothness of the generalized $\R$-Pareto process and its behaviour as $h\to\infty$ determines the extremal dependence regime.
Indeed, if the semi-variogram is bounded, as is the case for strictly stationary Gaussian processes, then $\pi_\R(h) > 0$ for any $h >0$, whereas if  $\gamma$ is unbounded then we obtain near-independence,  $\pi_\R(h) \rightarrow 0$,  for large $h$; see Figure~\ref{fig: properties vario}.
Use of a log-Gaussian $W$ implies that for any linear $\R$, $\Lambda(\partial\A_\R) = 0$, where $\partial\A_\R$ is the boundary of the set $\A_\R$ defined in~\eqref{eq: reference sets}.

An alternative model, for which {$\Lambda(\partial\A_\R) \neq 0$}, is the extremal-$t$ process \citep{Opitz2012}
$$
W(s) \propto \max\{G(s),0\}^{\nu}, \quad s \in S,\nu >0, 
$$
where $G$ is a strictly stationary Gaussian process with correlation function $C$.  The maximum in this definition induces non-zero measure on the boundary of $\Set^{\xi,a,b}$, making the model improper when $\xi < 0$, as then $\Pr\{X(s)  = -\infty\} > 0$.
Its extremogram, 
$$
2 \left(1 - t_{\nu+ 1}\left[(\nu + 1)^{1/2}\left\{\frac{1 - C(h)}{1 + C(h)}\right\}^{1/2} \right] \right),
$$
must be at least $2 \left[1 - t_{\nu+ 1}\left\{(\nu + 1)^{1/2} \right\} \right]$ for positive correlation functions, so when $\nu$ is low, the model can only produce strong dependence. This limitation weakens as $\nu$ increases; then the model approaches the Brown--Resnick model, which is often preferred for this reason. 

In the next section we describe an approach to joint inference on the complete parameter vector $\vartheta = (\xi, a_n', b_n', \theta_A, \theta_B, \theta_W)$.
Identifiability issues that may arise with the parametric models for $A$ and $B$  can be solved for instance by ensuring that $\R(A) = 1$ and, $\R(B) = 0$ for a linear functional; see \citet{Engelke}, for example.

\section{Statistical inference}\label{sec: stat inference}

In this section, we suppose for simplicity of exposition that the risk functional is linear; inference for general risk functionals essentially involves replacing the process $X$ by the shifted and scaled version $(X-b_n)/\R(a_n)$, with some further additional minor changes. Difficulties that might arise for general functionals are discussed in Appendix~\ref{app: stat inf details}.

Statistical inference for $\R$-exceedances of a stochastic process $X \in {\rm GRV}\left(\xi, a_n, b_n, \Lambda\right)$ is based on the approximation
\begin{align}\label{eq: approx r excess} 
\Prob\left(X \in \calR\right)  \nonumber & =  \Prob\left\{\R\left(X\right) \geqslant u_n\right\} \times \Pr\left\{ X \in \calR \mid \R\left(X\right) \geqslant u_n\right\},  \nonumber \\ & \approx  
 \Prob\left\{\R\left(X\right) \geqslant u_n\right\} \times   \Pr\left(P \in \calR \right),
\end{align}
where $\calR \subset \calR(u_n) = \{x \in \Set^{\xi,a_n,b_n} : \R(x) \geqslant u_n\}$ and $u_n=r(b_n)$ is a high quantile of $\R(X)$.

Let $x_1,\ldots, x_n \in {\Set}$ be independent realizations of a generalized regularly varying stochastic process $X$ observed at locations $s_1,\dots, s_L \in S$.
The log-likelihood function for~\eqref{eq: approx r excess}  based on the $\R$-exceedances over the threshold $u_n$ among $x_1,\ldots, x_n$ is
\begin{equation}\label{eq: gen pareto log lik}
\mathcal{L}_{\text{Thres}}(\vartheta) = \sum_{j \in K_{u_n}} \log \Prob\left\{\R\left(x_j\right) \geqslant u_n; \vartheta\right\}  + \sum_{j \in K_{u_n}} \log f^\R\left(x_j;\vartheta\right),
\end{equation}
where $K_{u_n}=\{j \in 1,\dots,n : \R(x_j) \geqslant u_{n}\}$ contains the indexes of the $n_{u_n}$ $\R$-exceedances over $u_n$, and $f^\R$ denotes the finite-dimensional density function of a generalized $\R$-Pareto process observed at  $s_1,\dots s_L$, i.e., 
\begin{equation}\label{eq: simple gen pareto dist}
\dfrac{\lambda_{\theta_W}\left(\left[ 1+ \xi \{x(s_{1:L}) - b_n(s_{1:L})\}/{a_n(s_{1:L})}\right]_+^{1/\xi}\right)}{\Lambda_{\theta_W}\left\{\Ar\right\}} \prod_{l = 1}^L a_n(s_l)^{-1}\left\{1 + \xi \dfrac{x(s_l) - b_n(s_l)}{a_n(s_l)}\right\}_+^{1/\xi - 1} ,
\end{equation}
where {$x(s_{1:L}) = \{x(s_1),\dots,x(s_L)\}$} and the $L$-dimensional intensity function $\lambda_{\theta_W}$ is given by 
\begin{equation}\label{eq: mvt intensity}
\Lambda_{\theta_W}\{\exceedanceSets_{\max}(z)\} = \int_{\RR^L \setminus (0,z]^L}  \lambda_{\theta_W}(y)\, \D{y}
\end{equation}
with $\exceedanceSets_{\max}(z) = \{y \in \Ar : \max_{ l =1,\dots,L} y(s_l)/z(s_l) \geq 1 \}$.  The second term of~\eqref{eq: simple gen pareto dist} is the Jacobian for the marginal transformations from the generalized Pareto scale used for the data to the unit Fr\'echet scale on which the dependence model is defined. 

A model for the probabilities that $\R\left(x_j\right) \geqslant u_n$ must be specified.  In similar contexts \cite{Wadsworth2013a} and \citet{Engelke2012b} use a Poisson distribution suggested by the relationship with block maxima,  which yields log-likelihood 
\begin{align}\label{eq: likh pp}
\mathcal{L}_{\rm Poiss}(\vartheta) = n_u \log \Lambda_{\theta_W}\left(\A_\R\right) - \Lambda_{\theta_W}\left(\Ar\right) +   \sum_{j \in K_{u_n}}  \log f^\R \left(x_j; \vartheta\right),
\end{align}
when the exceedance events are identically  distributed, 
but the Pareto methodology accommodates other possibilities.  \citet{Thibaud2013}, for instance,  suppose that the corresponding random variable $N_u$ is fixed and use a binomial distribution, which 
is easily linked to the Poisson point process model.
Such approaches presuppose that the probability of observing an exceedance does not depend on explanatory variables, but if it does then logistic regression could be used to model the probability of observing an extreme event; see Section~\ref{sec: strom and freq}.

Maximization of~\eqref{eq: gen pareto log lik} or~\eqref{eq: likh pp} can be difficult and we recommend first estimating the marginal parameters $\xi$, $a_n'$, $A$, $b_n'$ and $B$ and then fitting a dependence model by fixing them at their estimates. The marginal parameters can be estimated by maximizing the independence log-likelihood,
\begin{align}\label{eq: indep likh}
&\sum_{j=1}^{n}  \sum_{l = 1}^L \indicatorFun\{x_{j}(s_l) \geqslant b_n(s_l), \R(x_j) \geqslant u_n\}  \log \Pr\{x_{j}(s_l) \geqslant b_n(s_l)\} \times \nonumber\\
& \quad\quad\quad\quad \log \left[\dfrac{1}{a_n(s_l)} \left\{1 + \xi \frac{x_{j}(s_l) - b_n(s_l)}{a_n(s_l)} \right\}_+^{-1/\xi - 1}\right],
\end{align}
under the constraint $\R(b_n) = u_n$, with parameter uncertainty assessed by resampling the $x_j$. 
Any other inference procedure allowing a common value of $\xi$ could be used instead. 

One way to estimate the dependence parameters is to minimise the function
\begin{equation}
\label{ell_ext.eq}
\sum_{l,l' = 1,\dots, L} \left\{\widehat{\pi}(s_{l'},s_l) - \pi_{\theta_W}(s_{l'}, s_l)\right\}^2,
\end{equation}
where $\widehat{\pi}$ denotes an estimate of~\eqref{eq: cond prob}, such as that obtained by replacing exceedance probabilities by the corresponding frequencies \citep{Davis2013}, 
$$
\widehat{\pi}(s_{l'}, s_l) = \dfrac{\sum_{j = 1}^n\indicatorFun\{x_{j}(s_{l'}) \geqslant b_n(s_{l'}), x_{j}(s_l) \geqslant b_n(s_l), \R(x_j) \geqslant u_n\}}{ \sum_{j = 1}^n \indicatorFun\{x_{j}(s_l) \geqslant b_n(s_l), \R(x_j) \geqslant u_n\}}.
$$
This approach is robust and can be tailored to the situation at hand, for example by weighting summands to improve spatial prediction at ranges of particular interest or to reduce the computational burden when $L$ is very large.  It ensures that the fitted model has the same average number of locations jointly exceeding the location function $b_n$ as in the data, but uncertainty quantification for the resulting estimates typically involves resampling and may be time-consuming, though this allows uncertainty for both marginal and dependence aspects to be readily combined.

Maximum likelihood estimation of  $\theta_W$ has been been studied for specific risk functionals but can perform poorly because the limiting process is misspecified for finite $u_n$ \citep{Engelke2014, Huser}.
Alternatives involve censoring of low components \citep[e.g.,][]{Wadsworth2013a}, composite likelihoods \citep{Padoan2010, Huser2013a,Castruccio2014} or M-estimation using pairwise tail indexes \citep{Einmahl2016,Einmahl2016a}.
All are more robust to mis-specification but can be used only for specific risk functionals and are dimensionally limited, either by the computational burden due to the numerical evaluation of the normalising constant {$\Lambda_{\theta_W}\left(\A_\R\right)$} and the censoring, or, for pairwise procedures, by combinatorial considerations.
Efficient algorithms for censored likelihood are available \citep{Fondeville2016} and tractable for {$L$ up to a few hundred} for the Brown--Resnick and extremal $t$ models.  
Gradient scoring  \citep{Fondeville2017} can be applied to a large class of risk functionals and avoids the computation of {$\Lambda_{\theta_W}\left(\A_\R\right)$}, making inference tractable for {$L$ in the thousands}; for log-Gaussian random functions, its numerical complexity is that of matrix inversion.
These approaches could also be used to estimate the entire parameter vector $\vartheta$ simultaneously, thereby allowing a full quantification of the uncertainties, for instance by resampling.
More details about gradient scoring can be found in Appendix~\ref{app: score matching}.

\section{Model validation}\label{sec: model validation}

Suppose that we have an estimate $\widehat{\vartheta}$ of the parameters and a measure of its uncertainty and we wish to check the quality of the fitted model.

The marginal tail behaviour at each sampled location $s_1,\dots,s_L$ can be checked by comparing the observations with the fitted marginal model.  
If $u_q(s_l)$ denotes the empirical $q$ quantile of the $\R$-exceedances at $s_l$, estimated using only observations $x_j$ for which $\R(x_j) \geqslant u_n$, where $q$ has been chosen such that~\eqref{eq: gpp marginals} holds, and if $n_q$ denotes the number of $x_j$ exceeding $u_q(s_l)$, then we can check the marginal fits using the approximation
$$
\Pr\left\{X(s_l) - u_q(s_l) \geqslant x \mid X(s_l) \geqslant u_q(s_l) \right\} \approx H_{\widehat{\xi},\widehat{\sigma}(s_l)}(x), \quad x \geqslant 0,
$$
with {$\widehat{\sigma}(s_l) = \widehat{a_n}(s_l) + \widehat{\xi}\{u_q(s_l) - \widehat{b_n}(s_l)\}$}.
Pointwise confidence intervals for quantile-quantile plots can be obtained by resampling: we draw $m$ samples of size $n_q$,  $(Z_1^1, \dots, Z_{n_q}^1)$, $\dots, (Z_1^{m}, \dots, Z_{n_q}^{m})$ from the fitted distribution and let $Z_{(j)}^1,\dots,Z_{(j)}^{m}$ denote the $j${th} order statistic of each sample.
A $95\%$ confidence interval for the generalized Pareto fit is then defined as the $2.5$ and $97.5$ empirical percentiles of $Z_{(j)}^1,\ldots, Z_{(j)}^m$.
When the estimator used to obtain $\widehat{\vartheta}$ is asymptotically normal, estimation uncertainty can be taken into account to some extent by drawing the $m$ samples from different generalized Pareto distributions whose parameters $(\xi,\log\sigma)$ are normally distributed with mean $(\widehat{\xi},\log \widehat{\sigma}(s_l))$ and covariance matrix corresponding to the uncertainty of $\widehat{\vartheta}$.
When the risk functional is linear, a similar marginal check can be performed for the exceedances of $\R(X)$; see for instance Figure~\ref{fig: qq plots margins storm}.

The dependence model can be assessed by comparing the fitted extremogram with the corresponding empirical values of~\eqref{eq: cond prob}.
If the model is stationary and isotropic, then $\pi$ depends only the distance $h$ between two locations, and $\pi$ can be plotted as a function of the distance, and if relevant, the orientation, of pairs of locations.  For an anisotropic model  it is preferable to map how the dependence varies with the spatial coordinates, as in Figure~\ref{fig: dependence mode and checking}.  More general dependence measures based on aggregation \citep{Engelke} could also be considered.

Model comparison can be performed using the Akaike or composite likelihood information criteria \citep{Davison2011}, and formal comparison of nested models can be based on scoring rules \citep{Dawid2014, Fondeville2017}.
A relative root mean squared error or the continuous ranked probability score \citep{Gneiting2007b} can be used to assess the predictive performance of the model.
If $S$ has a temporal component, then an empirical probabilistic forecast is available by simulating from the fitted model at future times conditioned on currently available observations.  When the angular process is log-Gaussian, this is equivalent to conditional simulation of a Gaussian process, followed by a marginal transformation. 

We illustrate the application of these ideas in the next two sections.

\section{Modelling extreme European windstorms}\label{sec: windstorms}

\subsection{Motivation}

\begin{figure}[!t]
\begin{center}
\begin{tabular}{ccc}
\includegraphics[scale = 0.26]{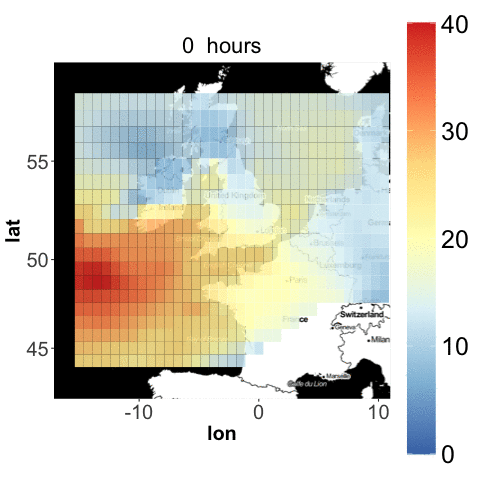} & \includegraphics[scale = 0.26]{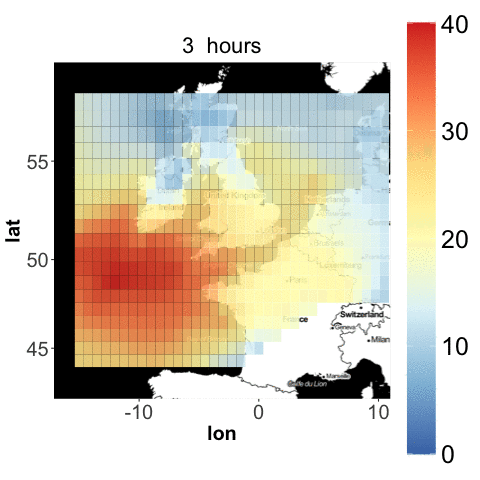} & \includegraphics[scale = 0.26]{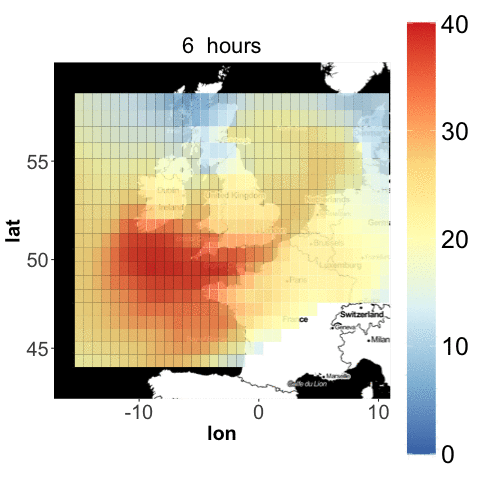} \\
\includegraphics[scale = 0.26]{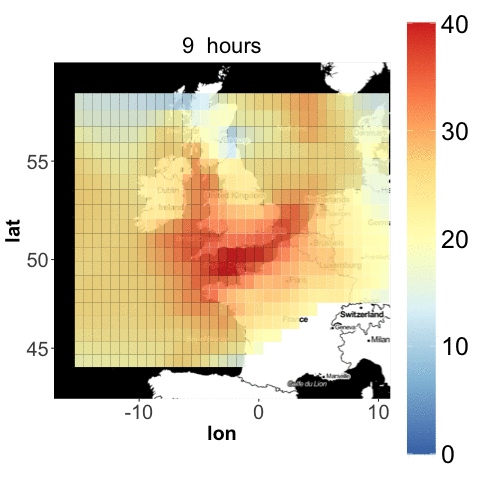} & \includegraphics[scale = 0.26]{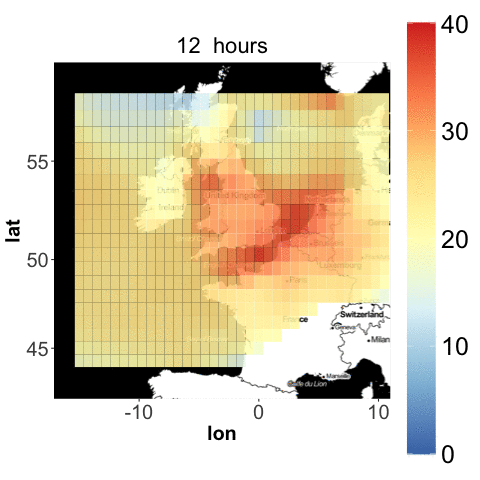} & \includegraphics[scale = 0.26]{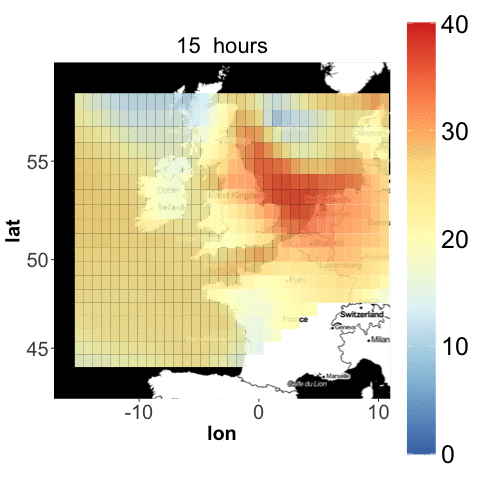} \\
\includegraphics[scale = 0.26]{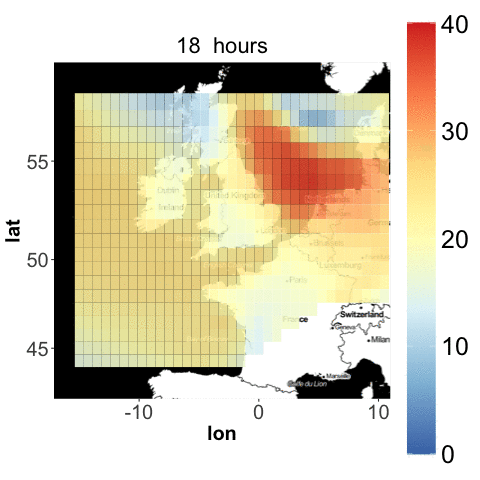} & \includegraphics[scale = 0.26]{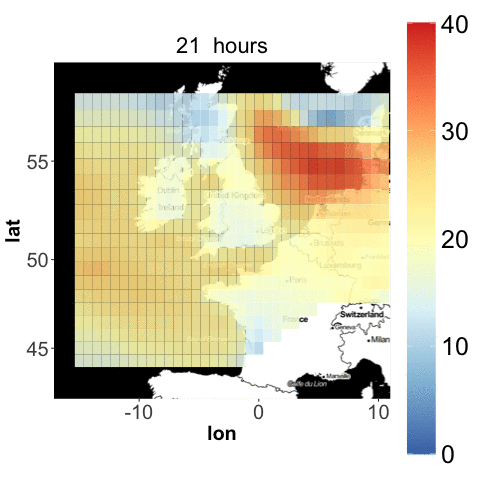} & \includegraphics[scale = 0.26]{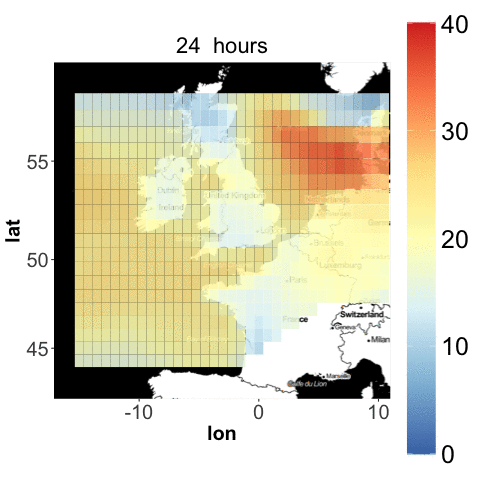}
\end{tabular}
\end{center}
\caption{Maximum speed (ms$^{-1}$) over the past $3$ hours of the wind gusts sustained for at least $3$s from ERA-Interim reanalysis during the peak of windstorm Daria, which swept over Europe during January $1990$.}
\label{fig: daria}
\end{figure}

One of the severest extra-tropical cyclones ever observed, windstorm Daria, struck the United Kingdom on $25$ January $1990$.  Over that day and the next, $97$ deaths were reported and damage valued at around $8.2$ billion US dollars occurred. The strongest measured gusts were $47.2$ ~ms$^{-1}$, equivalent to a category~$1$ hurricane. Figure~\ref{fig: daria} shows the maximum speed over  three-hour intervals of the wind gusts sustained for at least $3$s for the $24$ hours during which the storm peaked over the UK.  To give an idea of the severity of this storm, damaging windspeeds are considered to start at $25$ ~ms$^{-1}$ \citep{Roberts2014}.
About ten years later, on $26$ December 1999, storm Lothar swept across western and central Europe.  A wind speed of $46.9$ ~ms$^{-1}$ was recorded in Paris, and the weather station at the summit of `La Dole' in Switzerland recorded a maximum wind gust of $55.9$ ~ms$^{-1}$.  Lothar, equivalent to a category $2$ hurricane, caused losses of $8$ billion US dollars and more than $100$ deaths.

These two events illustrate why estimating the risk linked to such natural hazards has become a major question in recent decades, especially as the possible influence of global warming on them is far from understood.

\subsection{Risk estimation for extreme windstorms}

Risk estimation for extreme windstorms has generally been limited to the use of historical catalogues of events to test the resilience of infrastructure \citep{Haylock2011, Pinto2012}, but unfortunately such storms are rare and the catalogues usually span only a few decades. Further events can be generated by statistical perturbation of the wind field intensity, shape and location \citep{Hall2008} or by detecting storms in multiple numerical climate outputs \citep{Della-Marta2010}.  In both cases the same storms may be re-cycled but with differing climatological indexes because of different hypotheses and approximations used by the models. \citet{Yiou2014} proposed creating new storms from historical catalogues by reordering time steps based on spatial analogues. Uncertainties and bias linked to all these approaches may be large and difficult to estimate, and studies on climatological projections have stressed their inability to accurately reproduce extreme events \citep[e.g.,][]{Weller2013}.  All these methods generate storms whose tail behaviour cannot be extrapolated to still rarer events.  

Extreme value theory was applied to the problem by \citet{DellaMarta2009} and by \citet{Mornet2016}, who performed a POT analysis on univariate summaries characterizing extreme windstorms, but did not model spatial dependence. \citet{Ferreira2014} suggest how historical windstorm records might be up-scaled to higher intensities using Pareto processes, but their approach cannot generate new storms. \citet {Economou2009} used Bayesian hierarchical models of extra-tropical cyclones, but included dependence using covariates such as mean sea level pressure, which limits the capacity of the model to generate new patterns and intensities. The existing work closest to ours is by  \citet{Sharkey.Tawn.Brown:2020}, who use a Lagrangian approach to model the tracks and severity of European windstorms.  Their model for storm tracks is more detailed than ours, but their dependence structure uses a non-extremal model and neglects the temporal element.

We propose an approach based on generalized $\R$-Pareto processes, which extends the \citet{DellaMarta2009} approach to allow not only local risk estimation but also the generation of new extreme storms that are spatially and temporally consistent. 

\subsection{Data set and region of study}

To build our stochastic weather generator, we follow the methodology of the extreme windstorms (XWS) catalogue \citep{Roberts2014}, which provides historical records of the $50$ most extreme storms over Europe for winters from $1979$ to $2014$; more precisely it contains maps of $72$-hour maximum wind gusts over northern Europe. In this catalogue, the `extreme storms' are chosen to focus on events with high impact on infrastructure; indeed, the storms with the highest maximum wind speeds may not cause the most damage overall unless they cross inhabited areas.  To apply our methods we must define univariate summaries that characterise the most damaging events. 

\bRev{The XWS catalogue tracks storms in the ERA-Interim reanalysis \citep{Dee2011}, a real-time climate model whose records start in $1979$ and that provides time series for many climatological indexes. The model is run every six hours on a grid whose cells are squares with sides that can be chosen between $3\degree$ and $0.125\degree$; the native size is $0.75\degree$ and other resolutions are obtained by interpolation.
In addition to the 6-hourly fields obtained by data assimilation, which constrains the grid values to station measurements, $256$-hour forecasts are generated each day at $00$UTC and $12$UTC. These forecasts can be combined with the assimilated data to obtain a three-hourly database of the maximum speed of the wind gusts sustained for at least $3$s, as shown in Figure~\ref{fig: daria}. Most European winter storms evolve quickly and last only for a day or so, so fine time-resolution is necessary. The link between reanalysis outputs and station measurements is unclear, so for simplicity we treat observations as instantaneous rather than as temporal aggregates.}

\begin{figure}[!t]
\begin{center}
\begin{tabular}{ccc}
\raisebox{0.1\dimexpr\totalheight-\ht\strutbox}{\includegraphics[width = 0.25\textwidth]{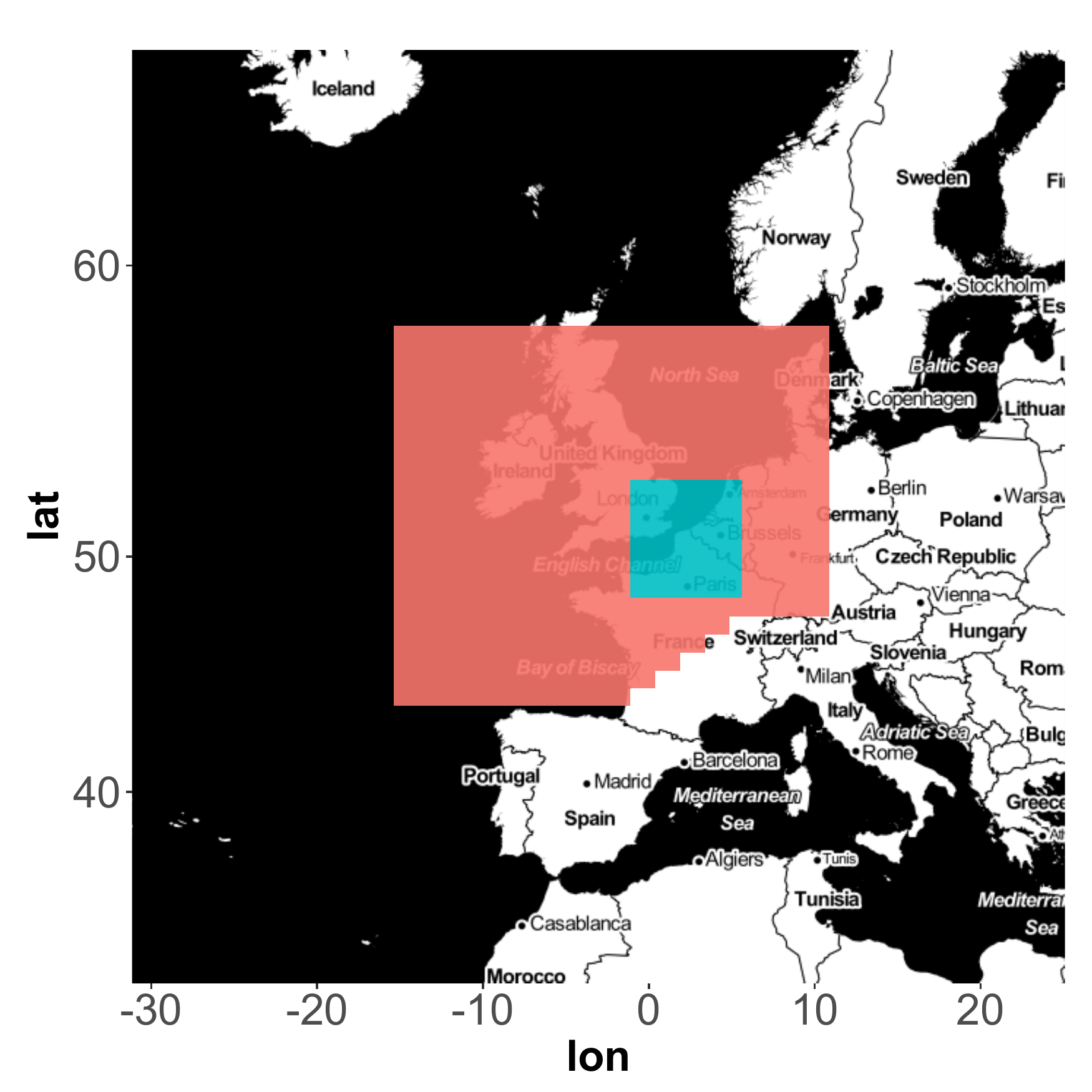}}&
\includegraphics[width = 0.3\textwidth]{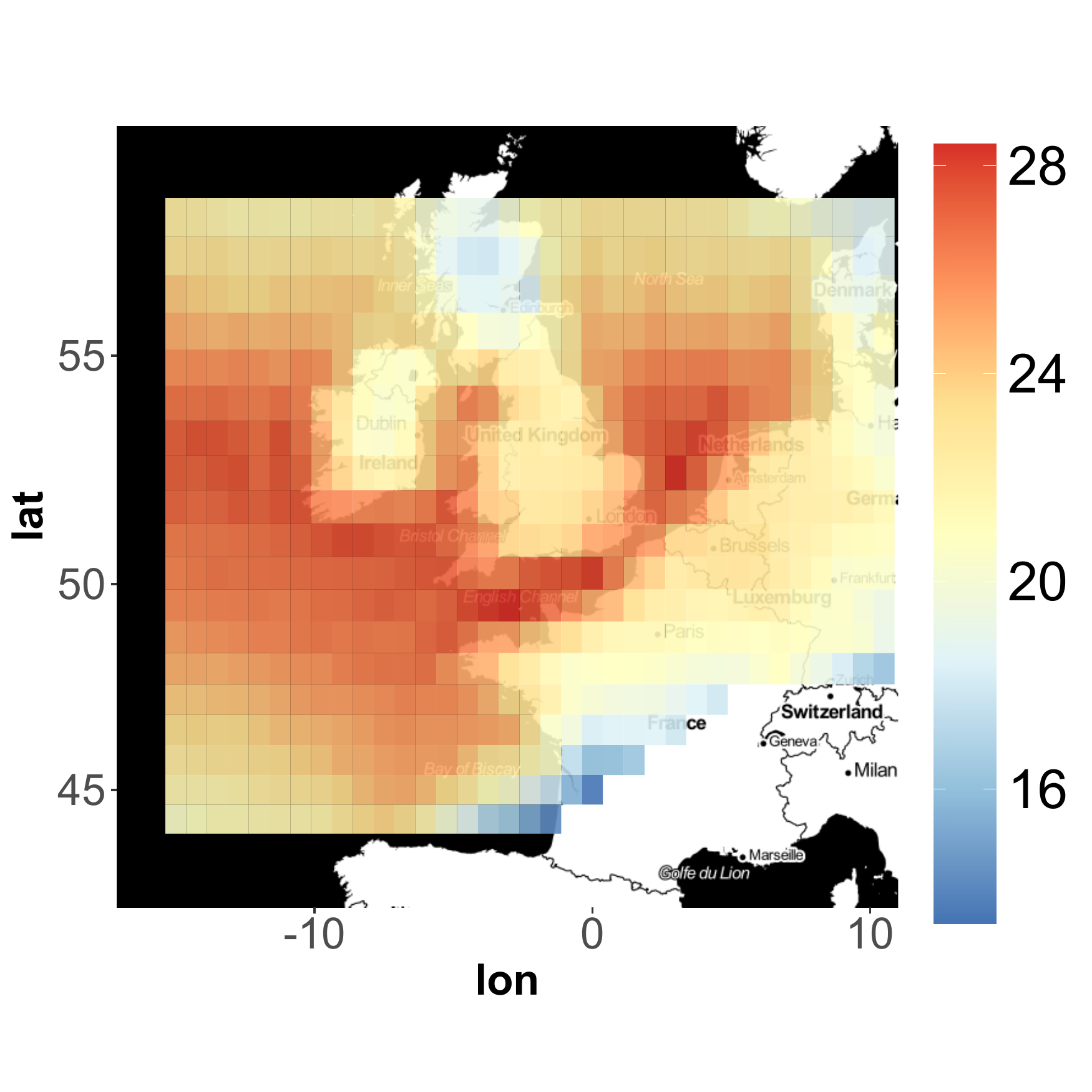} &
 \includegraphics[width = 0.3\textwidth]{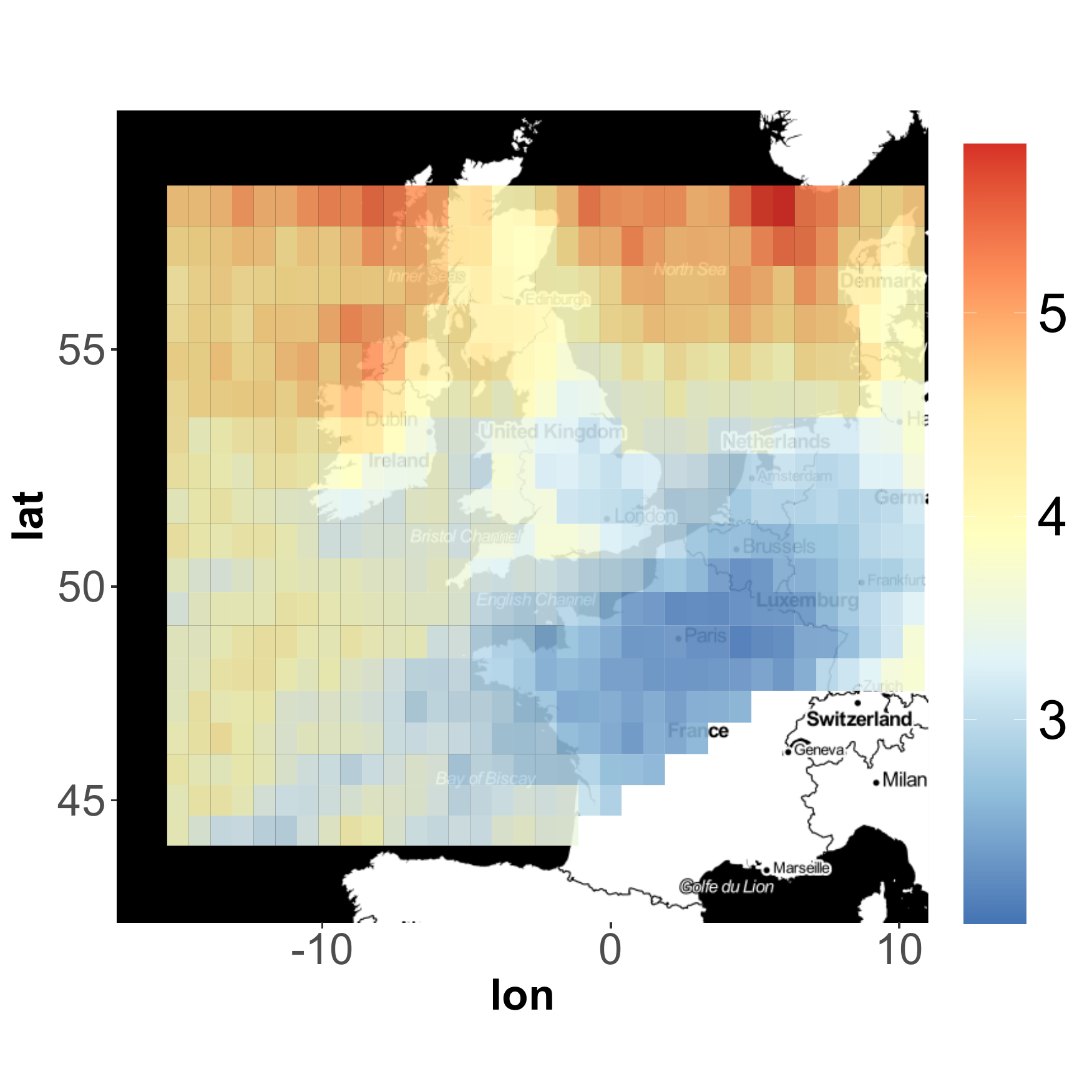} 
\end{tabular}
\end{center}
\caption{Left: Study region $E$ (coloured cells) for modelling extreme windstorms over Europe. Mountainous regions were removed to avoid the systematic bias of the reanalysis model. The green cells show the region $E_{\rm ABLP}$ containing Amsterdam, Brussels, London and Paris. Estimated location and scale functions $b_n$ (middle) and $a_n$ (right) (both in ms$^{-1}$) of the generalized $\R$-Pareto process for modelling extreme windstorms over Europe.}
\label{fig: location and scale}
\end{figure}

Our study focuses on the coloured region $E$ in the left-hand panel of Figure~\ref{fig: location and scale}.
The reanalysis model is known to be systematically biased and to have a different dependence regime over regions with rapid variations in altitude \citep{Donat2011}, so we exclude mountainous regions such as the Pyrenees and the Alps, leaving $605$ cells based on the native resolution of $0.75^{\circ}$. 
Similarly to the XWS catalogue methodology, we combined the maximum wind gust sustained for at least $3$s from the reanalysis with the forecasts to obtain a three-hourly spatial time series.  Extra-tropical windstorms over Europe occur only during the winter, so we take our study period $T$ to be the months  October--March over the years 1979 to 2014. 

\subsection{Storm definition and exceedance probability modelling}\label{sec: strom and freq}

Following \citet{Roberts2014} and \citet{Vautard2018}, we consider storms that give exceedances of the spatial average over a region with very dense infrastructure during a $24$-hour period. For this application we write $S=E\times [0,24]$ with $E \subset \RR^2$ denoting the region of Europe.  
The spatio-temporal process $X(s,t)$ represents the wind field at location $s\in E$ and time $t\in T$.   We take the risk functional $\R$ at time $t$ to be the spatial average of an observed wind field $x(s,t)$,
$$
\R(x)(t) = |E_{\rm ABLP}|^{-1}\int_{E_{\rm ABLP}} x(s,t) \, \D{s}, \qquad  t\in T, 
$$
where $E_{\rm ABLP}$ is the green region shown in Figure~\ref{fig: location and scale},
which includes Amsterdam, Brussels, London and Paris.
To suppress the temporal clustering of high values of $\R(x)(t)$, we centre the time frame on the largest spatial average for each event and keep only events that are at least $48$ hours apart, yielding $n = 1561$ observations. Storm Daria corresponds to a maximum intensity of $\R(x) = 32.1$~ms$^{-1}$. The choice of the declustering algorithm influences the distribution of the events and must be taken into account in the model and estimation procedures; in this work, the model described in Sections~\ref{marg.sect} and~\ref{sec: dep model windstorms} does not allow temporal variation of the dependence structure but ensures unbiased estimation.

The approximation~\eqref{eq: approx r excess} requires models for the probability that $\R\left(X\right) \geqslant u_n$, for the margins, including a tail index $\xi$ and the functions $a_n$ and $b_n$, and for the dependence structure of the generalized $\R$-Pareto process $P$.

\begin{figure}[t!]
\begin{center}
\includegraphics[width = \textwidth]{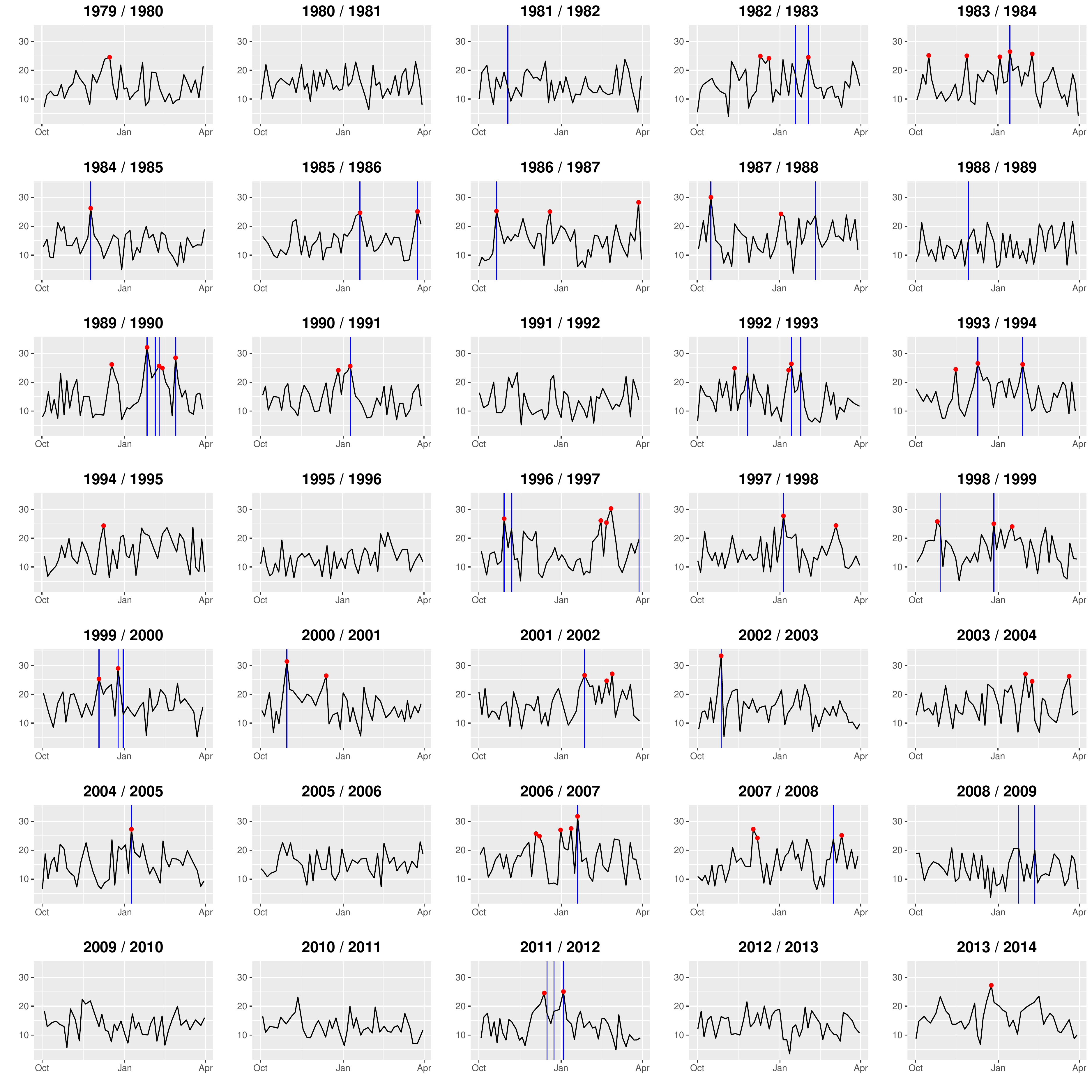}
\end{center}
\caption{Declustered risk functional $\R(X)(t) = |E_{\rm ABLP}|^{-1}\int_{E_{\rm ABLP}} x(s,t) \,\D{s}$ (ms$^{-1})$, computed on the ERA--Interim data set for each winter. $\R$-exceedances above the empirical $0.96$ quantile are represented by red dots and windstorm starting dates from the XWS catalogue are represented by blue vertical lines.}
\label{fig: univ risk with excesses}
\end{figure}

A natural choice for $u_n=\R(b_n)$ is a high quantile of the observed values $\R(x)$.  In order to include  most of the XWS storms in our set of exceedances, we take $u_n = q_{0.96}\{\R(x)\} = 24$~ms$^{-1}$, yielding $63$ events in the study period. The value 0.96 lies within a range of quantiles over which the estimated tail index for $\R(x)$ is stable.  The risk functional, $\R$-exceedances and XWS storms are shown in Figure~\ref{fig: univ risk with excesses}.  The $63$ events, depicted by the red dots, coincide with most of the windstorms from the XWS catalogue, depicted by the vertical lines, so the exceedances $\R(x)\geq u_n$ successfully characterise extreme windstorms that strike $E_{\rm ABLP}$.  The events in the catalogue that do not match large values of $\R(x)$  mostly pass over southern regions of Europe. 

Figure~\ref{fig: univ risk with excesses} shows that the temporal distribution of the selected events is non-stationary. \citet{Donat2010} and \citet{Pfahl2014} have established that climatic circulation patterns such as the North Atlantic Oscillation index (NAO) influence windstorms, and we use logistic regression to model this.  We extracted the $3$-hourly mean sea level pressure from the ERA--Interim reanalysis and computed the NAO using its definition in terms of empirical orthogonal functions (EOF) \citep{Blessing2005}, as the first eigenvalue of the mean sea level pressure anomaly at a given time $t$. We likewise computed the Antarctic Oscillation index (AAO)  and created indexes for temperature anomalies. Time was also included as a potential covariate.  Analysis of deviance reveals that the NAO index and the first and third eigenvalues of the temperature anomaly affect the occurrence of winter storms at the $0.1\%$ significance level.  Figure~\ref{fig: annual summary} summarises the fit of the resulting model.  Plots at a daily scale are shown in Appendix~\ref{app: sm winstorms}.

\begin{figure}[!t]
\begin{center}
\includegraphics[scale = 0.3]{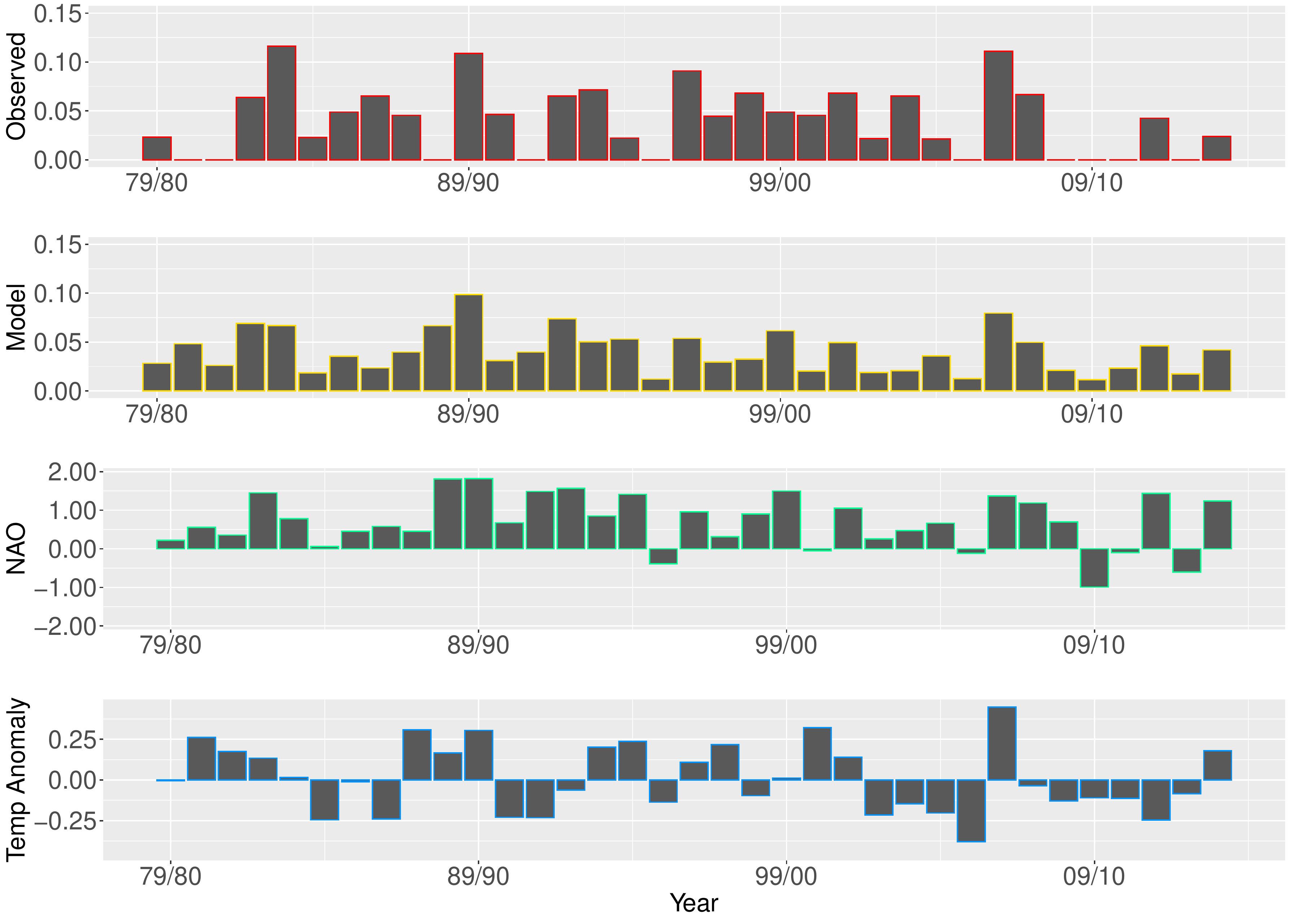}
\end{center}
\caption{Annual summary of the model for the probability of storm occurrence: Observed probability (top), modelled probability (second row), North Atlantic Oscillation index (third row) and aggregated temperature anomaly indexes (bottom).}
\label{fig: annual summary}
\end{figure}

\subsection{Marginal model}\label{marg.sect}
Fitting the marginal model involves the estimation of a tail index $\xi$ and the functions $a_n$ and $b_n$ under the assumptions of Section~\ref{sec: model form}.
In general, a parametric model for $a_n$ and $b_n$ might be necessary, as in \citet{Engelke}, but for simplicity we here set $a_n(s_l) = a_l > 0$ and $b_n(s_l) = b_l \in \RR$ for each of the $L=605$ locations $s_l$.

With the model for the probability of storm occurrence described in Section~\ref{sec: strom and freq}, the parameter $b_n' = \R(b_n)$ is fixed to the empirical $0.96$ quantile of the observed $\R(x)$.
The threshold-stability of generalized Pareto distributions does not allow us to identify the function $b_n$ without further assumptions, so we choose $b_l$ to equal the empirical $q'$ quantile $u_{q'}\{x(s_l)\}$ of the $\R$-exceedances above threshold $u_n$ at the location $s_l$, with $q'$ chosen so that $\R(b_n) = b_n'$.
We obtain $q' = 0.675$, yielding $184$ marginal excesses and estimated location function $\widehat{b}_n$ shown in the central panel of Figure~\ref{fig: location and scale}.

For tractability we first fit the marginal model, estimating the tail index $\xi$ and the positive scale parameters $a_1,\ldots,a_L$ by maximizing the independence log-likelihood~\eqref{eq: indep likh}.  For a given tail index $\xi$, the likelihood for the exceedances above the threshold $b_l$ is optimized independently for each location $s_l$. We treat storms as independent events, and account for strong temporal dependence within each of them by weighting each log-likelihood contribution inversely proportionally to the number of exceedances in the storm from which it arises, so that each storm affects the estimates roughly equally.
This yields the maximum independence likelihood estimate $\widehat{\xi} = -0.15_{0.01}$, close to the average of the  locally estimated tail indexes; the corresponding estimated scale function $\widehat a_n$ is shown in the right-hand panel of Figure~\ref{fig: location and scale}.
Standard errors for $\widehat{b}_n$, $\widehat{a}_n$ and $\widehat\xi$ are obtained by resampling.  Estimates of $A$ and $B$ can then be deduced using~\eqref{eq: loc scale fam} with $a_n' = \R(a)$.

\begin{figure}[p]
\begin{tabular}{ccc}
\includegraphics[scale = 0.15]{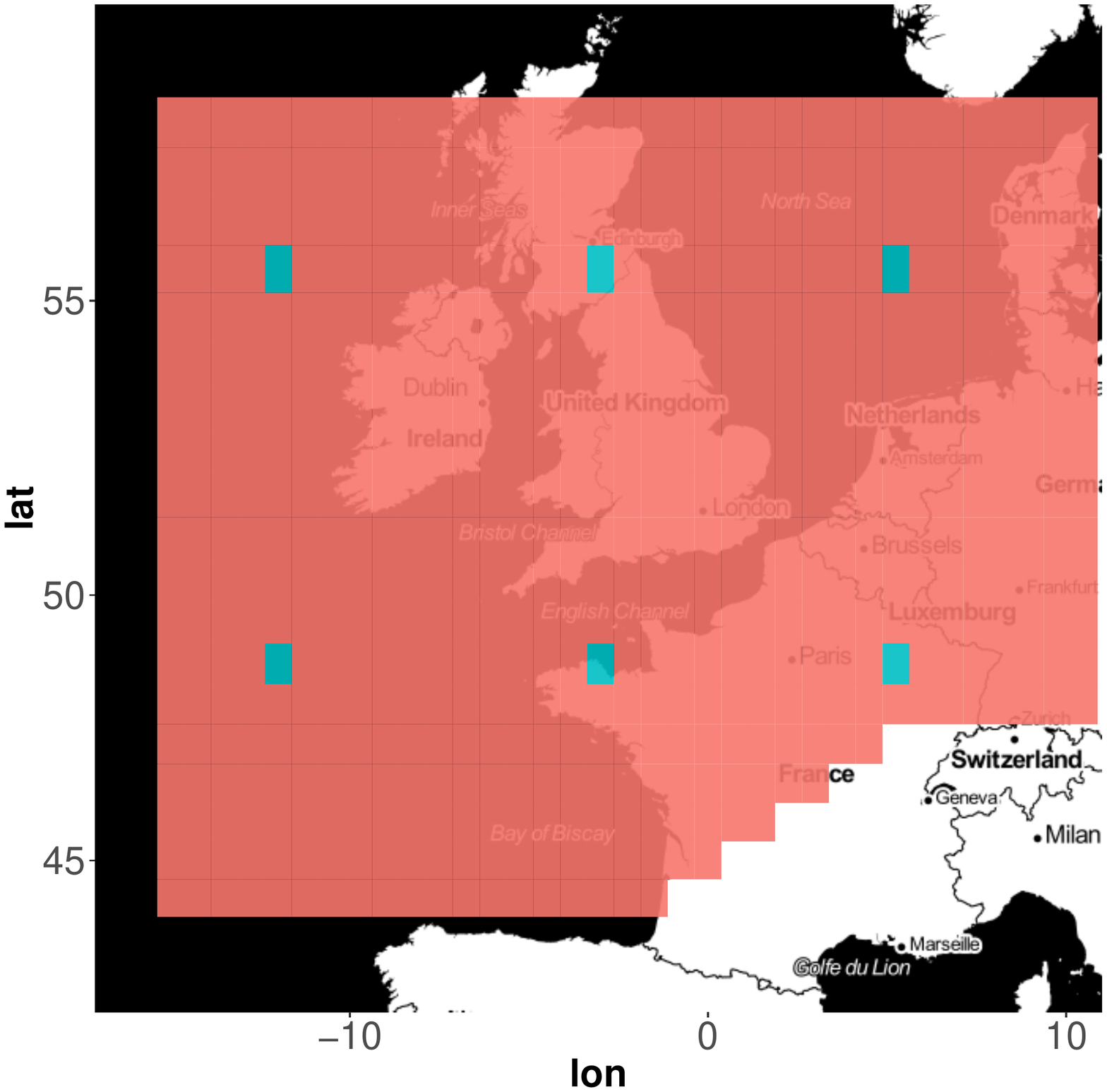} & 
 \multicolumn{2}{c}{\includegraphics[height = 3.4cm]{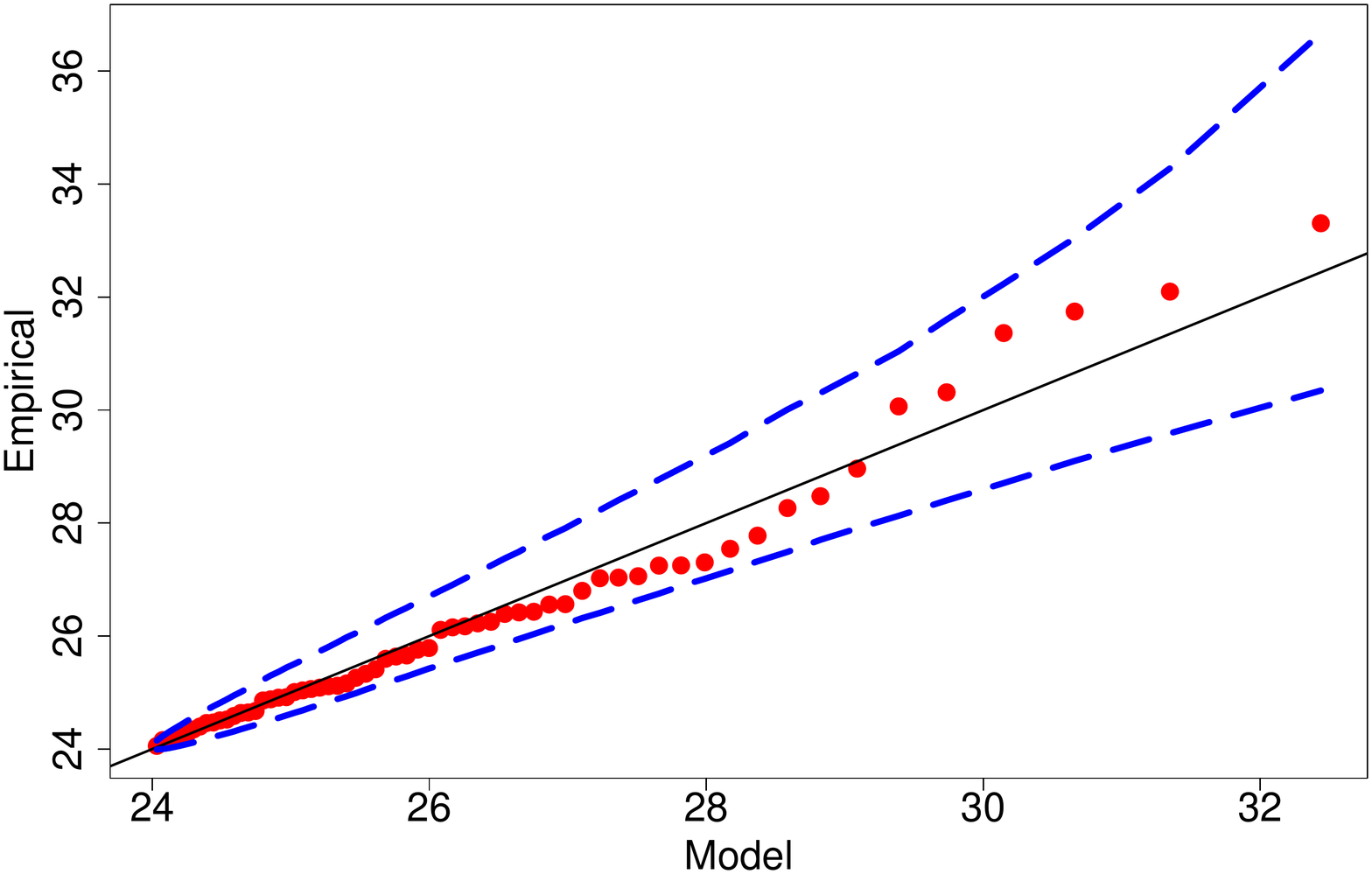}}\\
\includegraphics[scale = 0.16]{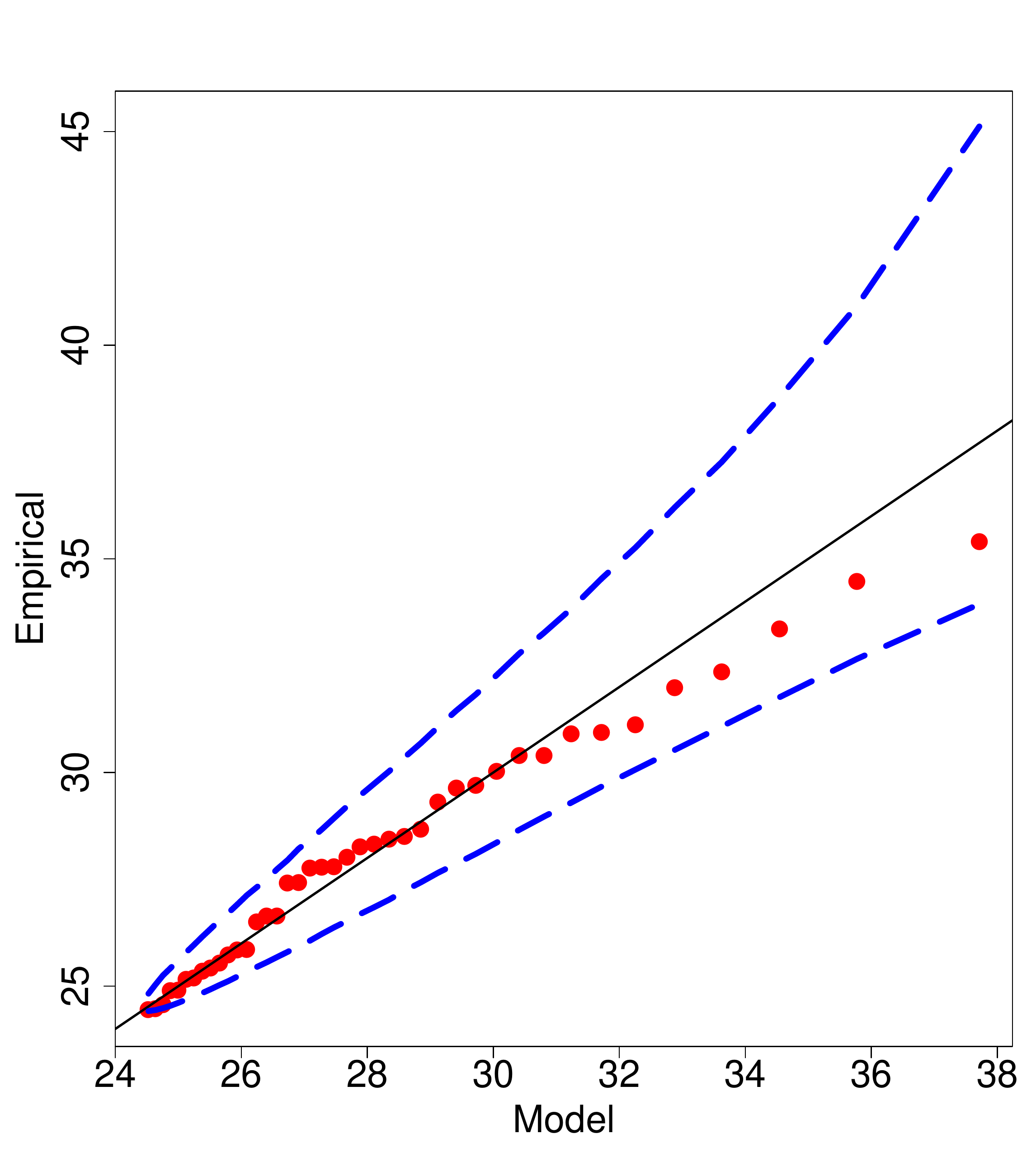} & \includegraphics[scale = 0.16]{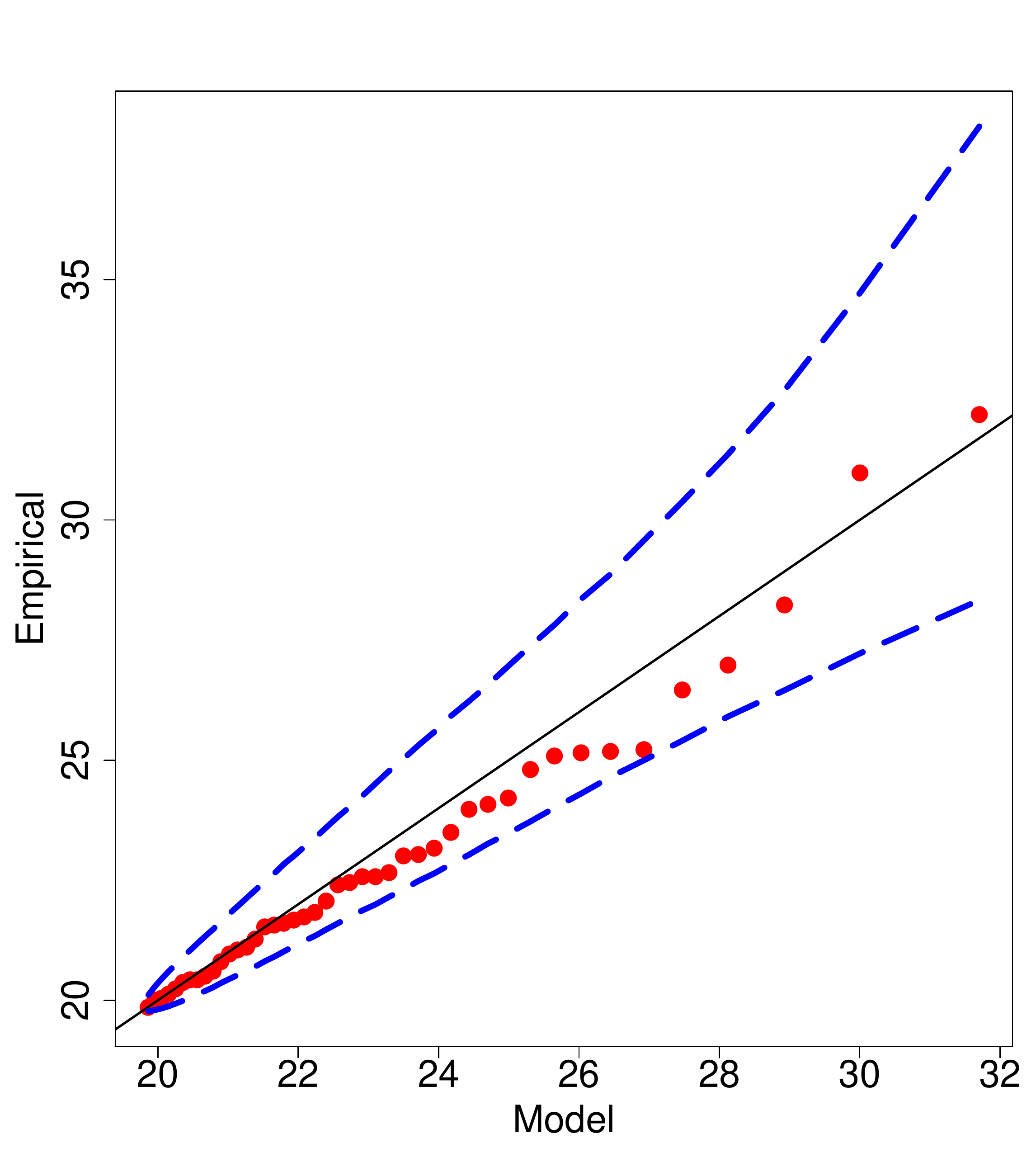} &
\includegraphics[scale = 0.16]{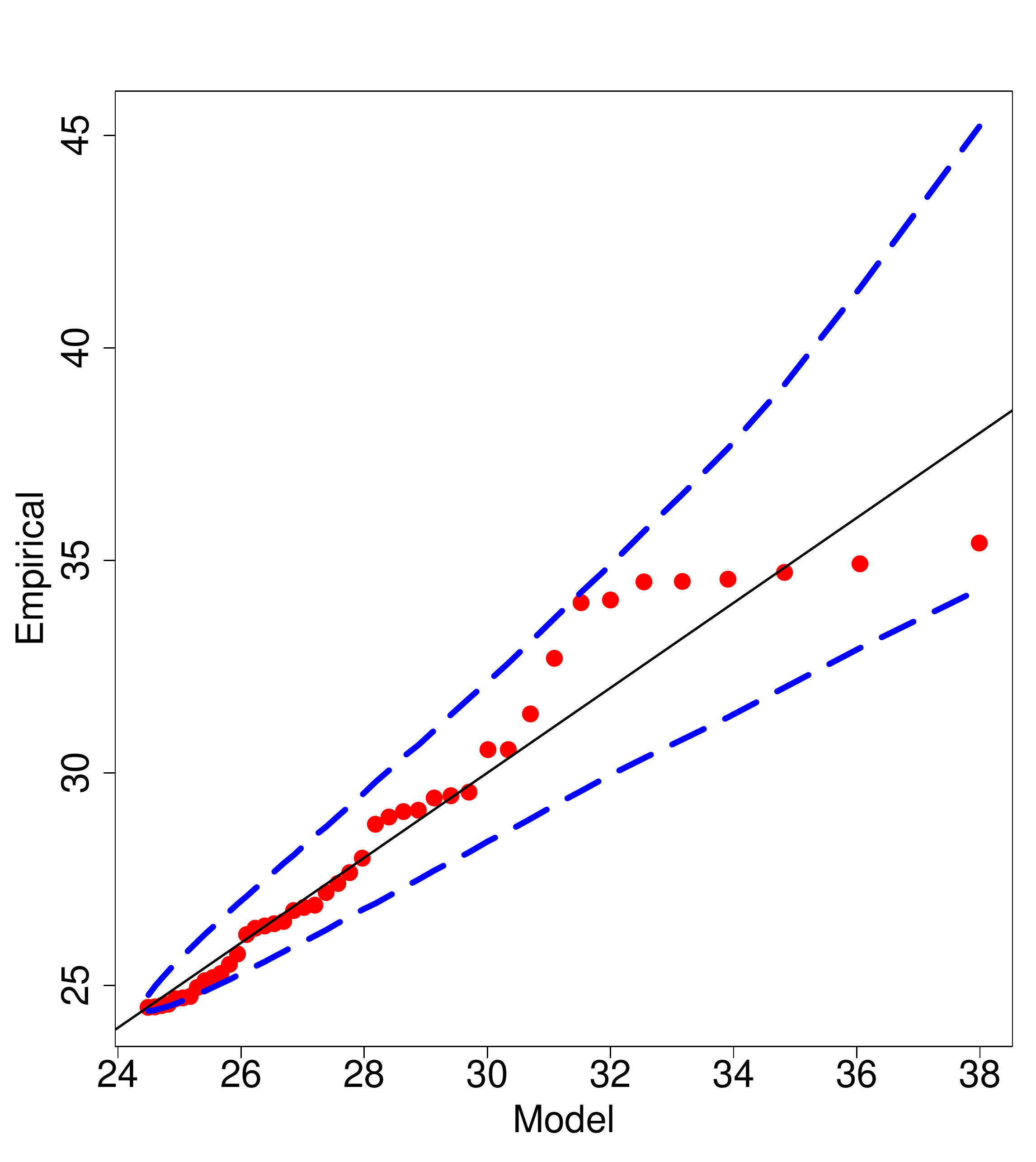} \\
\includegraphics[scale = 0.16]{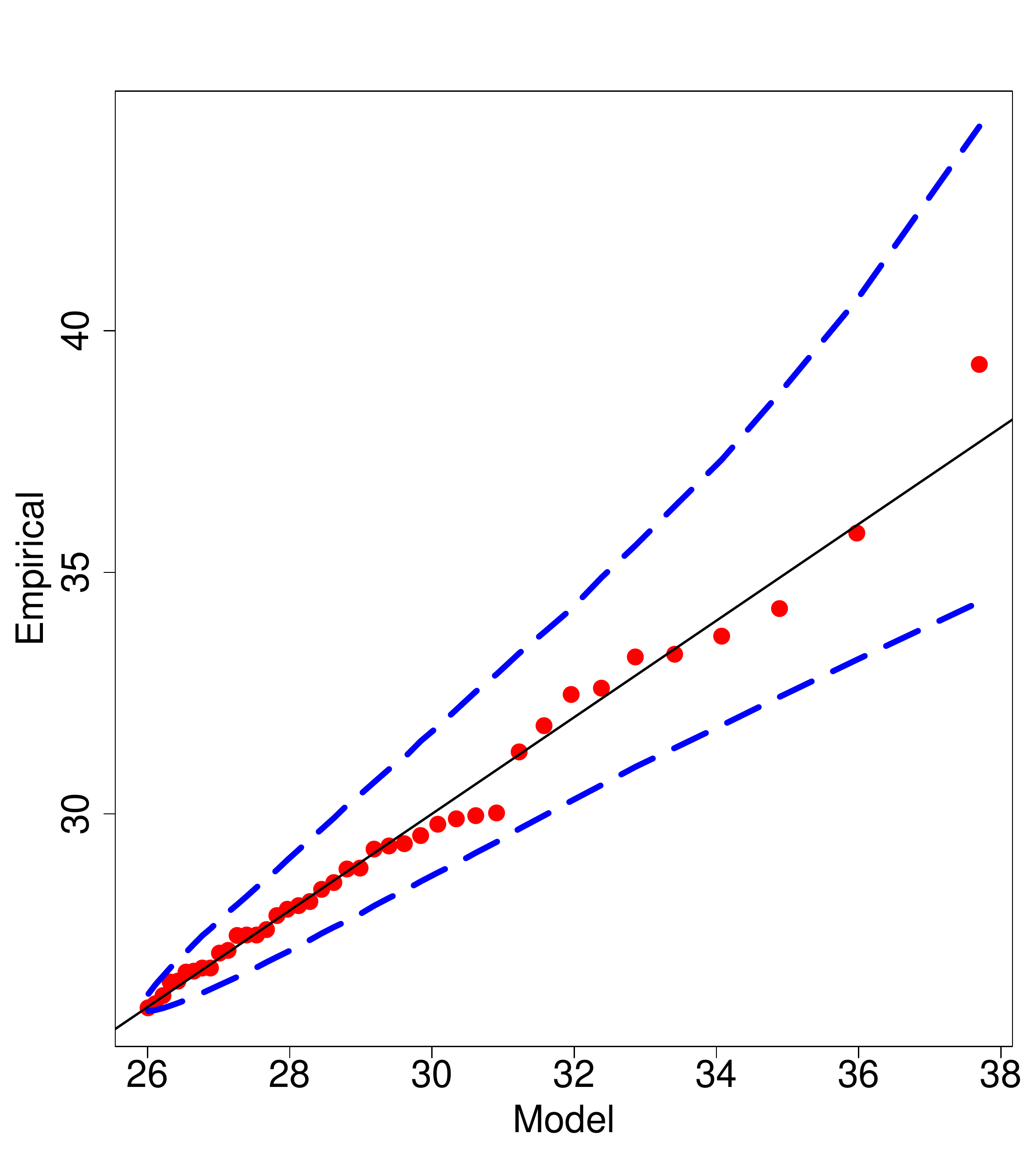} & \includegraphics[scale = 0.16]{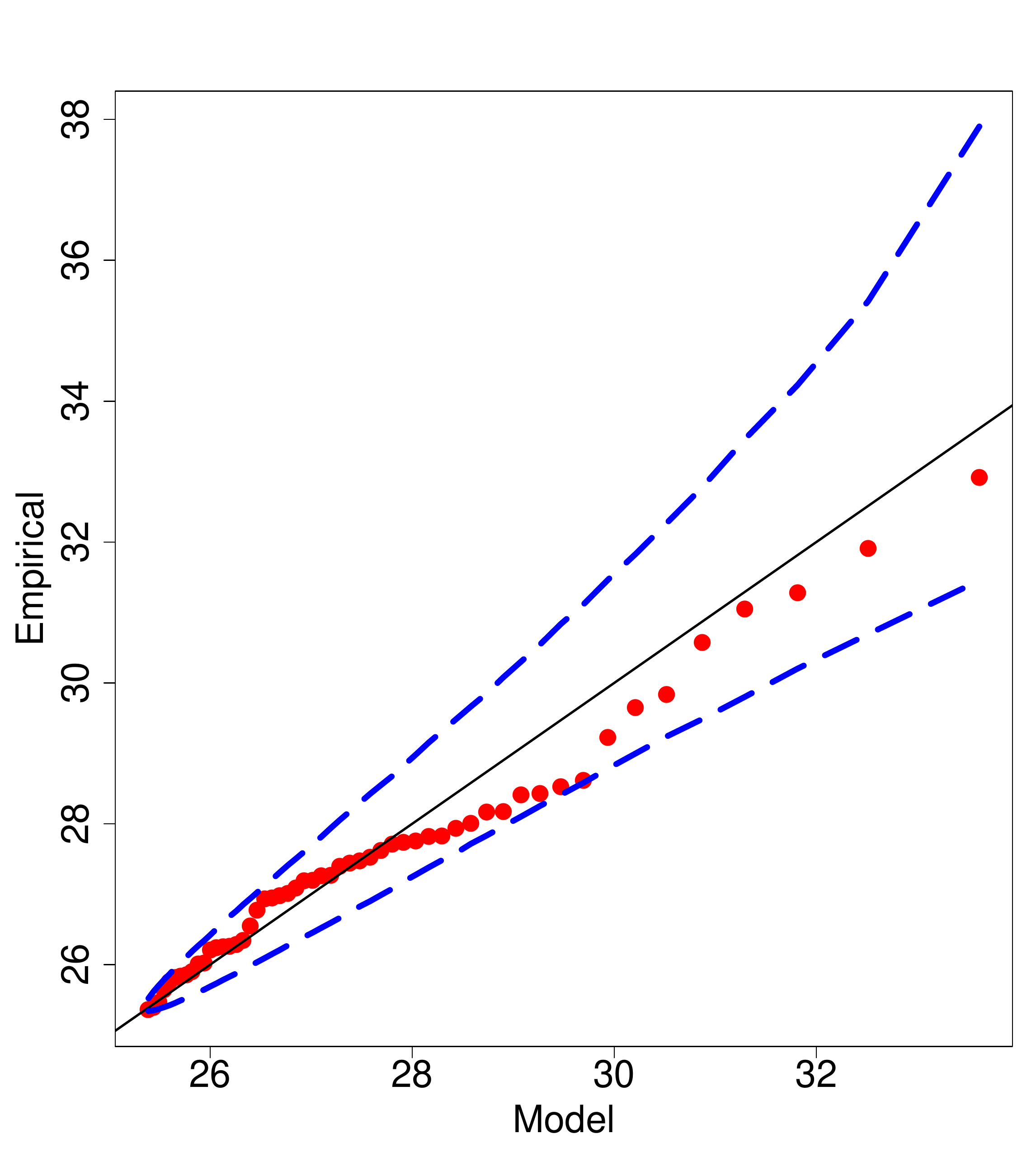} &
\includegraphics[scale = 0.16]{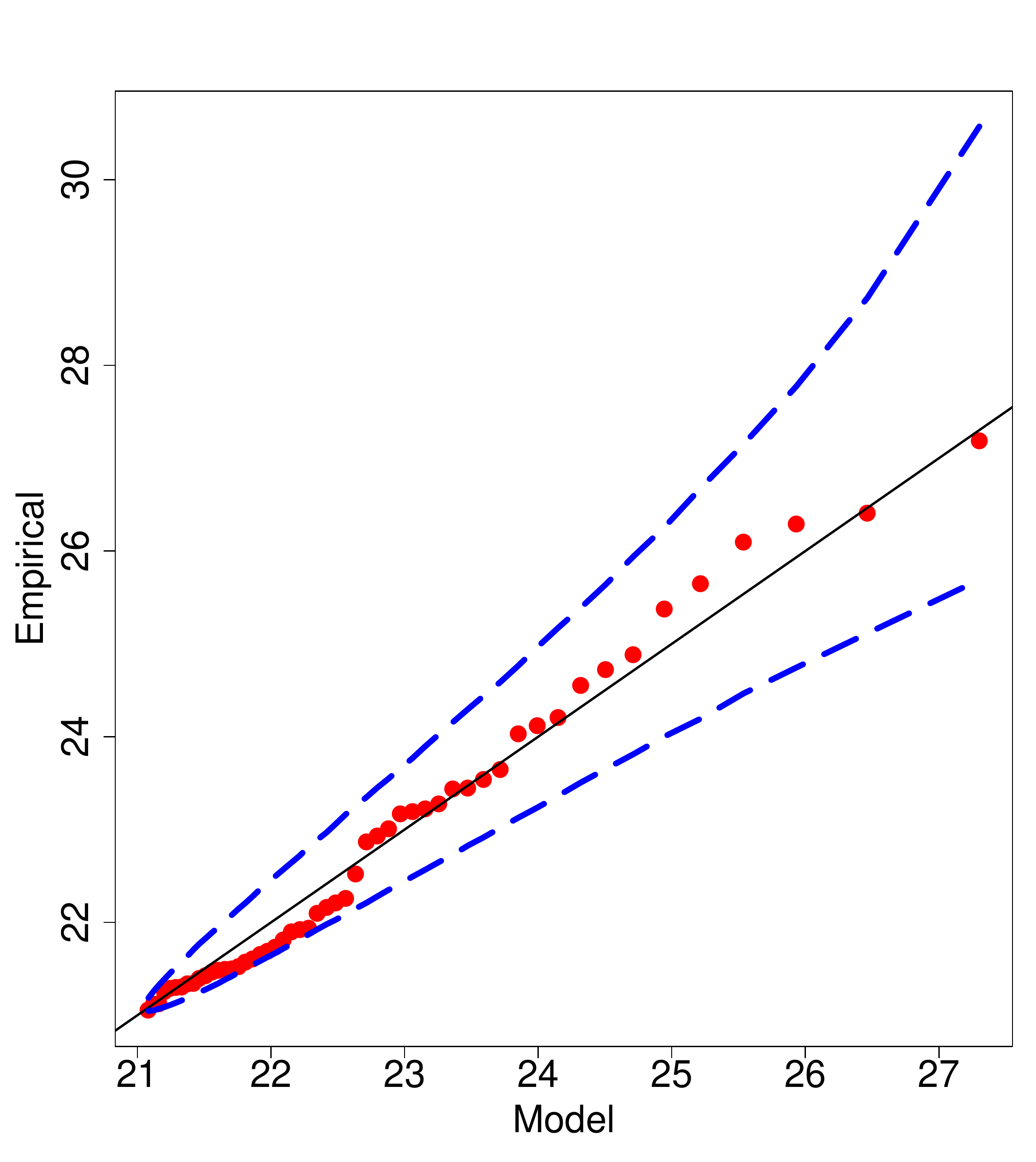} \\
\end{tabular}
\caption{Model assessment for the windstorm data.  The lower six panels show QQ-plots of the local tail distributions for the locations represented by the green cells in the map at the upper left.  The thresholds correspond to the local $0.675$ quantiles of the $\R$-exceedances, yielding $184$ excesses for each cell.  The upper right panel QQ-plot is for exceedances of $\R(x)$ above the threshold $u_n = 24$~ms$^{-1}$ modelled by a generalized Pareto distribution with scale $\widehat{a}_n'$ and tail index $\widehat{\xi} = -0.15$. The blue dashed lines correspond to pointwise $95\%$ confidence intervals. }
\label{fig: qq plots margins storm}
\end{figure}

Figure~\ref{fig: qq plots margins storm} displays QQ-plots {for} the tail distributions at six locations. The overall fit of the marginal model is convincing, and the quality of the fit for $\R(x)$ above the threshold $u_n$ also seems to be adequate.  

\subsection{Dependence model}\label{sec: dep model windstorms}
Following equation~\eqref{eq: approx r excess},  we model the storms by a generalized $\R$-Pareto process with state space $S = E \times [0,24]$ and whose dependence structure must be specified.  For the angular component $W$, we choose a process with log-Gaussian random functions and Whittle--Mat{\'e}rn  \citep{Trust2017a,Whittle1963,Matern1986} semi-variogram
\begin{equation}\label{eq: vario model}
	\gamma(s,s',t,t') = \kappa \left\{1 -  \|h\|^\nu K_\nu(\|h\|) \right\}, \quad \kappa, \nu > 0, 
\end{equation}
where  $K_\nu$ is the modified Bessel function of the second kind of order $\nu$, and \citep[pp.~428, 432]{Gelfand2010}
\begin{equation}\label{eq: model norm}
\|h\| = \|h(s,s',t,t')\| = \left\{\left\| \dfrac{\Omega (s' - s) - V(t' - t)}{\tau_s} \right\|_2^2 + \left|\dfrac{t' - t}{\tau_t}\right|^2 \right\}^{1/2}, 
\end{equation}
for $s,s' \in E$ and $t,t' \in [0,24]$, with positive scale parameters $\tau_s$ and $ \tau_t$ for the space and time dependence, a wind vector  $V \in \mathbb{R}^2$ that models the average displacement of the storm in a three-hour period, and an anisotropy matrix
\begin{equation*}\label{eq: anisotropic matrix}
\Omega = 
\left[\begin{array}{cc}\cos \eta & -\sin \eta \\ a \sin \eta & a \cos \eta\end{array} \right], \quad \eta \in \left(-\dfrac{\pi}{4}, \dfrac{\pi}{4}\right], \quad a > 0, 
\end{equation*}
that allows the spatial dependence in~\eqref{eq: model norm} to decrease faster in a  direction determined by the angle $\eta$. 
Estimation of $\nu$ is known to be difficult, so we set $\nu = 1$, to foreshadow our planned use of more flexible non-stationary models such as that of \citet{Fuglstad2015a}. Indeed, further exploratory analysis reveals that dependence varies over space, so more complex models would ideally be considered.

\begin{figure}[!h]
\begin{center}
\begin{tabular}{cccc}
& Empirical & Least Squares & Gradient Score \\
\rotatebox[origin=l]{90}{~~~~~~~~0 hours} &\includegraphics[width = 11em,height = 8em]{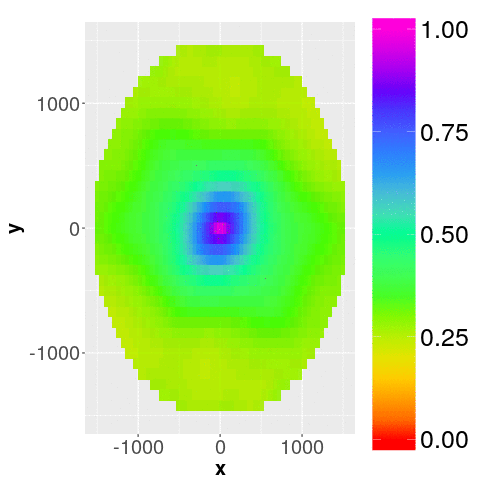} &\includegraphics[width = 11em,height = 8em]{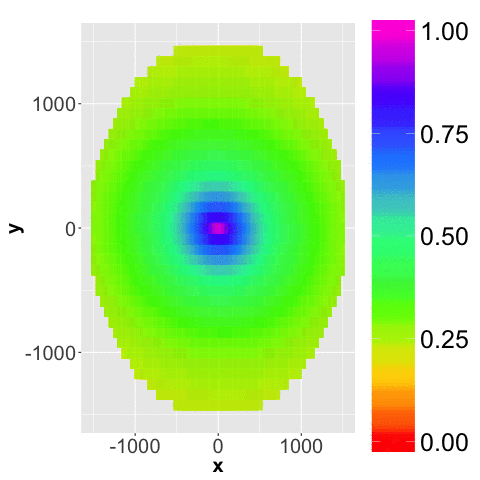} &\includegraphics[width = 11em,height = 8em]{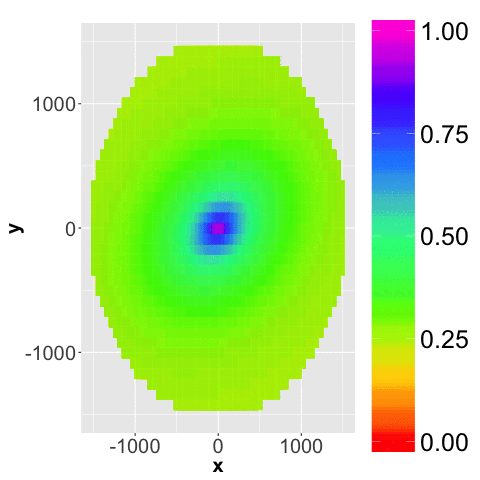}\\
\rotatebox[origin=l]{90}{~~~~~~~~3 hours} & \includegraphics[width = 11em,height = 8em]{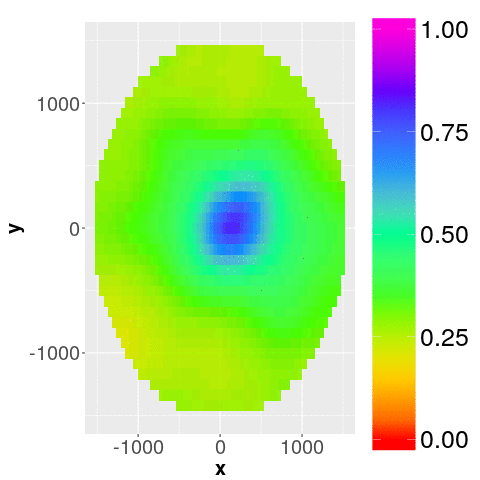} & \includegraphics[width = 11em,height = 8em]{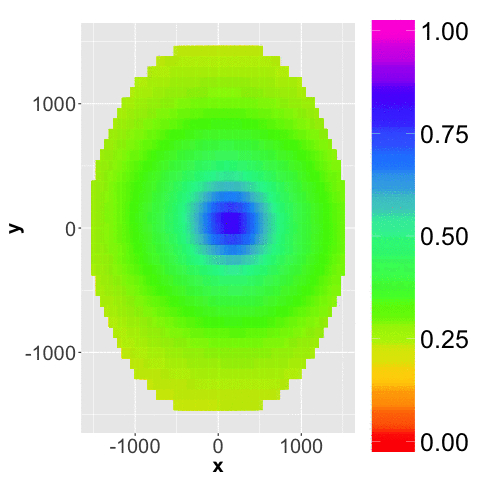} &\includegraphics[width = 11em,height = 8em]{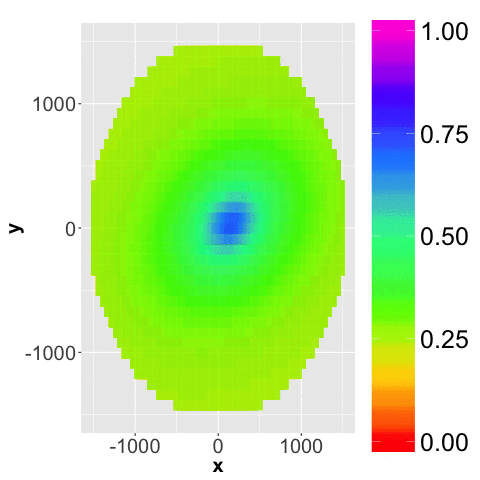}\\
\rotatebox[origin=l]{90}{~~~~~~~~6 hours} & \includegraphics[width = 11em,height = 8em]{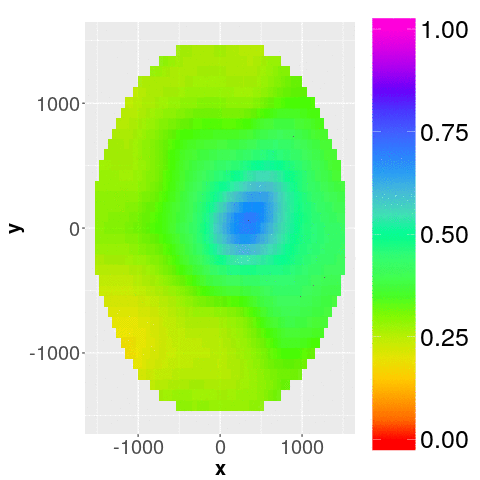} & \includegraphics[width = 11em,height = 8em]{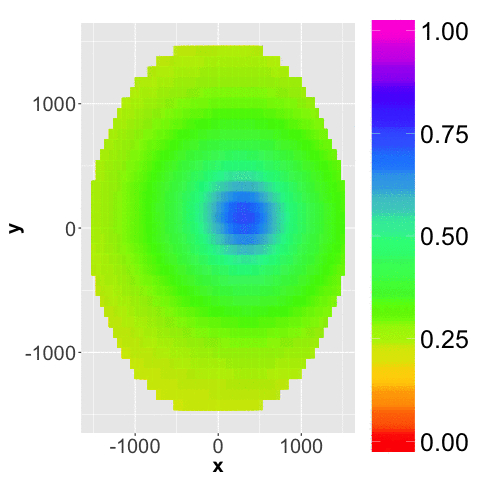} &\includegraphics[width = 11em,height = 8em]{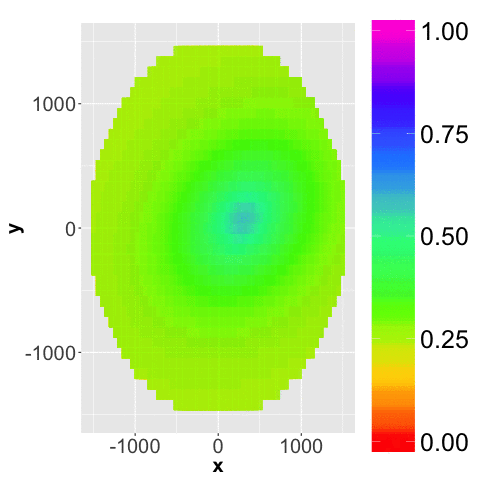}\\
\rotatebox[origin=l]{90}{~~~~~~~~9 hours} & \includegraphics[width = 11em,height = 8em]{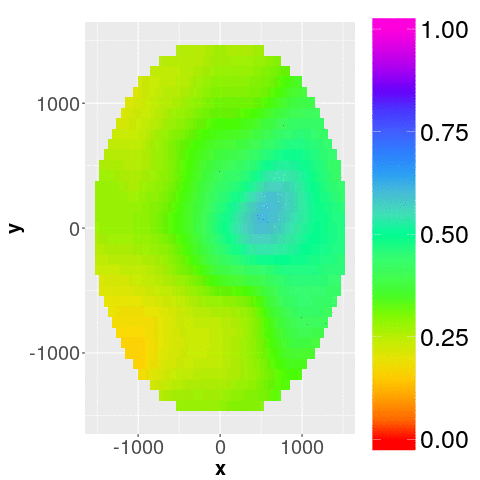} & \includegraphics[width = 11em,height = 8em]{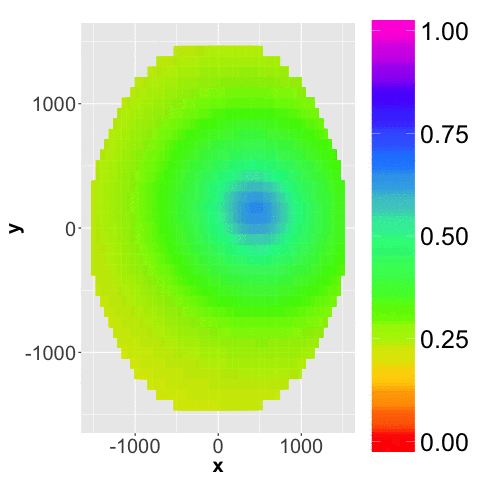} &\includegraphics[width = 11em,height = 8em]{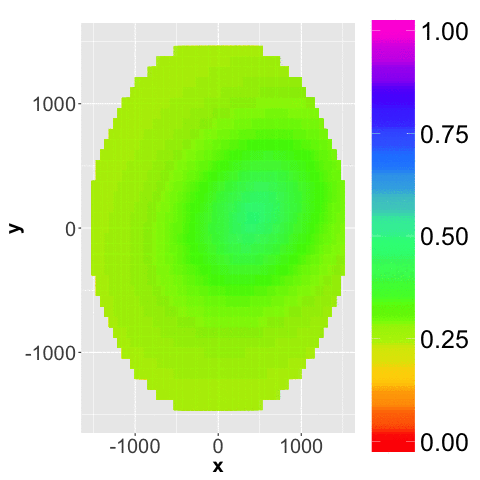}\\
\end{tabular}
\end{center}
\caption{Extremograms as functions of distance (km): empirical estimates (left), fitted values obtained using the parameters from least squares (middle) and gradient scoring (right) estimates. Each row represents a $3$-hour time step.}
\label{fig: dependence mode and checking}
\end{figure}

The semi-variogram function~\eqref{eq: vario model} is motivated by an exploratory analysis in which the space-time extremogram
$$
\pi(h_s,h_t) = \Prob\{X(s',t') \geqslant u' \mid X(s,t) \geqslant u\}, \quad h_s = s' - s, \quad h_t = t'- t,
$$
with thresholds $u$, $u'$ at local $0.675$ empirical quantiles of the set of $\R$-exceedances, is estimated as described in Section~\ref{sec: stat inference}; see the first column of Figure~\ref{fig: dependence mode and checking}.

We used both least squares and gradient scoring procedures to estimate the parameters of~\eqref{eq: vario model}, the latter  using a composite approach with $100$ random subsets and the same $50$ locations for every storm, since we found this to be more robust than including all locations.  In general we recommend using subsets whose size roughly equals the number of chosen events.

Table~\ref{tab: models parameters} and Figure~\ref{fig: dependence mode and checking} summarise the resulting fits, which agree on the strength of dependence at long distances and are overall consistent with the empirical values, but differ for the anisotropy: least squares picks out the long-range north-east anisotropy but the gradient score fit captures the short-range south-east anisotropy.  This change in direction cannot be captured by our over-simple model.   The estimated wind vectors $\widehat{V}$ for the two fits are similar and agree with the observation that storms are born over the Atlantic and usually move towards the North Sea in an east-north-easterly direction.  The fits look reasonable, though the scoring approach may slightly  under-estimate the temporal dependence.  

\begin{table}
\caption{Semi-variogram parameter estimates obtained by minimizing~\eqref{ell_ext.eq} and using the 
gradient score. The standard errors (subscripts) are obtained using a block jackknife.
\label{tab: models parameters}}
\centering
\fbox{%
\begin{tabular}{*{8}{c}}
& $\kappa$ & $\tau_s$(km) & $\tau_t$(h) & $a$ & $\eta$($^{\circ}$) & $V_1$(km.h)$^{-1}$ & $V_2$(km.h$^{-1}$)\\ \hline
Least squares & $3.5$ & $614$ & $23.8$ & $1.41$ & $-4.12$ &  $51.3$ & $14.4$\\
Gradient score & $2.85_{0.01}$ & $337_{11.6}$ & $9.6_{2.8}$ & $1.32_{0.01}$ & $21.2_{0.1}$ &  $50.4_{2.9}$ & $12.5_{1.7}$\\
\end{tabular}}
\end{table}

\subsection{Simulations}

The usefulness of our model can be checked by simulating extreme storms from it, using a version of Algorithm~\ref{algo1} modified to ensure that the maximum spatial average occurs at $t = 12$ hours, consistent with our definition of an extreme storm.  We first simulate the  angular component of the spatial process at time $t=12$, and then simulate the remaining time steps by successively generating the spatial process at times $t = 9,6,3,0,15,18,24$ conditionally on the variables already simulated. If a new time step yields a spatial average greater than its value at time $t=12$, the sample is rejected and the procedure is repeated until a suitable candidate is found.

\begin{figure}[t]
\begin{tabular}{ccc}
\includegraphics[scale = 0.26]{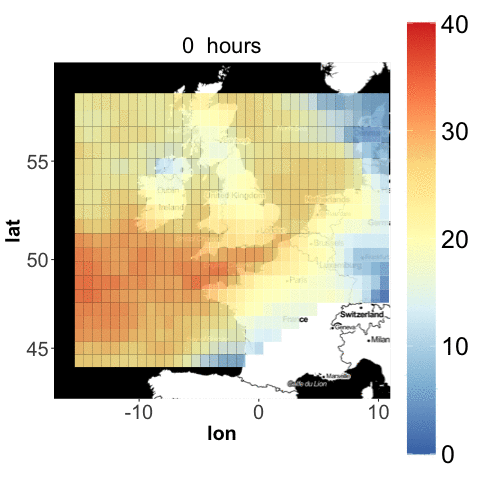} & \includegraphics[scale = 0.26]{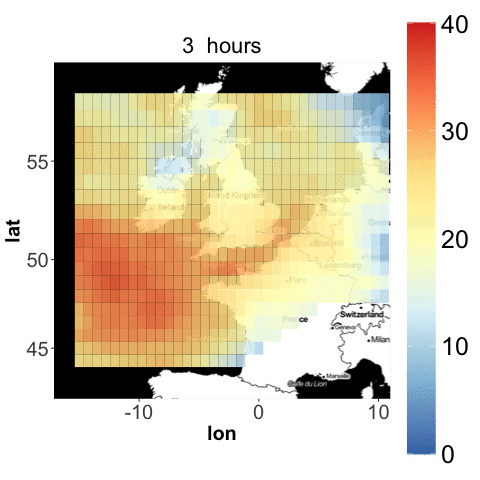} & \includegraphics[scale = 0.26]{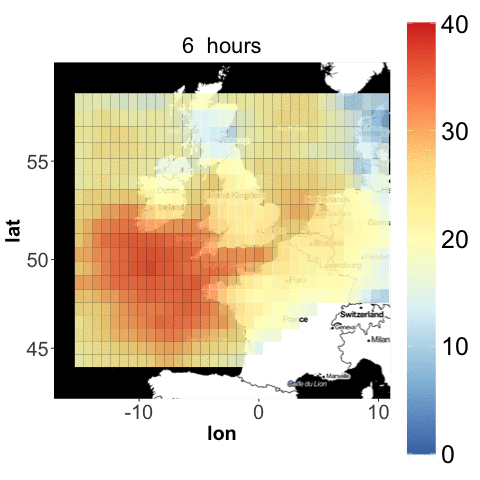} \\
\includegraphics[scale = 0.26]{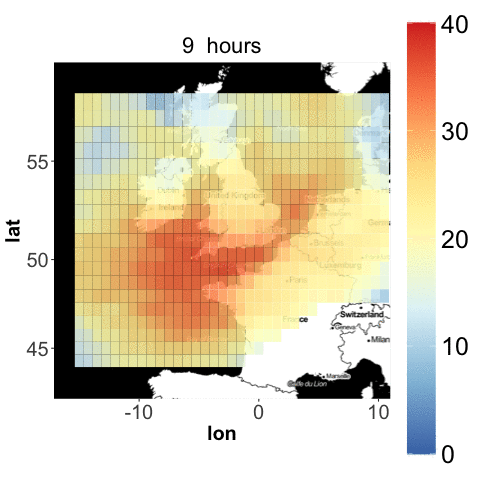} & \includegraphics[scale = 0.26]{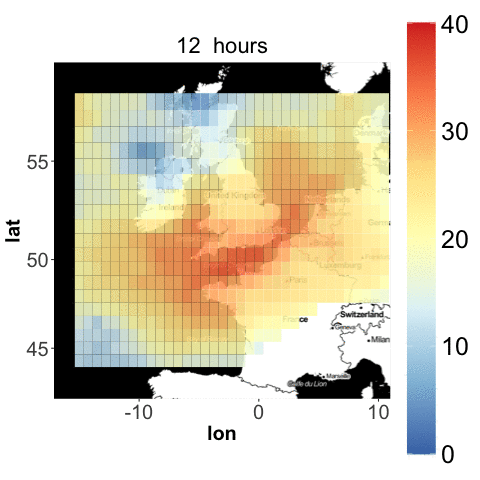} & \includegraphics[scale = 0.26]{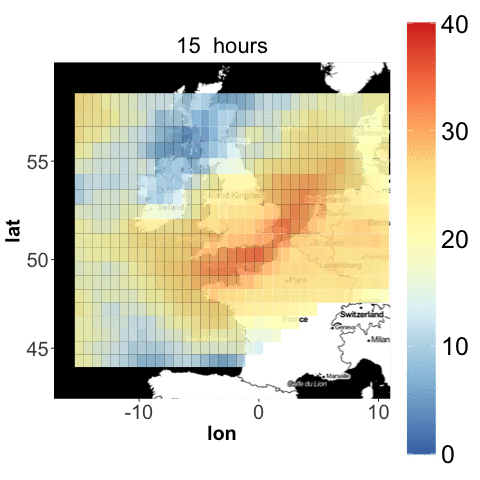} \\
\includegraphics[scale = 0.26]{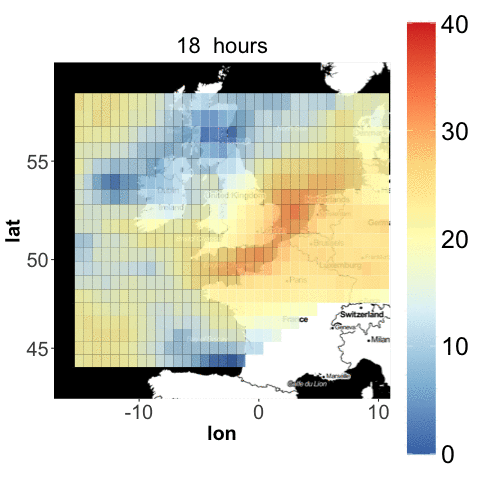} & \includegraphics[scale = 0.26]{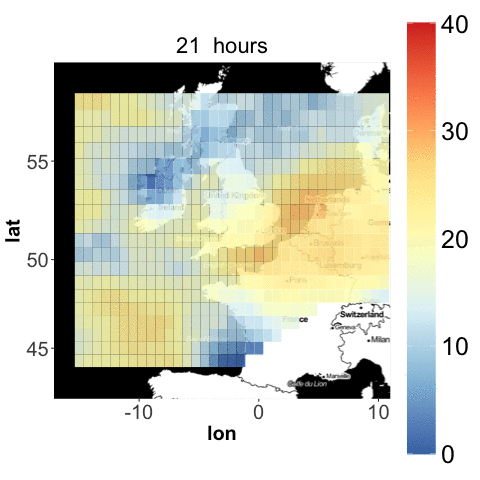} & \includegraphics[scale = 0.26]{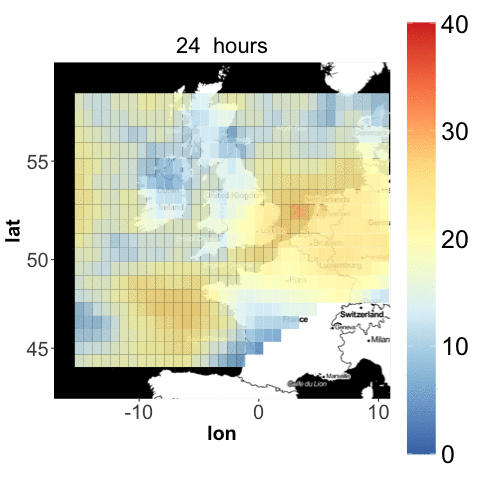}
\end{tabular}
\caption{Simulated maximum speed (ms$^{-1}$) over the past $3h$ hours of wind gusts sustained for at least $3$s. The storm has an intensity $\R(x) = 29.1$ ~ms$^{-1}$.}
\label{fig: simulated storm}
\end{figure}

For an angular process with log-Gaussian random functions, such a simulation algorithm is equivalent to conditional simulation of multivariate Gaussian random vectors.  Figure~\ref{fig: simulated storm} shows a simulated storm with intensity $\R(x) = 29.1$~ms$^{-1}$, similar to that of Daria.  The images are rougher than those in  Figure~\ref{fig: daria} but nevertheless the higher windspeeds at sea, the general scale of spatial dependence and the directionality seem credible.

\section{Flood risk assessment}\label{sec: rainfall}

\subsection{Motivation}

In August $2005$, the city of Zurich suffered from heavy floods that led to estimated property damage of around $3$ billion Swiss francs and six deaths \citep{Bezzola2007}.
Zurich is especially risk-prone, as it lies at the foot of a lake and is traversed by several rivers, including the Sihl, which flows under the city's main railway station. 
Although the $2005$ event was not caused by an unusually high level of the Sihl \citep{Jaun2008}, it triggered an overall assessment of flood risk for the city.
An extreme discharge of this river could cause hundreds of millions of francs of losses by damaging infrastructure and by preventing half a million commuters from travelling.  A good understanding of the risk related to high levels of the Sihl  is thus crucial when considering potential mitigation measures.
Below we use our ideas to construct a stochastic generator of extreme rainfall over the Sihl river basin, in order to create a catalogue of events for input to hydrological models.  \citet{Cloke2009} review similar approaches based on climate models. 

\subsection{Data set and region of study}

\begin{figure}[t]
\begin{center}
\includegraphics[width=0.8\textwidth]{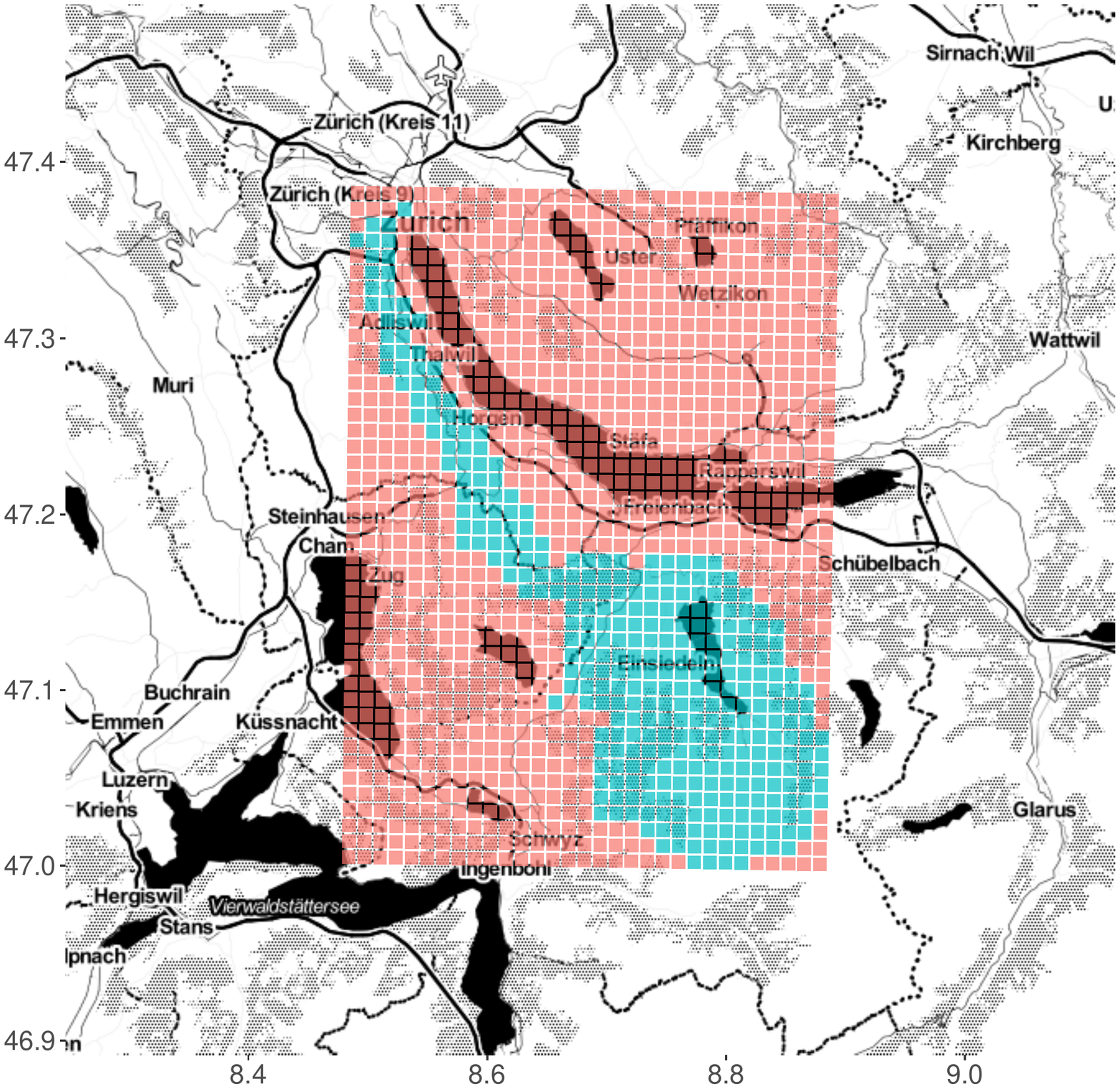} 
\end{center}
\caption{Sihl river basin (green) and study region (red).}
\label{fig: sihl river bassin}
\end{figure}

Figure~\ref{fig: sihl river bassin} shows the region of study, a rectangle south-west of Zurich that includes the Sihl river basin.  Any rain falling in the green area can be expected to flow under Zurich main station. Rainfall is the result of various physical processes, including cyclonic and convective regimes, which can usually only be distinguished using high resolution data such as radar measurements.   In this study we use the CombiPrecip data set produced by MeteoSwiss \citep{Sideris2014a,Gabella2017,Panziera2018}, which estimates the  hourly accumulated rainfall for a grid over Switzerland from $2013$ to $2018$. Owing to changes in 2013, earlier measurements are  inconsistent with more recent data, but even with this reduced period the data set includes $n=52,413$ radar images. The Sihl river basin is orographically homogeneous and is located at a reasonable distance from the radar, so the estimated rain fields are thought to be fairly reliable. CombiPrecip provides discrete measures of rain accumulation that result from post-processing, and this particularity would require specific treatment, for instance using a discrete generalized Pareto distribution \citep{Anderson1970, Krishna2009, Prieto2014}.  Here we aim to illustrate the flexibility and advantages of functional peaks-over-threshold analysis, so we leave the discreteness to future work. To ensure good behaviour of rank-based procedures such as the computation of the empirical extremogram, the original discrete measurements are jittered by adding a small noise.

\subsection{Risk definition and model formulation}

Following \citet{Fondeville2017}, we model both locally intense and large spatial accumulations of rainfall, but rather than use unnatural risk functions based on standardized data, we here first defined the risk in terms of the jittered measurements $x_1,\ldots, x_n$ through the functionals
\begin{equation*}
\R_1(x) = |S|^{-1} \int_S x(s)\, \D{s}, \quad \R_2(x) = \max_{s \in S} x(s), 
\end{equation*}
where $\R_1$ represents a volume of water and thus has a direct hydrological interpretation.  In order to use $\R_1$ and $\R_2$ to entirely separate these different types of events, we must choose the thresholds $u_1,u_2 >0$ so high that only the eleven most intense events are used for inference; see Figure~\ref{fig: rainfall observations}.  In order to use more events and to illustrate the flexibility of the functional POT methodology, we instead study a modified spatial average risk functional, 
$$
\R_1'(x) =  |S|^{-1} \int_S x(s)\, \D{s} \times \dfrac{\|\widehat{x}_1\|}{\|\widehat{x}\|},
$$
where $\|\widehat{x}\|$ denotes the norm of the two-dimensional discrete Fourier transform of $x$ and $\|\widehat{x}_1\|$ denotes the norm of its first component.
This focuses the risk on events with large spatial average rainfall that are also spatially widespread and discards `hybrid' events: deeper exploratory analysis suggests that more than two types of rain are encountered in this region.

The series $\R_1'(x_i)$ is  highly correlated with $\R_1(x_i)$, especially in the tail, but using the former allows us to lower the threshold enough to retain $132$ events.
It also illustrates the use of a risk functional that has a non-linear part and shows that image processing ideas can help to characterize extreme rain types. Another way to discriminate between types of extremes would be to project the database onto specific weather regimes obtained via EOF analysis \citep{Braud1993} or via a methodology tailored for extremes \citep{Cooley2019}, which could help in studying weather patterns such the North American winter dipole \citep{Wang2015}.

When building a model for rainfall, it is important to be able to handle dry grid cells, for which $x(s)=0$.
Below we use a log-Gaussian generalized $\R$-Pareto process,   as presented in Section~\ref{sec: model form}, but the model must accommodate zeros. 
In our region we can treat the distribution of dry cells as homogeneous, so we suppose that zero rainfall corresponds to a negative value of the process, which we treat as left-censored at zero.
A simple modification to allow for variation in the distribution of dry cells would be to construct a new data set $x'$ by adding a positive function $c\equiv c(s)$ to the original data $x$, to treat $x'(s)$ as left-censored if it equals $c(s)$, and to let $c(s)$ increase with the frequency of dry events. This does not affect the model fit, as $x' - b_n' = x - b_n$ if $b_n$ and $b_n'$ are local empirical quantiles of $x$ and $x'$, though the censoring must be accommodated.

To fit the model, we first estimate the marginal tail behaviour and then the dependence model.
For the margins we proceed as in Section~\ref{marg.sect}: $\widehat{b_n^1}$ and $\widehat{b_n^2}$ are defined for $\R_1$ and $\R_2$ separately as local empirical quantiles of the exceedances for these risk functionals, with the levels chosen such that $\R'_1(b_n^1) = u_1$ and $\R_2(b_n^2) = u_2$.
The corresponding tail indexes and scale parameters are then estimated using the independence likelihood~\eqref{eq: indep likh}.

We considered the Mat{\'e}rn and the Bernstein \citep{Schlather2017} semi-variogram models for the dependence.
Mat{\'e}rn semi-variograms are bounded above, whereas the Bernstein model bridges the two regimes illustrated in Figure~\ref{fig: properties vario}.
In the case of the spatial maximum functional $\R_2$, we use censored likelihood estimation with thresholds $\widehat{b_n^1} > c$. For the spatial average functional $\R'_1$, no dry cells were observed for any of the $132$ events yielding exceedances, and we can use a gradient score approach to estimate the dependence model unhampered by dry cells.
In both cases, we found composite approaches to be more stable, so we estimated the dependence using {$1000$} random sets of $30$ locations for $\R_1'$ and $100$ random sets of $10$ locations for $\R_2$, for which the number of subsets was reduced for tractability. We again observed that composite procedures with subsets of size roughly $n$ gave fairly stable estimates.

\subsection{Estimated models}

\begin{figure}[t]
\begin{tabular}{cccc}
Tail Index & Dependence Models  & Observations & Simulations \\ \hline
~\\

$\widehat{\xi} = -0.2_{0.04}$   & \raisebox{-.5\height}{\includegraphics[scale = 0.1]{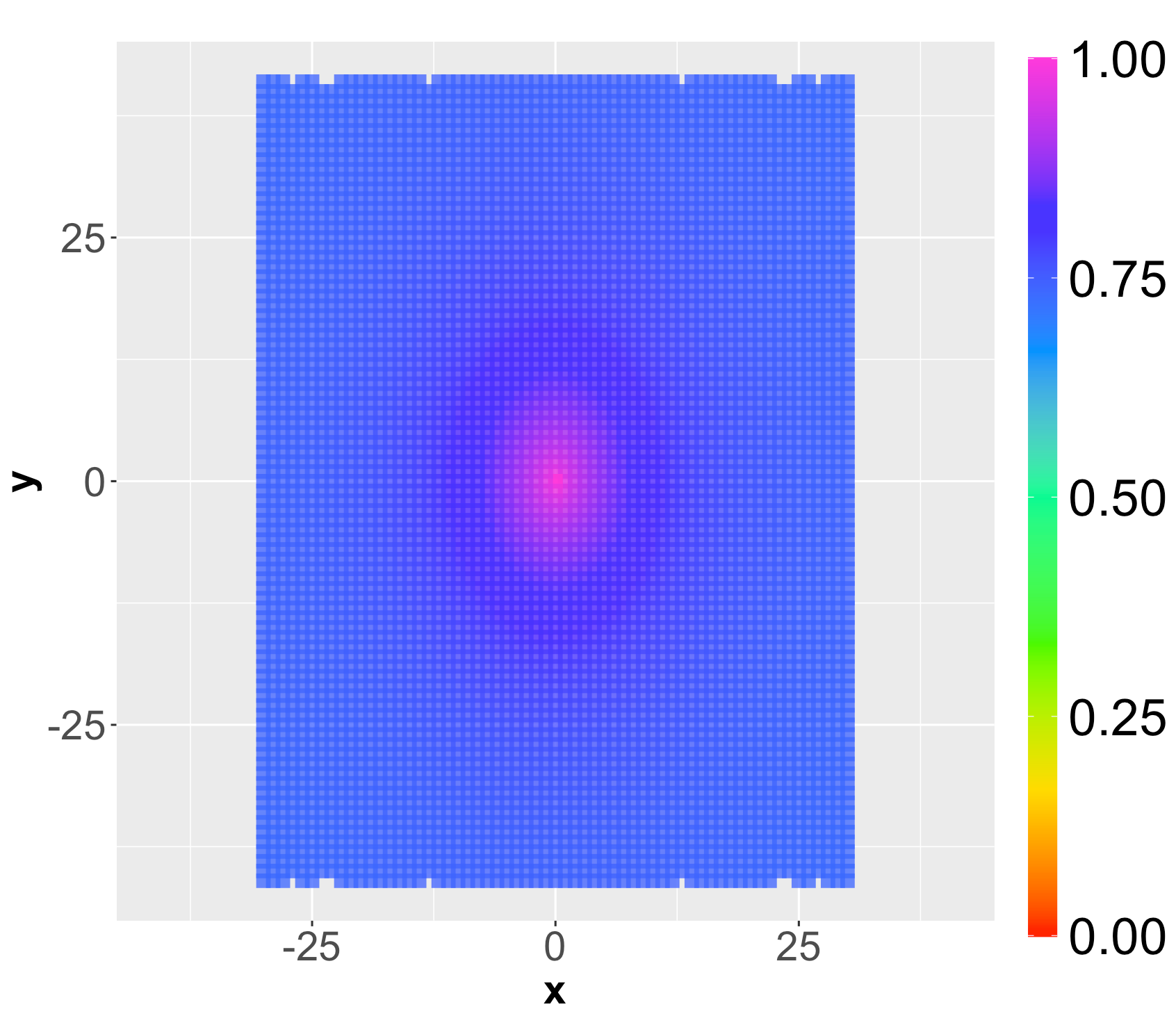}}   &
\raisebox{-.5\height}{\includegraphics[scale = 0.13]{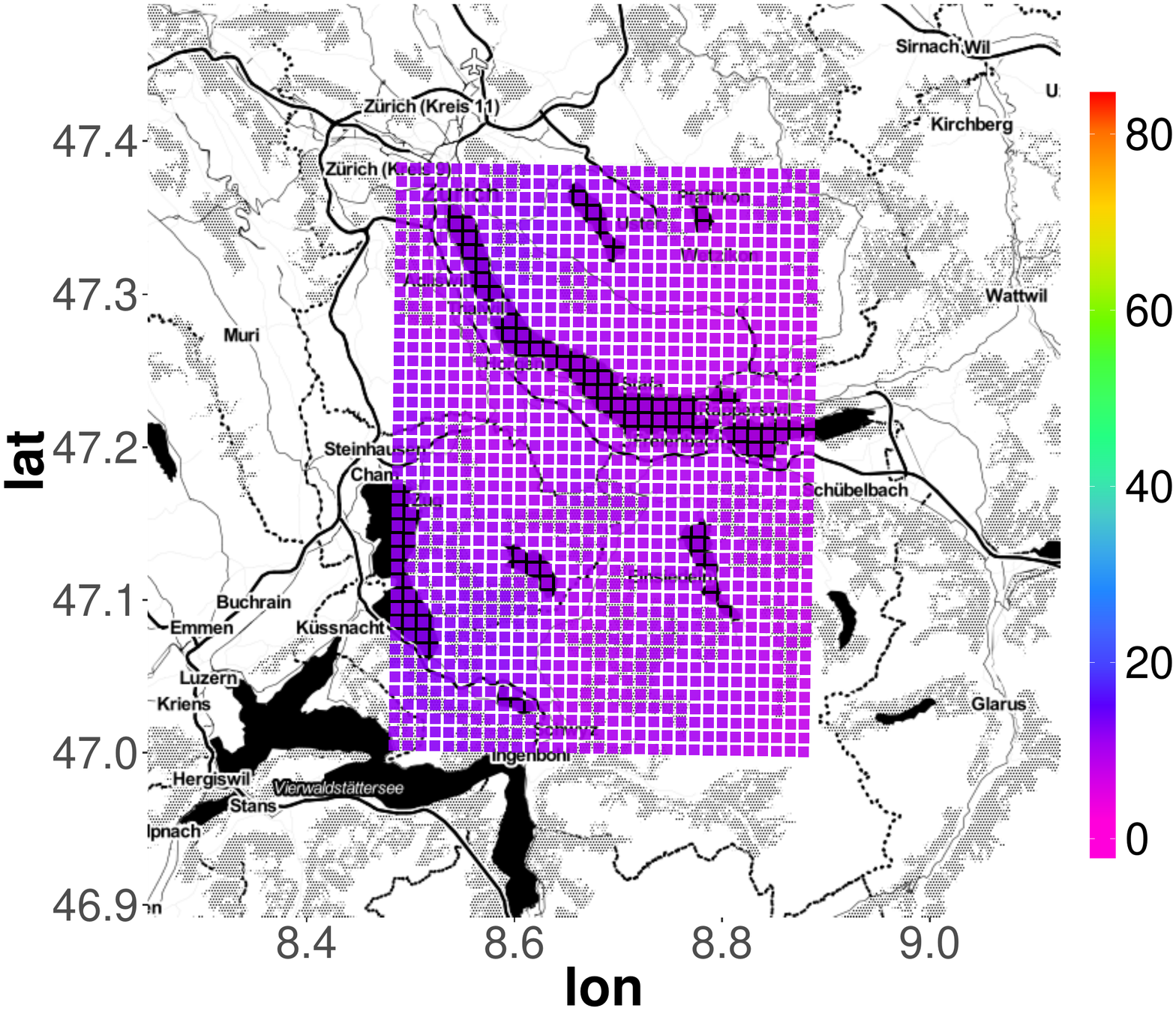}} &   \raisebox{-.5\height}{\includegraphics[scale = 0.13]{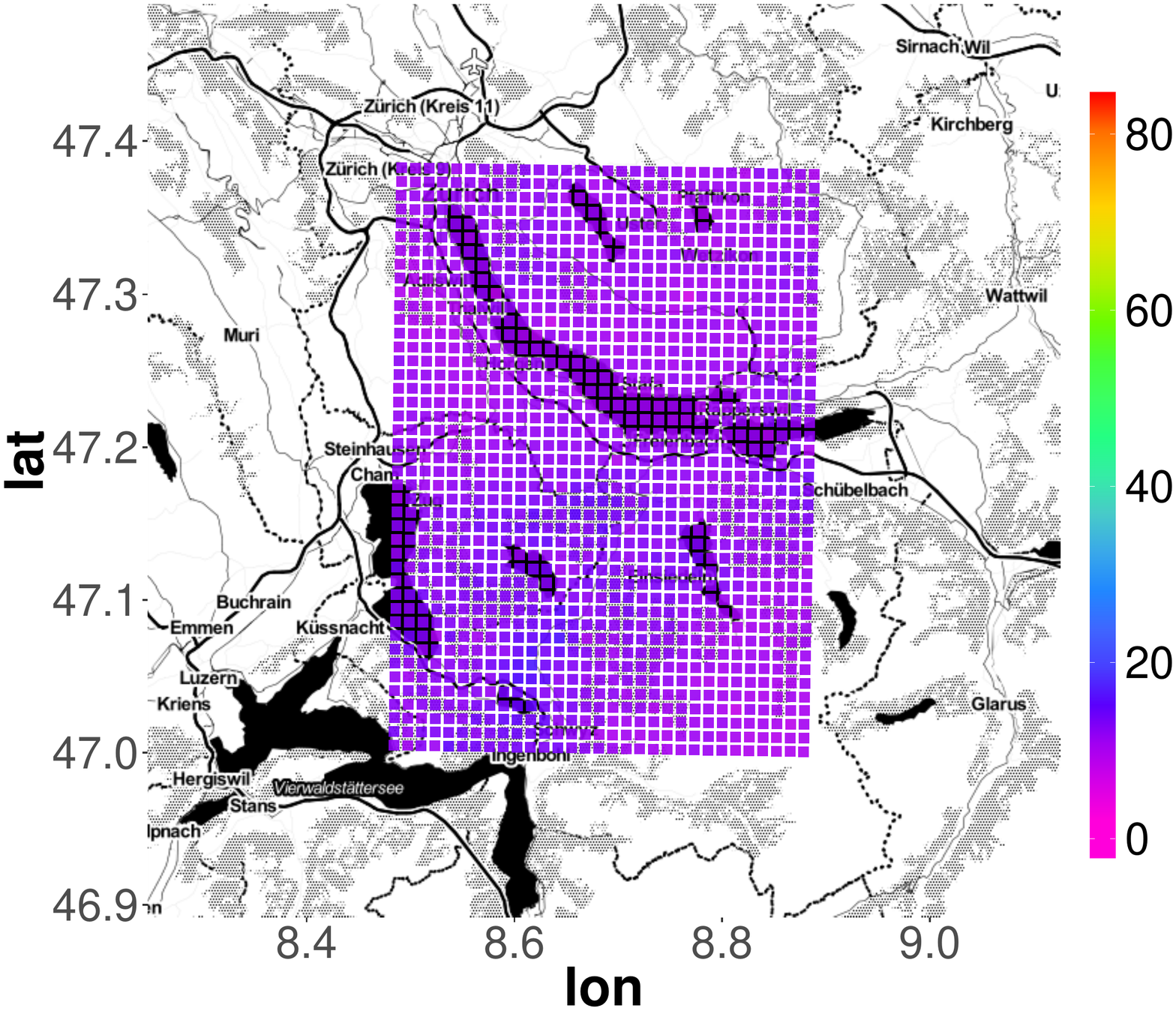}} \\
~\\
 \multicolumn{4}{c}{$\R_1'(x) =  |S|^{-1} \int_S x(s)\, \D{s} {|\widehat{x}_1|}/{|\widehat{x}|}$} \\
 ~\\
 $\widehat{\xi} = 0.2_{0.02}$ &  \raisebox{-.5\height}{\includegraphics[scale = 0.1]{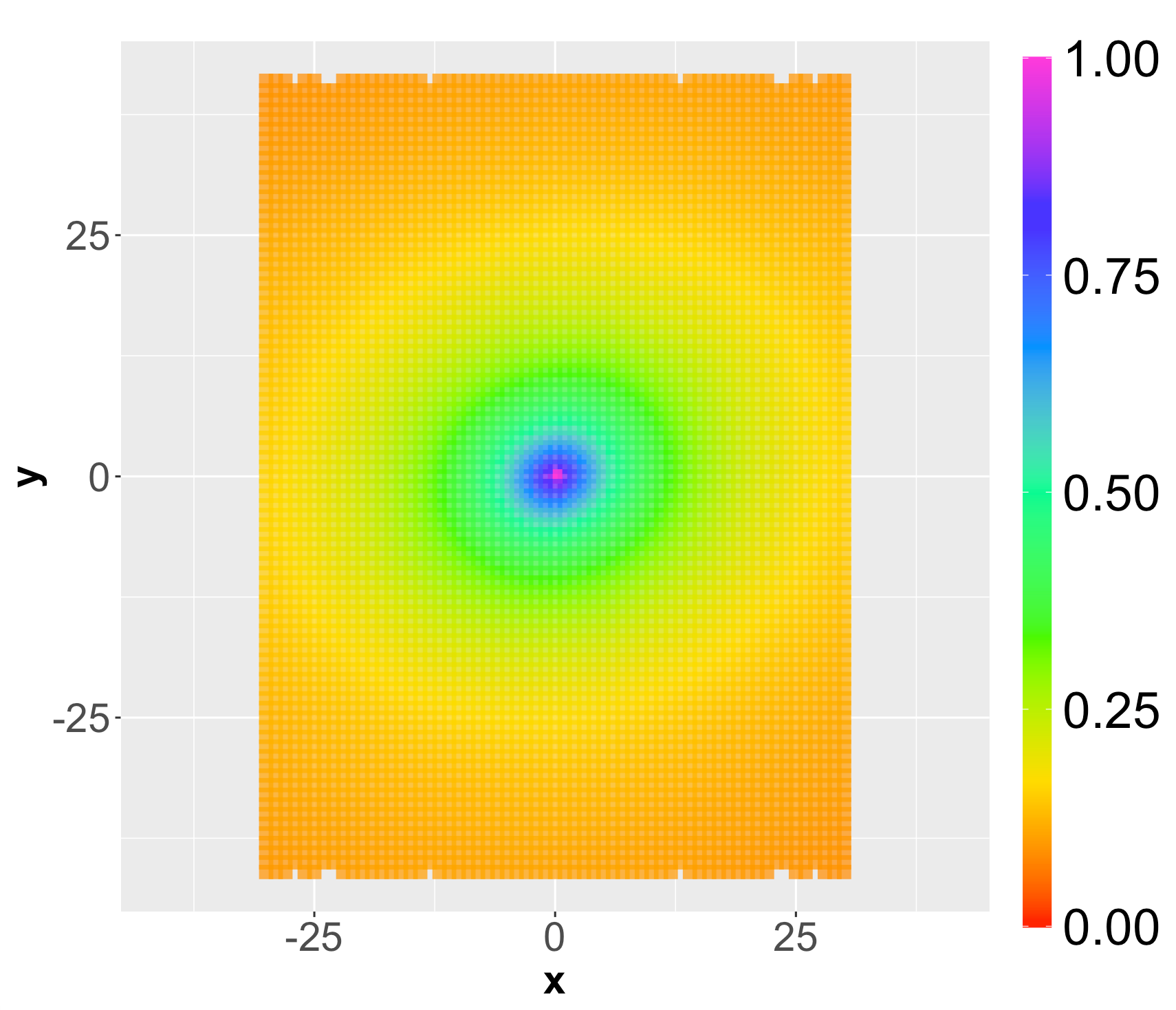}}   & 
 \raisebox{-.5\height}{\includegraphics[scale = 0.13]{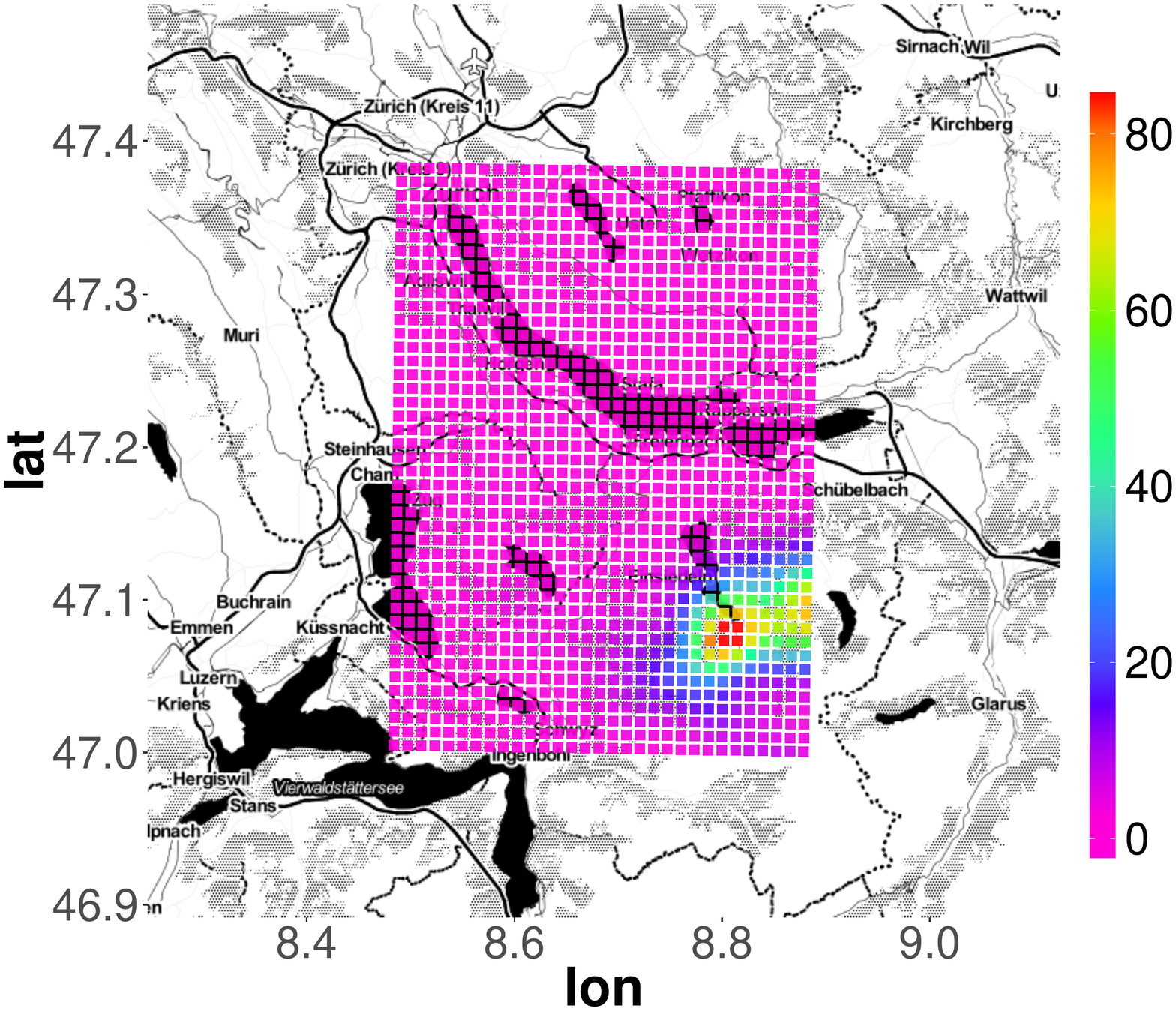}}  &   \raisebox{-.5\height}{\includegraphics[scale = 0.13]{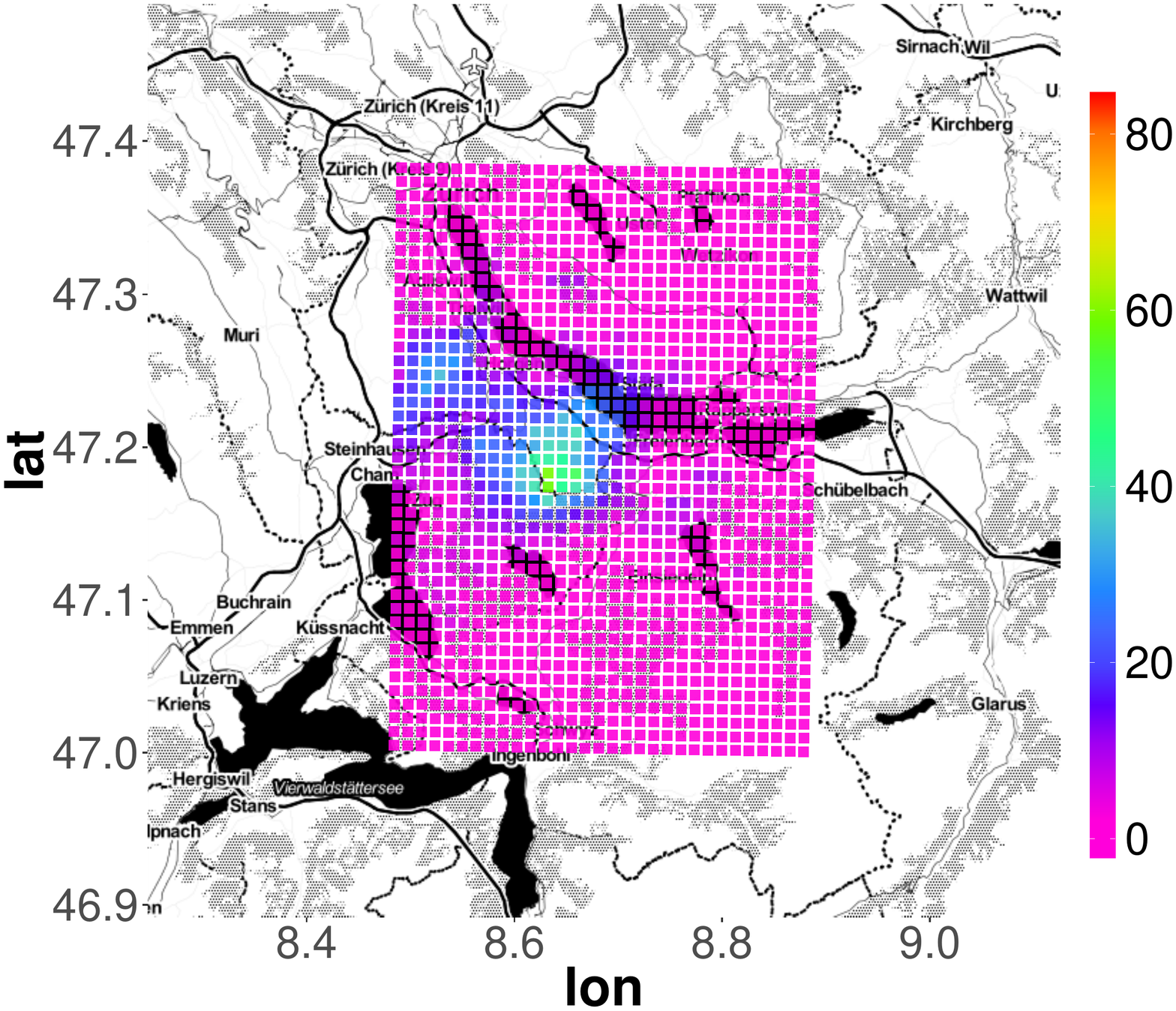}} \\
 ~\\
 \multicolumn{4}{c}{$\text{r}_2(x) = \max_{s \in S} x(s)$} \\
\end{tabular}
\caption{Fitted models for extremes of the modified spatial average (top) and spatial maxima (bottom).  Left: estimated tail index and fitted extremogram.  Center: largest observed events.  Right: simulated events.}
\label{fig: rainfall models}
\end{figure}

The marginal model fits for both risk functionals were checked using QQ-plots and were found to be good everywhere.  The estimated models, summarized in Figure~\ref{fig: rainfall models}, have different tail behaviours.  Events corresponding to exceedances for $\R'_1$ have estimated tail index $\widehat{\xi_1} = -0.2_{0.04}$, and those for $\R_2$ have $\widehat{\xi_2} = 0.02_{0.02}$; the rough standard errors shown as subscripts were obtained by resampling.  The estimates suggest that spatially widespread accumulations of rainfall are bounded above, whereas the tail decay estimate for locally heavy rain lies in the Fr\'echet regime, which gives no upper bound.  While one could argue that events for  $\R_2$-exceedances will dominate in the limit,  other types of event are  nevertheless of interest, especially if we consider more complex definitions of extremes.  In this application there appears to be a worst-case scenario for large widespread rainfall over the Sihl river basin that could be used in deriving mitigation procedures, above which we need \bRev{to} focus only on locally intense rainfall events. 

For $\R'_1$ the lower score was obtained with a Mat{\'e}rn model, while the Bernstein semi-variogram gave a higher likelihood for $\R_2$-exceedances.
The fitted models show much weaker extremal dependence for $\R_2$, whereas the theoretical extremogram does not drop below $0.7$ for $\R'_1$, highlighting the importance of  suitable definitions of risk.   The illustrative simulations in Figure~\ref{fig: rainfall models} appear consistent with the data. The model estimated for $\R_2$ seems to over-estimate extremal dependence compared to the data: as the threshold increases, the estimated extremogram decreases.  This decrease in dependence at high levels is not accommodated by our model.   \citet{Huser2017} and \citet{Huser2019} have proposed spatial models in which dependence decreases that could be extended to our setting. 

\section{Discussion}

Peaks-over-threshold methods are widely used for modelling the tails of univariate distributions, but a more general setting is needed to take advantage of complex data. 
This paper extends peaks-over-threshold analysis to continuous stochastic processes. Exceedances are defined in terms of a real-valued functional $\R$, and modelled with the generalized $\R$-Pareto process, which appears as the limit for $\R$-exceedances of a properly rescaled process and is the functional generalization of generalized Pareto variables. 
We derive construction rules for such processes, give simulation algorithms, highlight their link to max-stable processes, and propose inference and model validation procedures.   The ideas are illustrated by applications to  windstorms and spatial rainfall.

The minimal assumptions under which one can derive the convergence of conditional $\R$-exceedances are quite weak: if the marginal distributions are assumed to have generalized Pareto tails, then the existence of a non-zero joint limit should naturally be considered. If the assumptions are unrealistic, then the need for a functional model might be questioned. If one assumes that such a limit is a continuous function, then generalized $\R$-Pareto processes arise naturally.  A consequence is that the convergence results presented here do not allow asymptotic independence throughout $S$, which would involve the appearance of discontinuous functions in the limit. More general convergence notions, as yet undeveloped, are needed to provide a fully unified peaks-over-threshold analysis for functions.

The stochastic windstorm generator obtained in Section~\ref{sec: windstorms} produces events consistent with historical records.  The simulated windstorms peak at the centre of the 24-hour time window, but  this is not accommodated in the estimation process;  ideas of  \citet{Coles.Tawn.Smith:1994} or \citet{Smith.Tawn.Coles:1997} might be adapted to deal with this in future work.  Moreover, the underlying model does not capture the full complexity of the spatio-temporal structure of extreme windstorms, whose dependence changes over space.  \citet{Oesting2013} show that the potential types of non-stationarity are limited, but models with varying local anisotropy, such as in \cite{Fuglstad2015a} or \cite{Fouedjio2016}, would be a natural extension.  The realism of simulated storms might be improved by using the methodology of \cite{Lindgren2011a}  to build physically-inspired  non-stationary spatio-temporal dependence structures, using for instance the diffusion equation, and this would be computationally efficient and perhaps more realistic.  Our windstorm model introduces non-stationarity by allowing the probability that a windstorm will occur to depend on explanatory variables, but the distribution of conditional $\R$-exceedances does not vary, and this may be too restrictive.  The methodology is flexible enough to allow explanatory variables to influence the generalized $\R$-Pareto process, if necessary.  

The rainfall application in Section~\ref{sec: rainfall}  highlights the importance of an appropriate definition of risk by illustrating how it impacts the tail behaviour of the selected events and showing how $\R$-exceedances allow one to disentangle mixtures of extremes. The approximation provided by the asymptotic framework would be questionable if, for instance,  marginal shape parameter estimates varied strongly over the region and disagreed with that for the risk functional. Sub-asymptotic models for which extremal dependence diminishes with intensity would then be preferable, as this phenomenon is commonly observed with rainfall.

Another notion of complexity for extremes is linked with compound events.  Let $(X^1,X^2)$ be a bivariate continuous stochastic process and let $\R^1$ and $\R^2$ be suitable risk functionals.  Then under conditions similar to those above, the functional
\begin{equation*}
\R(X^1,X^2) = \min \left\{ {\R^1(X^1)}-{u^1}, {\R^2(X^2)}-{u^2}\right\}
\end{equation*}
can be used to characterize extremes of both types and could be applied when studying infrastructure that is vulnerable to different sources of risk.  This differs from~\eqref{mix-risk.eq}, which concerns multiple risks for a single process.

\subsection*{Acknowledgement}
We thank the Swiss National Science Foundation for financial support and reviewers for exceptionally helpful comments.  Alexis Berne and Gionata Ghiggi kindly gave us access to the Swiss radar rainfall data, which were originally provided by Meteoswiss.

\bibliographystyle{apalike}
\bibliography{library}

\newpage\appendix

\section{Limit tail distribution for linear risk functionals}\label{app: general risk}

In this appendix we derive the limiting distributions of $\R$-exceedances when $\R$ is linear, i.e., $\R(x + y) = \R(x) + \R(y)$ for all $x,y \in \Set$.
If $\xi = 0$, we further suppose that $\R$ is an {evaluation functional}, i.e., $\R(x) = x(s)$ for some $s\in S$.
In both cases, $\R$ is a valid linear risk functional.
Let $\xi $ be a real-valued shape parameter and let $a\equiv a(s) > 0$ and $b\equiv b(s)$ be continuous functions defined for $s\in S$.
We again consider the sets $\Set^{\xi,a,b}$ defined in~\eqref{F.eq}.
Given $a$ and $b$ and threshold $u\geq 0$, a linear risk functional over $\Set^{\xi,a,b}$ can take values only in the intervals 
$$
\radialSet^{\xi}_\R(u) = \begin{cases} [u,\infty), &\xi \geqslant 0,\\ \left[u,\R(b)-\xi^{-1}\R(a)\right), & \xi < 0.
\end{cases}
$$

As in the general case, we assume that the sequence of functions $\{a_n\}$ satisfies~\eqref{eq: asym cond}. 
Linearity of $\R$ ensures that the rates of convergence of $\R(a_n)$ and $a_n'$ are the same.

\begin{theorem}\label{th: linear functional}
Let $X$ be a stochastic process with sample paths in {$\Set$} and let $\R$ be a valid linear risk functional.
If $u \geq 0$, $X \in {\rm GRV}\left(\xi, a_n, b_n, \Lambda\right)$ {and the functions $a_n$} satisfy~\eqref{eq: asym cond}, then 
\begin{equation}
\Prob\left.\left\{\left\lfloor\dfrac{X - b_n}{\R(a_n)}\right\rfloor \in \; \cdot \; \right| \R(X) \geqslant u_n \right\} \rightarrow \Prob(P \in  \cdot ), \quad n \rightarrow \infty,
\end{equation}
where $\lfloor \cdot \rfloor$ is defined in \eqref{floor.eq}, $u_n = \R(a_n)u + \R(b_n) \in \radialSet^{\xi}_\R \{\R(b_n)\}$, and $P$ is a generalized $\R$-Pareto process with tail index $\xi$, scale function $A$, location function zero and measure $\Lambda$.
\end{theorem}
 
\subsection{Generalized $\R$-Pareto processes: definition and properties}

For a scale function $A > 0$ and a linear risk functional $\R$, we define the simplex in the function space $\Set_+$ to be 
$$
\simplex_{r}^{\xi,A} = \left\{\begin{array}{ll} \{ y \in \Set_+ : \R(Ay^\xi) \geqslant 1, \|y\| = 1\}, & \xi > 0, \\
\{ y \in \Set_+ : \R(A\log y)  = 0\}, & \xi = 0, \\
\{ y \in \Set_+ : \R(Ay^\xi) \leqslant 1, \|y\| = 1\}, & \xi < 0.
\end{array}
\right.
$$
and describe the corresponding generalized $\R$-Pareto processes hereafter.
In the linear case definition~\eqref{eq: reference sets} simplifies to
\begin{equation}
\label{mathcalA.eq}
\Ar = \left\{ \begin{array}{ll}
\left\{ y \in \Set_+ : \R\left(Ay^\xi\right) \geqslant 1\right\}, & \xi > 0, \\
\left\{ y \in \Set_+ : \R\left(A\log y\right) \geqslant 0\right\}, & \xi = 0, \\
\left\{ y \in \Set_+ : \R\left(Ay^\xi\right) \leqslant 1\right\}, & \xi < 0.
\end{array}\right.
\end{equation}

\begin{definition}\label{def: generalized pareto process with linear risk functional}
Let $\xi$ be a tail index, let $a(s) > 0$ and $b(s)$ be continuous functions on $S$, let $\R: \Set^{\xi,a,b} \rightarrow \radialSet^{\xi}_\R$ be a valid linear risk functional, let $\Lambda$ be a $(-1)$-homogeneous measure on $\Set_+$ and let $A = a / \R(a)$; note that $\R(A) = 1$.
The generalized $\R$-Pareto process $P$ associated to the measure $\Lambda$ and tail index $\xi $ is the stochastic process on $\{x \in \Set^{\xi,a,b} : \R(x) \geqslant \R(b)\}$ defined by
\begin{equation}\label{eq: linear gen R pareto}
P = \left\{ \begin{array}{ll}  \dfrac{\R(a)}{\xi}R^{\xi} \dfrac{W_{\xi,A}}{\R(W_{\xi,A} )} + b - \xi^{-1}a, & \xi \neq 0,\\
\R(a)\log (R  W_{0,A})  + b, & \xi = 0,
\end{array}\right.
\end{equation}
where $R$ is a scalar unit Pareto random variable independent of $W_{\xi,A}$, and the latter is a stochastic process with state space $S$, taking values in  $\simplex_{r}^{\xi,A}$ and having probability measure 
\begin{equation}\label{eq: angular measure}
\sigma_{\R}^\xi (\cdot) = 
\left\{\begin{array}{ll} \dfrac{\Lambda\left\{y \in \Ar  :  Ay^\xi/\|Ay^\xi \| \in \cdot \right\}}{\Lambda(\Ar)}, & \xi \neq 0, \\
\dfrac{\Lambda\left\{y \in \Ar :  A\log y - \R(A \log y) \in \cdot \right\} }{ \Lambda(\Ar)} , & \xi = 0. \\
 \end{array}\right. 
\end{equation}
\end{definition}

This construction relies on a pseudo-polar decomposition: the process is the product of a radial component, namely a univariate Pareto variable representing the intensity of the process, and an angular component that determines how the process varies over $S$. 

Similarly to the general case, generalized $\R$-Pareto processes with linear risk functionals are closely related to stochastic processes $Y_\R$ defined on $\Ar$ with probability measure $\Lambda(\cdot) / \Lambda(\mathcal{A}^\R)$, where $\Lambda$ is a $(-1)$-homogenous measure on $\Set_+$.
The process $Y_\R$ admits a pseudo-polar decomposition, 
\begin{equation}\label{eq: polar Y lin}
Y_\R \bRev{\Deq} RW \;\left|\; \R[A\xi^{-1}\{(RW)^\xi - 1\}]\right. \geq 0,
\end{equation}
where $R$ is a unit Pareto random variable independent of the stochastic process $W$ with state space $S$ taking values in $\simplex= \{y \in \Set_+: \|y\|_{1} = 1\} $, with probability measure~\eqref{eq: angular Y}. 
Following~\eqref{eq: angular measure}, the angular process $W_{\xi,A}$ can be constructed as
\begin{equation}\label{eq: angle gen Pareto}
W_{\xi,A} =  \left\{\begin{array}{ll}   \dfrac{AY_\R^\xi}{\left\| AY_\R^\xi\right\| },&  \xi \neq 0, \\
\exp \left\{ A\log Y_\R - \R(A\log Y_\R) \right\}, & \xi = 0, \\
 \end{array}\right.
 \end{equation}
which allows simulation of generalized $\R$-Pareto processes for linear $\R$.

In contrast to the general case, the distribution of the risk $\R(P)$ over the threshold $\R(b)$ is generalized Pareto with tail index $\xi$ and scale $\R(a)$; see Appendix~\ref{app: gen r pareto marginals}. Marginal conditional distributions above a sufficiently high threshold are also generalized Pareto.

\subsection{Simulation algorithm for linear risk functionals}\label{app: sim linear}

Let $\R$ be a valid linear risk functional and let $P$ be the corresponding  generalized $\R$-Pareto process with measure $\Lambda$, positive tail index $\xi$, positive  scale function $a$ and location function $b$.
If we can find a threshold $u > 0$ such that
$$
\left\{y \in \Set_+ : \R\left(Ay^{\xi}\right) \geqslant 1 \right\} \subset \left\{y \in \Set_+ : \|y\|_{1} \geqslant u \right\},
$$
then Algorithm~\ref{algo2} enables the simulation of generalized $\R$-Pareto processes for a given value of the risk functional by replacing $R_2$ therein by any  desired risk level. In the algorithm, every unit Pareto variable is independent of every other, and all have survivor function $1/v$ for $v>1$.

A similar algorithm can be derived for $\xi <0$ simply by replacing $<1$ by $>1$ in the while condition, and for $\xi = 0$ by replacing $Y_r^\xi$ by $\log Y_r$ throughout Algorithm~\ref{algo2}.

\begin{algorithm}
\SetAlgoLined
Input: scaling functions $a$ and $b$, threshold function $u$, and scalar shape parameter $\xi$.

Set $Y_\R = 0$ and  $A = a / \R(a)$\;
 
  \While{$\R\left(AY_\R^{\xi} \right) < 1$}{
  generate a unit Pareto random variable  $R_1$\;
  generate $W_1$ with probability measure $\sigma_0$ given in~\eqref{eq: angular Y}\;
  set $Y_\R = uR_1W_1$;
 }
 set $W_2 = {AY_\R^{\xi} }/{\|AY_\R^{\xi} \|}$.
 
Generate a unit Pareto random variable  $R_2$.

Return $P = \R(a)\xi^{-1} R_2^{\xi} W_2/ \R(W_2) + b - \xi^{-1}a$.
 \caption{Simulation of generalized $\R$-Pareto process, $P$, with $\xi > 0$ and linear $\R$}
 \label{algo2}
\end{algorithm}

\section{Generalized $\R$-Pareto and max-stable processes}\label{sec: ms process}

In univariate extreme-value theory the marginal assumptions of equation~(\ref{eq: pot}) are equivalent to convergence of rescaled block maxima toward the generalized extreme value (GEV) distribution, i.e., for each $s\in S$ we have 
$$
\lim_{n \rightarrow \infty} \Pr\left\{\dfrac{\max_{j = 1,\dots,n} X_j(s) - b_n(s)}{a_n(s)} \leqslant z\right\} = 
\left\{\begin{array}{ll}
\exp\left\{ - \left(1 + \xi z \right)^{-1/\xi}_+ \right\}, & \xi \neq 0, \\
\exp\left\{ - \exp \left( -z\right) \right\} , & \xi = 0.
\end{array}\right.
$$
There is a similar relation between generalized $\R$-Pareto processes and max-stable processes. The latter have a variety of  stochastic representations; we use that of \citet{DeHaan1984}, which relies on Poisson point processes. 

Consider the Poisson process $(R_j,W_j)_{j = 1,\dots}$ on $(0,\infty) \times \simplex_0$ with intensity measure $r^{-2}\D{r} \times \sigma_0(\D{w})$, where $\sigma_0$ is given by~\eqref{eq: angular Y}.
Then the process
\begin{equation}\label{eq: max stable}
M(s) =\left\{\begin{array}{ll}  \sup_{j \geqslant 1} a(s)\dfrac{\{R_jW_j(s)\}^{\xi} - 1}{\xi} + b(s), & \xi \neq 0, \\
 \sup_{j \geqslant 1}  a(s)\log \{R_jW_j(s)\} + b(s), & \xi = 0,
\end{array}\right. \quad s \in S,
\end{equation}
is max-stable with exponent measure $\Lambda \circ T_{\xi,a,b}$ \citep[Proposition 3.7]{Resnick1987}, where $T_{\xi,a,b}(z)$ is the non-atomic map 
$$
T_{\xi,a,b}(z) = \left\{
\begin{array}{ll}
\left\{1 + \xi (z - b)/{a}\right\}_+^{1/\xi}, & \xi \neq 0, \\
\exp \left\{ (z - b)/{a} \right\},&\xi = 0.
\end{array}\right.
$$

The finite-dimensional distribution function of $M(s)$ at locations $s_1, \dots, s_L \in S$ is
\begin{equation}
\label{max-s.eq}
\Pr\left\{M(s_l) < z_l , l = 1,\dots, L\right\} = \exp\left\{- \Lambda \circ T_{\xi,a,b} \left({\mathcal A}_z\right)\right\},
\end{equation}
where ${\mathcal A}_z = \{x \in \Set^{\xi,a,b} : \max_{l = 1,\dots,L} x(s_l)/z_l \geq 1\}$.  The exponent in expression~\eqref{max-s.eq} contains the measure of a generalized $\R$-Pareto process with risk functional $\R(x) = \max_{l = 1,\dots,L} x(s_l)$.
According to representation~\eqref{eq: max stable}, the max-stable process $M(s)$ is constructed using infinitely many single events of a Poisson process, and the $\R$-exceedances of these events above a threshold $u$ correspond to a generalized $\R$-Pareto process; 
the latter also arises as the limit of $\R$-exceedances for its corresponding max-stable process. 
The intensity measure of the Poisson process, which is necessary to model the occurrence of single events in the max-stable process, can be transformed to a Pareto distribution by conditioning on the $\R$-exceedance.  Outside the max-stable framework the number of exceedances need not be Poisson; for instance, seasonality or trend can be incorporated, as in the windstorm generator of Section~\ref{sec: strom and freq}. 

\section{Statistical inference}\label{app: stat inf details}
Statistical inference for generalized $\R$-Pareto processes with non-linear risk functional follows the same principle as in Section~\ref{sec: stat inference} and relies on the approximation
\begin{equation}\label{eq: inference non linear}
\Prob\left\{\dfrac{X - b_n}{\R(a_n)} \in \calR\right\} \approx \Prob\left[\R\left\{\dfrac{X - b_n}{\R(a_n)}\right\} \geqslant 0 \right] \times \Prob(P \in \calR),
\end{equation}
where $\calR \subset \calR(0) = \{x \in \Set^{\xi,A,0}: \R(x) \geqslant 0\}$ for sufficiently large $n$.
Likelihood-based inference using~\eqref{eq: inference non linear} is delicate in general.
Indeed, estimation of the marginal parameters jointly with the dependence parameters is typically numerically unstable if the set of observed $\R$-exceedances, $\mathcal{E}_\R = \{x_j : \R\{(x_j - b_n) / \R(a_n)\} \geqslant 0, j = 1,\dots, n\}$, depends on $a_n$ and $b_n$.
Thus it is necessary to either restrict the functionals to those for which $\mathcal{E}_\R$ is independent of the rescaling, as is the case for linear functionals, or to use a two-step procedure.
In the latter, we first estimate the marginal parameters $\widehat{a}_n$, $\widehat{b}_n$ and $\widehat{\xi}$ and then fix them while estimating the dependence model. 
To have a marginal model tailored to the $\R$-exceedances and thus to disentangle any mixtures in the tail, we propose an iterative procedure.  The underlying principle is, if necessary, to refine a different risk functional $\R'$ until the set of $\R'$-exceedances of $(x_j-\widehat{b}_n)/\R(\widehat{a}_n)$ equals the set of $\R$-exceedances of $\bRev{x_j}$. To do so, we 
\begin{enumerate}
\item set $a = 1$, $b = 0$, and $\mathcal{E}_{\R} = \{x_j : \R(x_j) \geqslant 0, j = 1,\dots, n\}$;
\item define or refine $\R'$, for example by applying a filter as in Section~\ref{sec: rainfall}; 
\item \bRev{estimate marginal parameters $\widehat a_n$, $\widehat b_n$ and $\widehat\xi$ using 
$$
\mathcal{E}_{\R'} = \{x_j : \R'\{(x_j - b) / \R'(a)\} \geqslant 0, \, j = 1,\dots, n\}
$$
where $a$ and $b$ remain fixed};
\item set $a = \widehat{a}_n$ and $b = \widehat{b}_n$;
\item return to step (b) if $\mathcal{E}_{\R'} \neq \mathcal{E}_\R$;
\item \bRev{use $\widehat a_n$, $\widehat b_n$ and $\widehat \xi$ to rescale the exceedances and use them to estimate the dependence model as described in Section~\ref{sec: stat inference}.}
\end{enumerate}
An example of functional refinement inspired by the application of Section~\ref{sec: rainfall} consists of modifying the frequency domain of a Fourier filter until $\mathcal{E}_\R$ and $\mathcal{E}_{\R'}$ are equal.

Identifiability issues caused by the conditional nature of generalized $\R$-Pareto processes might also arise.
A natural idea is to set $b_n$ equal to local empirical quantiles estimated from $\mathcal{E}_\R$.
Apart from these considerations,  the inference procedures described in Section~\ref{sec: stat inference} can be used in the same way.

\section{Gradient scoring inference}\label{app: score matching}
This section summarizes the background behind score-matching inference as presented in \citet{Fondeville2017}. 

Let $x_1,\ldots, x_n \in \Set$ be independent realizations of a generalized regularly varying stochastic process $X$ observed at locations $s_1,\dots, s_L \in S$ with asymptotic measure $\Lambda$ parametrized by ${\theta}_W$.
For simplicity, we suppose that $\xi$, $a_n$ and $b_n$ are known and can used to obtain the rescaled process $Y = \{1 + \xi(X - b_n)/a_n  \}_+^{1/\xi}$, which has unit tail index and sample space in $\Set_+$.
In practice, the re-scaling parameters must be estimated, for example using the independence likelihood as in Section~\ref{sec: stat inference}. Joint estimation of $\vartheta = (\xi, \theta_{a_n},\theta_{b_n},\theta_W)$ by score matching is also possible by accounting for the rescaling in the following formulae, using the chain rule for composite derivatives.

The log-likelihood function based on the asymptotic model of $\R$-exceedances among $x_1,\ldots, x_n$ is given at~\eqref{eq: gen pareto log lik}. In terms of $\theta_W$ it is necessary to compute terms of the form  
$$
\dfrac{\lambda_{\theta_W}( y)}{\Lambda_{\theta_W}\left(\Ar\right)}, \quad y \in \RR^L_+ \setminus \{0\}
$$ 
where $\Lambda_{\theta_W}$ is given by \eqref{eq: mvt intensity}. Classical likelihood inference minimizes the Kullback--Leibler divergence, which is equivalent to maximizing
$$
-n_0\log {\Lambda_{\theta_W}\left(\Ar\right)} + \sum_{i = 1}^{n_0} \log{\lambda_{\theta_W}(y_i)} 
$$
with respect to $\theta_W$, requiring either evaluation or simplification of the scaling constant $\Lambda(\A_{\R})$, whose complexity increases with the number of dimensions. Efficient algorithms have been developed only for the maximum function $\R = \max$, and they are computationally demanding when $L$ is larger than a few hundred.

Score matching, based on the gradient scoring rule \citep{Hyvarinen2005}, uses {only} the derivative $\nabla_{x} \log f^\R(x)$, making the scaling constant $\Lambda_{\theta_W}(\A_{\R})$ vanish.
\citet{Hyvarinen2007} adapted this scoring rule for strictly positive variables, and \citet{Fondeville2017} extended the methodology to domains such as $\A_{\R}$.
The inference procedure minimizes the divergence measure
\begin{equation*}\label{eq: gradient divergence}
 \int_{A_{\R}} \| \nabla_{y} \log  \lambda_{{\theta}_W}(y) \otimes {w}(y) - \nabla_{y} \log  \lambda(y) \otimes {w}(y) \|^2_2 \: \dfrac{\lambda(y)}{\Lambda\left(\Ar\right)} dy,
\end{equation*}
where ${\lambda}$ is the underlying intensity of angular process $W$ with measure $\Lambda$, $\lambda_{{\theta}_W}(y)$ is  differentiable for all $\theta_W \in \Theta_W$ on { $\A_{\R} \setminus \partial \A_{\R}$, $\partial \A$ denotes the boundary of $\A$}, $\nabla_y$ is the gradient operator, ${w}: \A_{\R} \rightarrow \mathbb{R}_+^L$ is a positive weight function,  and $\otimes$ denotes the Hadamard product.
If ${w}$ is differentiable on $\A_{\R}$ and vanishes on $\partial \A_{\R}$, then minimizing
\begin{equation*}\label{eq: grad score new}
\sum_{i = 1}^{n_0}\sum_{l =1}^L \left(2w_l(y_i)\frac{\partial w_l(y_i)}{\partial y_l} \frac{\partial \log {\lambda_{{\theta}_W}}(y_i)}{\partial y_l} + w_l(y_i)^2 \left[ \frac{\partial^2 \log {\lambda_{{\theta}_W}}(y_i)}{\partial y_l^2} + \frac{1}{2} \left\{\frac{\partial \log {\lambda_{{\theta}_W}}(y_i)}{\partial y_l} \right\}^2 \right] \right)
\end{equation*}
yields an asymptotically normal estimator \citep[][Appendix~D]{Fondeville2017}. 

The gradient score for a log-Gaussian Pareto process satisfies the necessary regularity conditions for normality. The formulae for the Brown--Resnick model, used in Section~\ref{sec: dep model windstorms} to estimate windstorm dependence, can be found in Appendix~B of \citet{Fondeville2017}.

\section{Proofs}\label{app: proofs}
\subsection{Theorem \ref{th: generalized convergence}}\label{app: proof th1}

Recall that $S \subset \mathbb{R}^D$ $(D \geqslant1)$ is a compact metric space, let $\Set$ denote the Banach space of real-valued continuous functions on $S$ with norm $\|x\|$ and let $\Set_+ $ denote the subset of $\Set$ containing only non-negative functions that are not everywhere zero; thus $\Set_+$ excludes the zero function $\{0\}$. 
In the statistics of extremes of positive functions, the cones $\{0\}$ or $\{x \in \Set_+ : \inf_{s \in S} x(s) = 0\}$ are generally excluded from the set of continuous non-negative functions over $S$ in order that points with infinite mass do not appear in the limiting measure. 
Let $M_{\Set_+}$ denote the  class of {Borel} measures on the Borel sigma-algebra $\mathcal{B}(\Set_+)$ associated to $\Set_+$.  We say that a set $\A \in \mathcal{B}(\Set_+)$ is bounded away from $\{0\}$ if $d(\A,\{0\}) = \inf_{x\in \A} \|x\| > 0$.

We first transform the process $X$, which takes values in $\Set$, using the mapping
$$
T_{\xi,a_n,b_n}(x) = \left\{
\begin{array}{ll}
\left\{ 1 + \xi (x - b_n)/{\R(a_n)}\right\}_+^{1/\xi}, & \xi \neq 0, \\
\exp \left\{\bRev{x} - b_n)/{\R(a_n)} \right\},&\xi = 0,
\end{array}\right.
$$
to ensure that $T_{\xi,a_n,b_n}(X)$ takes values in $\Set_+$.
We then apply classical theory on regularly varying processes on $\Set_+$: 
a sequence of measures $\{\Lambda_n\}\subset M_{\Set_+}$ is said to converge to a limit $\Lambda\in M_{\Set_+}$ if
$\lim_{n \rightarrow \infty} \Lambda_n(\A) = \Lambda(\A)$, 
for all $\A \in \mathcal{B}(\Set_+)$ bounded away from $\{0\}$ and for which  $\Lambda(\partial \A) = 0$ on the boundary $\partial \A$ of $\A$. This mode of convergence is called $\hat{w}$-convergence and written $\Lambda_n \xrightarrow{\widehat{w}} \Lambda$ \citep{Hult2005}; 
see \citet{Lindskog2014} for further details and equivalent definitions.

The essential idea is as follows. Generalized regular variation of $X$, as defined in~\eqref{eq:gen rv}, is equivalent to regular variation of the transformed process $Y_n=T_{\xi,a_n,b_n}(X)$ on $\Set_+$.  For very large $n$ we can argue heuristically that $\{1+ \xi(X-b_n)/a_n\}^{1/\xi}_+\approx  Y$, where $Y$ is a random function in $\Set_+ $ having measure $\Lambda$, and therefore $(X-b_n)/a_n \approx (Y^\xi-1)/\xi$. Hence
$$
{X-b_n\over r(a_n)} = {a_n\over r(a_n)} \times {X-b_n\over a_n} \approx A\times {Y^\xi-1\over \xi}, 
$$
with appearance of $A$ following from assumptions~\eqref{eq: asym cond} and~\eqref{eq:marginal conditions 2}, and the continuity of $\R$ then implies that
$$
\R\left\{{X-b_n\over r(a_n)} \right\} \approx  \R\left( A  {Y^\xi-1\over \xi} \right).
$$
Conditional on $\R\left\{(X-b_n)/ r(a_n)\right\}\geq 0$, therefore, the limiting process takes values in the set $\A_r$ defined at~\eqref{eq: reference sets}, and thus the limiting measure of $(X-b_n)/ r(a_n)$ conditional on $\R\left\{(X-b_n)/ r(a_n)\right\}\geq 0$ is given by
$$
\dfrac{\Lambda\{y \in \Ar} :  A(y^\xi - 1)/\xi \in \cdot\}{\Lambda(\Ar)}.
$$

To make the above heuristic reasoning more precise, we proceed similarly to \citet{Engelke}. Let $X \in {\rm GRV}(\xi, a_n, b_n, \Lambda)$ be as defined in Section~\ref{sec: generalized functional pot} and suppose first that  $\xi >  0$, in which case we can choose \citep[Proposition 1.11]{Resnick2007}
$$
b_n \equiv 0, \quad a_n(s) = \inf\{ x \in \RR : \Pr\{X(s) \leq x\} \bRev{\,\geq\,} 1 - 1/n\},
$$ 
and $\R(a_n) \rightarrow \infty$ as $n \rightarrow \infty$.
The continuous function $A$ in assumption \eqref{eq: asym cond} is strictly positive and thus is bounded away from zero on the compact set $S$.
Hence, for any $\varepsilon > 0$,  $|\R(a_n)^{-1} a_n(s) - A(s)| < \varepsilon A(s)$ for all $s \in S$ and sufficiently large $n$.
We assumed that  for $\xi>0$ the support of $X$ has a finite lower endpoint, so there exists $b_0 \in \RR$ such that $X(s) \geq b_0$ for all $s \in S$.
For a sufficiently large $n$, therefore,
\begin{align}
\dfrac{X - b_n}{\R(a_n)}  ={} \dfrac{a_n}{\R(a_n)} \dfrac{X - b_n}{a_n}  & \geq (1-\varepsilon)  \left\lfloor\dfrac{A(X - b_n)}{a_n}\right\rfloor  - \sup_{s\in S}\left(\frac{-X}{\R(a_n)},0\right) \nonumber \\
& \geq (1-\varepsilon)  \left\lfloor\dfrac{A(X - b_n)}{a_n}\right\rfloor  - \left|\frac{b_0}{\R(a_n)}\right| \nonumber \\ & \geq (1-\varepsilon)  \left\lfloor\dfrac{A(X - b_n)}{a_n}\right\rfloor   - \varepsilon,\label{eq: ineq 1}
\end{align}
and likewise
\begin{align}\label{eq: ineq 2}
\dfrac{X - b_n}{\R(a_n)} \leq{}& (1+\varepsilon)\left\lfloor\dfrac{A(X - b_n)}{a_n}\right\rfloor.
\end{align}
The bounds~\eqref{eq: ineq 1} and~\eqref{eq: ineq 2} apply for any $\varepsilon$, so as $\xi>0$ and $A>0$ and $X \in {\rm GRV}(\xi, a_n, b_n, \Lambda)$, 
 \begin{eqnarray*}
\lim_{n \to \infty} n \Pr  \left\{\left\lfloor\dfrac{X - b_n}{\R(a_n)} \right\rfloor\in \cdot \right\}
&=& \lim_{n \to \infty} n  \Pr\left\{  \left\lfloor\dfrac{A(X - b_n)}{a_n}\right\rfloor  \in \cdot \right\} \\
&=& \lim_{n \to \infty} n  \Pr\left\{  \max\left\{\dfrac{A(X - b_n)}{a_n},-A/\xi\right\}  \in \cdot \right\} \\
&=& \lim_{n \to \infty} n  \Pr\left\{  \left(1+\xi\dfrac{X - b_n}{a_n}\right)^{1/\xi}_+ \in (1+\xi \cdot/A)^{1/\xi} \right\} \\
&= &\Lambda\left\{y \in \Set_+ :  A \dfrac{y^\xi - 1}{\xi} \in \cdot \right\}.
\end{eqnarray*}
Monotonicity of $\R$ likewise implies that 
$$
\R\left\{(1+\varepsilon) \left\lfloor\dfrac{A(X - b_n)}{a_n}\right\rfloor  \right\}\geq \R\left\{\dfrac{X - b_n}{\R(a_n)} \right\} \geq \R\left\{(1-\varepsilon) \left\lfloor\dfrac{A(X - b_n)}{a_n}\right\rfloor  - \varepsilon\right\}, 
$$
so the events $\R[\lfloor\{A(X - b_n)\}/a_n\rfloor ] \geq 0$ and $\R\{(X - b_n)/\R(a_n)\} \geq 0$ coincide when $n\to\infty$.

As the risk functional $\R$ is valid, $\R( - A\xi^{-1}) < 0$ and $\R$ is continuous at $ - A\xi^{-1}$, so $
d_\infty\left(\Ar, \{0\}\right) > 0$, 
i.e.,  $\Ar$ is bounded away from the singleton $\{0\}$.  Thus $\widehat{w}$-convergence can be applied on any $\mathcal{A} \subset \Ar$, yielding 
\begin{equation*}
\lim_{n \to \infty} \Pr\left.\left[ \left\lfloor\dfrac{X - b_n}{\R(a_n)}\right\rfloor  \in \cdot \, \right|\, \R\left\{\dfrac{X - b_n}{\R(a_n)}\right\} \geqslant 0\right] = 
\dfrac{\Lambda\{y \in {\Ar} :  A\xi^{-1}(y^\xi - 1) \in \cdot\}}{\Lambda(\Ar)}.
\end{equation*}

For $\xi \leqslant 0$, the operator $\lfloor \cdot \rfloor$ simplifies to the identity, making inequalities such as~\eqref{eq: ineq 1} and~\eqref{eq: ineq 2} superfluous, and the hypothesis $\R(x) \rightarrow -\infty$ as $x \rightarrow -\infty$ ensures that 
$\Ar$
is also bounded away from $\{0\}$.
The argument when $\xi <0$ is analogous to that for $\xi>0$,
and for $\xi = 0$ we obtain
 \begin{equation*}
    \lim_{n \to \infty} n\Prob\left\{\left\lfloor \dfrac{X - b_n}{{\R(a_n)}}\right\rfloor \in \cdot \right\} = 
      \Lambda\left\{ y \in \Set_+ : A\log y  \in \cdot \right\},
 \end{equation*}
which proves the theorem. \hfill $\Box$

\subsection{Theorem \ref{th: linear functional}}\label{app: linear functional}
We start from the conclusion of Theorem \ref{th: generalized convergence}. For $\xi \neq 0$, we use the pseudo-polar decomposition centered at $-A/\xi$, i.e., 
  $$
  \rho = \R(x) + \xi^{-1}, \quad w = \text{sign}(\xi) \dfrac{x + \xi^{-1}A}{\|x + \xi^{-1}A\|}.
  $$

For $\xi >0$, let $\rho' \geqslant \xi^{-1}$,  let $\simplexSub \subset \simplex_\R^{\xi,A}$ and let 
$$
\A_{\rho',\simplexSub}=\{y\in\Set_+: y = \rho w\mbox{ where } \rho\geq \rho', w\in\simplexSub\}.
$$
Then the linearity of the risk functional $\R$, the fact that $\R(A)=1$ and the $(-1)$-homogeneity of $\Lambda$ yield
 \begin{align*}
    \Lambda\left( \A_{\rho',\simplexSub} \right) &=\Lambda\left\{y \in \Set_+: \R_{}\left\{\xi^{-1}A(y^\xi - 1)\right\} + \xi^{-1} \geqslant \rho', \text{sign}(\xi)\dfrac{\xi^{-1}Ay^\xi}{\|\xi^{-1}Ay^\xi \|_{}} \in \simplexSub \right\}  \\
    & ={}\left(\xi \rho'\right)^{-1/\xi} \Lambda\left\{y \in \Set_+ : \R_{}\left(Ay^\xi\right) \geqslant 1,  {Ay^\xi}/{\|Ay^\xi \|_{}} \in \simplexSub \right\}  , \\
      & ={}(\xi \rho')^{-1/\xi} \Lambda\left\{y \in \Set_+ : \R_{}\left(Ay^\xi\right) \geqslant 1 \right\}  \times \sigma_{\R}^\xi (\simplexSub),
 \end{align*}
 where 
 \begin{equation*}
\sigma_{\R}^\xi (\simplexSub) =  \dfrac{ \Lambda\left\{y \in \Set_+  : \R\left(Ay^\xi \right) \geqslant 1, {Ay^\xi}/{\|Ay^\xi\|_{}} \in \simplexSub \right\} }{\Lambda\left\{y \in \Set_+  : \R_{}\left(Ay^\xi \right) \geqslant 1 \right\} }, \quad \xi>0.
 \end{equation*}
 
 For $\xi <0$, we proceed similarly, but with $0 \geqslant \rho' \geqslant \xi^{-1}$: 
  \begin{align*}
    \Lambda\left( \A_{\rho',\simplexSub} \right)  &={}  \Lambda\left\{y \in \Set_+: \R_{}\left\{\xi^{-1}A(y^\xi - 1)\right\} + \xi^{-1} \geqslant \rho', \text{sign}(\xi)\dfrac{\xi^{-1}Ay^\xi}{\|\xi^{-1}Ay^\xi \|_{}} \in \simplexSub \right\}  \\
    & ={}\left(\xi \rho'\right)^{-1/\xi} \Lambda\left\{y \in \Set_+ : \R_{}\left(Ay^\xi\right) \leqslant 1,  {Ay^\xi}/{\|Ay^\xi \|_{}} \in \simplexSub \right\} \\
          & ={}(\xi \rho')^{-1/\xi} \Lambda\left\{y \in \Set_+ : \R_{}\left(Ay^\xi\right) \leqslant 1 \right\}  \times \sigma_{\R}^\xi (\simplexSub),
 \end{align*}
 where
 \begin{equation*}
\sigma_{\R}^\xi (\simplexSub) =  \dfrac{ \Lambda\left\{y \in \Set_+  : \R\left(Ay^\xi \right) \leqslant 1, {Ay^\xi}/{\|Ay^\xi\|_{}} \in \simplexSub \right\} }{\Lambda\left\{y \in \Set_+  : \R_{}\left(Ay^\xi \right) \leqslant 1 \right\} }, \quad \xi<0.
 \end{equation*}

For $\xi = 0$, we use the change of variables $\rho = \R(x)$, $w = \exp\{ x - \R(x)\}$. 
As $\R$ is an evaluation function,   $\exp\{\R(\log x)\} = \R(x)$, and for any $\rho' \geqslant 0$, 
 \begin{equation*}
  \Lambda\left\{ (\rho', \simplexSub) \right\} = \Lambda\left[ y \in \Set_+ : \R\left(y \exp A \right) \geqslant e^{\rho'}, \exp \left\{ A\log y- \R(A\log y) \right\}\in \simplexSub \right],
  \end{equation*}
  and the $(-1)$-homogeneity of $\Lambda$ yields
 \begin{align*}
       \Lambda\left\{ (\rho', \simplexSub) \right\}     &={}  e^{-\rho'} \Lambda\left[y \in \Set_+: \R\left(y \exp A \right) \geqslant 1,  \exp \left\{A\log y - \R(A\log y)\right\} \in \simplexSub \right]  \\
            &={}  e^{-\rho'} \Lambda\left\{y \in \Set_+ : \R\left( y \exp A  \right) \geqslant 1 \right\}  \times \sigma_{\R}^\xi (\simplexSub),
 \end{align*}
 with
  \begin{equation*}
\sigma_{\R}^{0} (\simplexSub) = \dfrac{ \Lambda\left[y \in \Set_+ : \R\left(y \exp A \right) \geqslant 1,  \exp\left\{ A\log y - \R(A\log y)\right\} \in \simplexSub \right] }{\Lambda\left\{y \in \Set_+ : \R\left( y \exp A \right) \geqslant 1 \right\} },
 \end{equation*}
which proves the theorem. \hfill  $\Box$
 
\subsection{Marginal properties of generalized $\R$-Pareto processes}\label{app: gen r pareto marginals}

For~\eqref{eq: gpp marginals}, we use the representation of generalized $\R$-Pareto processes for non-linear functionals.
Let $s_0 \in S$, and suppose that  $u_0 \geqslant b(s_0) $ is such that for any $t \geq 1$, $x \in \A_{u_0}$ implies that $tx \in \A_{u_0}$, 
where $\A_{u_0} = \left\{ x \in \Set^{\xi,a,b} : x(s_0) \geq u_0, \R \left\{(x - b) / \R(a)\right\} \geqslant 0\right\}$, which implies that $\A_{u_0} = t_0^{-1}\A_{t_0u_0}$ for any $t_0\geq 1$.
Then for any $x_0 \geq u_0$,
\begin{align*}
\Pr\{P(s_0)> x_0\} & = \Pr\left(P\in \A_{u_0} \right) \\
 & = \dfrac{  \Lambda\left\{y \in \Ar  : a(s_0)\{y^\xi(s_0) - 1\}/\xi + b(s_0) \geqslant x_0\right\}}{\Lambda\left(\Ar \right)} \\
 & = \dfrac{  \Lambda\left\{y \in \Ar  : y^\xi(s_0) \geqslant 1 + \xi \{x_0 - b(s_0)\}/a(s_0)\right\}}{\Lambda\left(\Ar \right)} \\
 & = \dfrac{  \Lambda\left\{y \in \Ar  : y^\xi(s_0) \geqslant t_0[1 + \xi \{u_0 - b(s_0)\}/a(s_0)]\right\}}{\Lambda\left(\Ar \right)} \\
 & = \dfrac{  \Lambda\left\{t_0^{-1/\xi}y \in \Ar  : y^\xi(s_0) \geqslant 1 + \xi \{u_0 - b(s_0)\}/a(s_0)\right\}}{\Lambda\left(\Ar \right)} \\
 & = \left[1 + \xi\dfrac{(x_0 - u_0)}{a(s) + \xi \{u_0 - b(s_0)\}}\right]^{-1/\xi}  \times \\ 
 &\quad\qquad \dfrac{  \Lambda \left\{y \in \Ar  : y^\xi(s_0) \geqslant 1 + \xi \{u_0 - b(s_0)\}/a(s_0)\right\}}{\Lambda\left(\Ar \right)},
\end{align*}
where $t_0 = [a(s_0) + \xi \{x_0 - b(s_0)\}] / [a(s_0) + \xi \{u_0 - b(s_0)\}]$ and we used the $(-1)$-homogeneity of $\Lambda$ for the last equality.
So for any $x_0 \geqslant u_0$, 
\begin{align*}
\Pr\left\{ P(s_0) > x_0 \mid  P(s_0) > u_0 \right\}
& = \left\{1 + \xi \dfrac{x_0 - u_0}{\sigma(u_0)}\right\}^{-1/\xi} ,
\end{align*}
where  $\sigma(u_0) = a(s_0) + \xi \{u_0 - b(s_0)\}$.

For a linear risk functional and if $\rho' > \R(b)$ and $\xi >0$, 
\begin{align*}
\Pr\{\R(P) > \rho'\} & = \Pr\left[P\in\left \{x \in \Set^{\xi,a,b} : \R(x) > \rho'\right\}\right] \\
 & = \dfrac{  \Lambda\left[y \in \Ar  : \R\left(A\dfrac{y^\xi - 1}{\xi}\right) > \dfrac{\rho' - \R(b)}{\R(a)} \right]}{\Lambda\left(\Ar\right)} \\
 & = \dfrac{  \Lambda\left[y \in \Ar  : \R\left(A{y^\xi}\right) >1 + \xi \dfrac{\rho' - \R(b)}{\R(a)} \right]}{\Lambda(\Ar)} \\ 
 & = \left\{1 + \xi\dfrac{\rho' - \R(b)}{\R(a)}\right\}^{-1/\xi},
\end{align*}
with the same conclusion if $\xi\leq 0$.
\hfill $\Box$

\subsection{Derivation of (\ref{eq: simple gen pareto dist})}\label{app: simple gen pareto dist}
Consider a set $\{s_1,\dots,s_L\}$ of locations in $S$ and sets of the type
$$
\calR_{\max}(x') = \left\{x \in \Set^{\xi,a,b} : \max_{l = 1,\dots,L}\dfrac{x(s_l) -b(s_l)}{x_l' -b(s_{l})} \geq 1, \R\left\{(x' - b)/\R(a)\right\}  \geqslant 0 \right\}
$$ 
where each component of $x'=(x_1', \ldots, x_L')$ satisfies $x_l'>b(s_l)$.
Other sets characterized by $x'$ could also be considered, but we focus on $\calR_{\max}$ for easier comparison with articles such as \citet{Wadsworth2013a}.  For economy of notation write $a(s_l)=a_l$, $b(s_l)=b_l$ and so forth. Then
\begin{align*}
\Pr\left\{P \in \calR_{\max}(x') \right\}  &= {  \Lambda\left\{y \in \Ar : a\dfrac{y^{\xi} - 1}{\xi} + b  \in \calR_{\max}(x') \right\}}\div
{ \Lambda\left(\Ar\right)} \\
 &= {\Lambda \left[ y \in \Ar : \underset{l = 1,\dots,L}{\max} \dfrac{y_l}{\{1 + \xi(x_l' -b_l)/a_l\}^{1/\xi}} \geq 1 \right]}\div{\Lambda\left(\Ar\right)}, 
\end{align*}
with the appropriate changes when $\xi=0$. Differentiation of the resulting expression using the chain rule gives
\begin{align*}
\MoveEqLeft
\dfrac{\partial \Pr\{ P \in \calR_{\max}(x')\} }{\partial x'} =  
  \dfrac{\lambda\left[\left\{1 + \xi(x'- b)/a\right\}^{1/\xi}\right]}{\Lambda\left(\Ar\right)} 
  \prod_{l = 1}^L a_l^{-1}\left(1 + 
  \xi \dfrac{x' - b_l}{ a_l}\right)^{1/\xi - 1},
\end{align*}
where $\lambda$ is the $L$-dimensional intensity function given by
$$
\Lambda\{\exceedanceSets_{\max}(y')\} = \int_{\RR^L \setminus (0,y']^L}  \lambda(y)\, \D{y}, 
$$
with $\exceedanceSets_{\max}(y') = \{y \in \Ar : \max_{ l =1,\dots,L} y_l/y'_l \geq 1 \}$,  giving~\eqref{eq: simple gen pareto dist}.

\section{Windstorm model validation plots}
\begin{figure}[!h]
\begin{center}
\includegraphics[width=\textwidth]{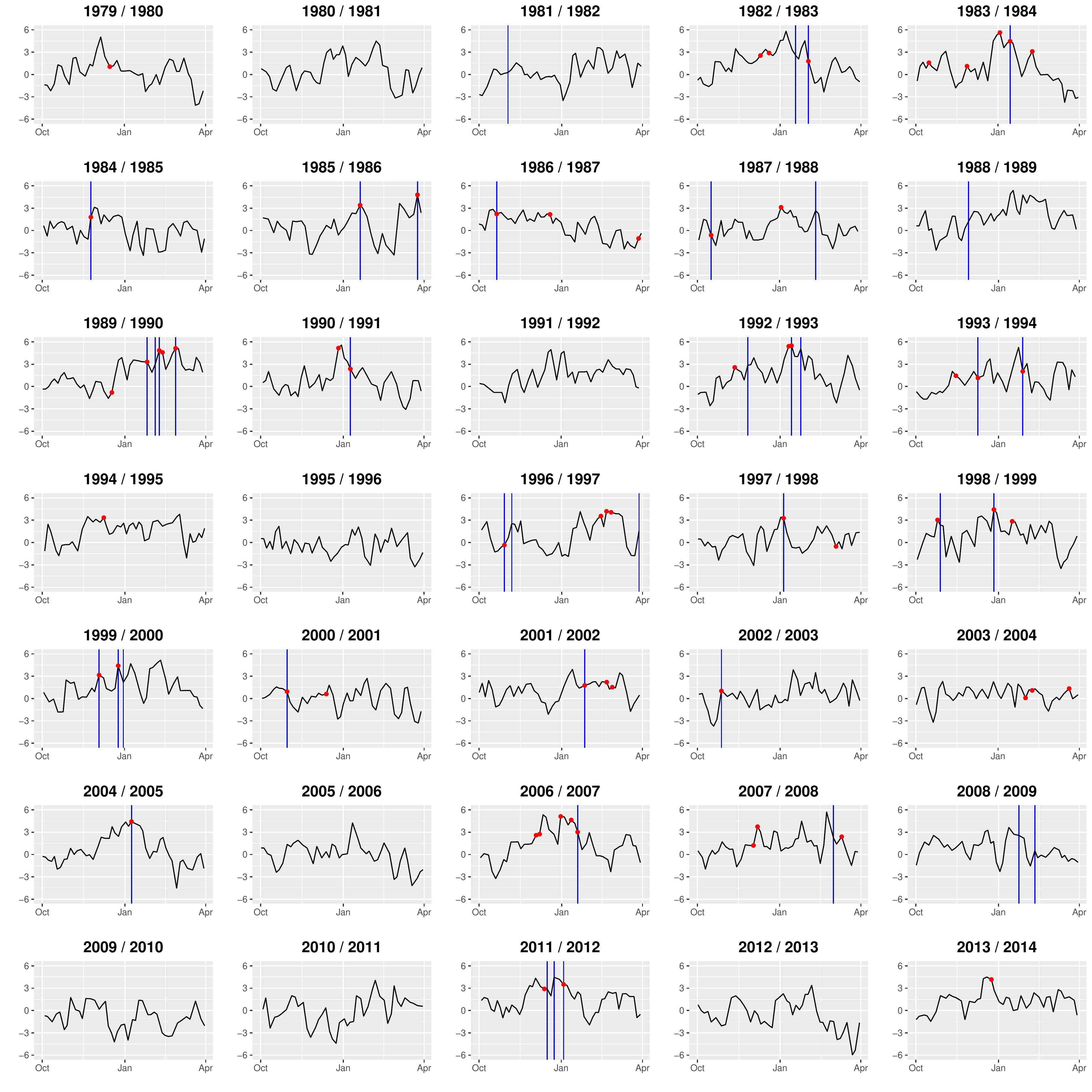}
\end{center}
\caption{Three-hourly North Atlantic Oscillation (NAO) index computed on the ERA--Interim data set for each winter. $\R$-exceedances above the $0.96$ empirical quantile are represented by red dots and windstorms starting dates from XWS catalogue are represented by blue vertical lines.}
\label{fig: nao}
\end{figure}

\begin{figure}[!h]
\begin{center}
\includegraphics[width=\textwidth]{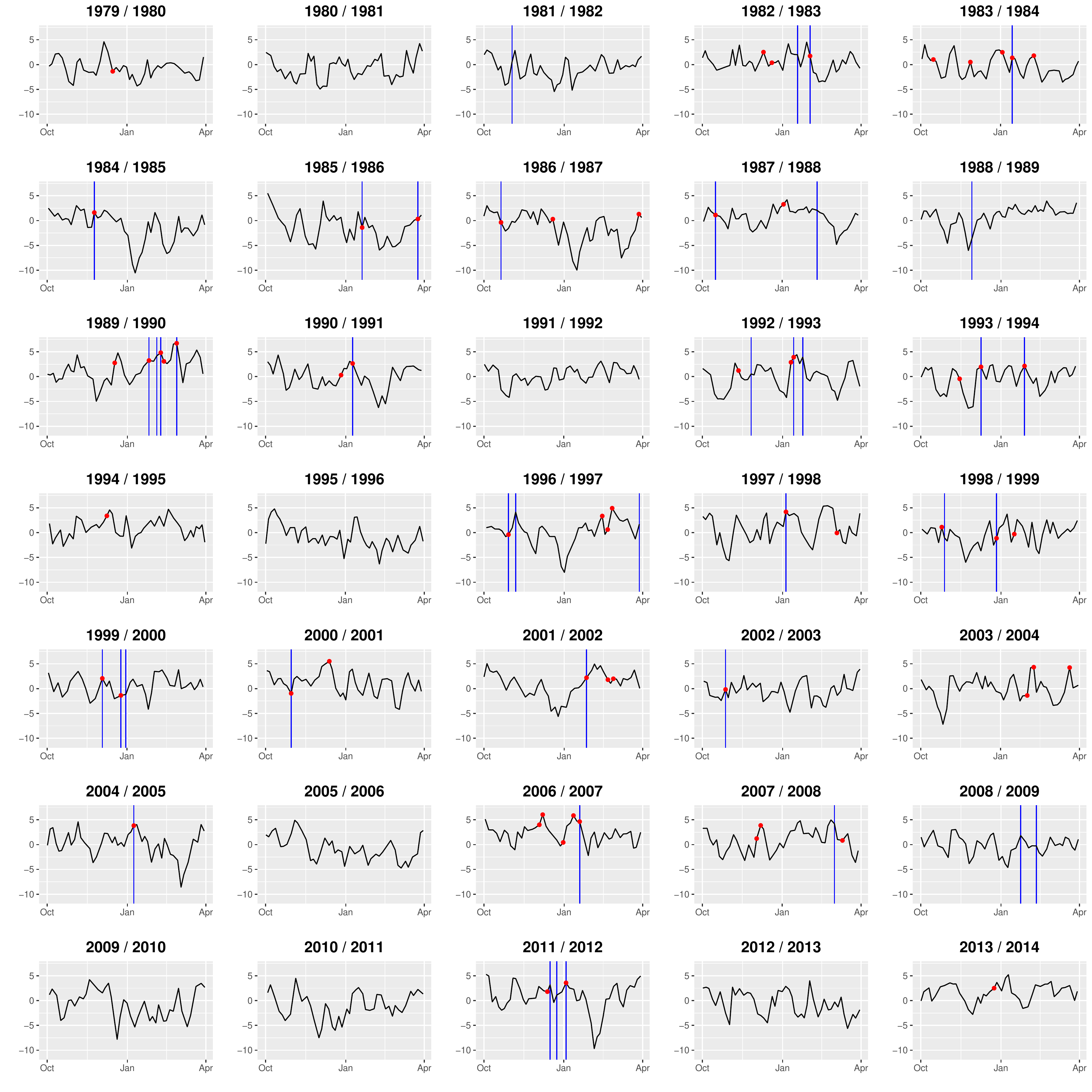}
\end{center}
\caption{{Three-hourly first eigenvalue} of the spatial EOF decomposition of the temperature anomaly computed on the ERA--Interim data set for each winter.  $\R$-exceedances above the $0.96$ empirical quantile are represented by red dots and windstorms starting dates from XWS catalogue are represented by blue vertical lines.}
\label{fig: first}
\end{figure}

\begin{figure}[!h]
\begin{center}
\includegraphics[width=\textwidth]{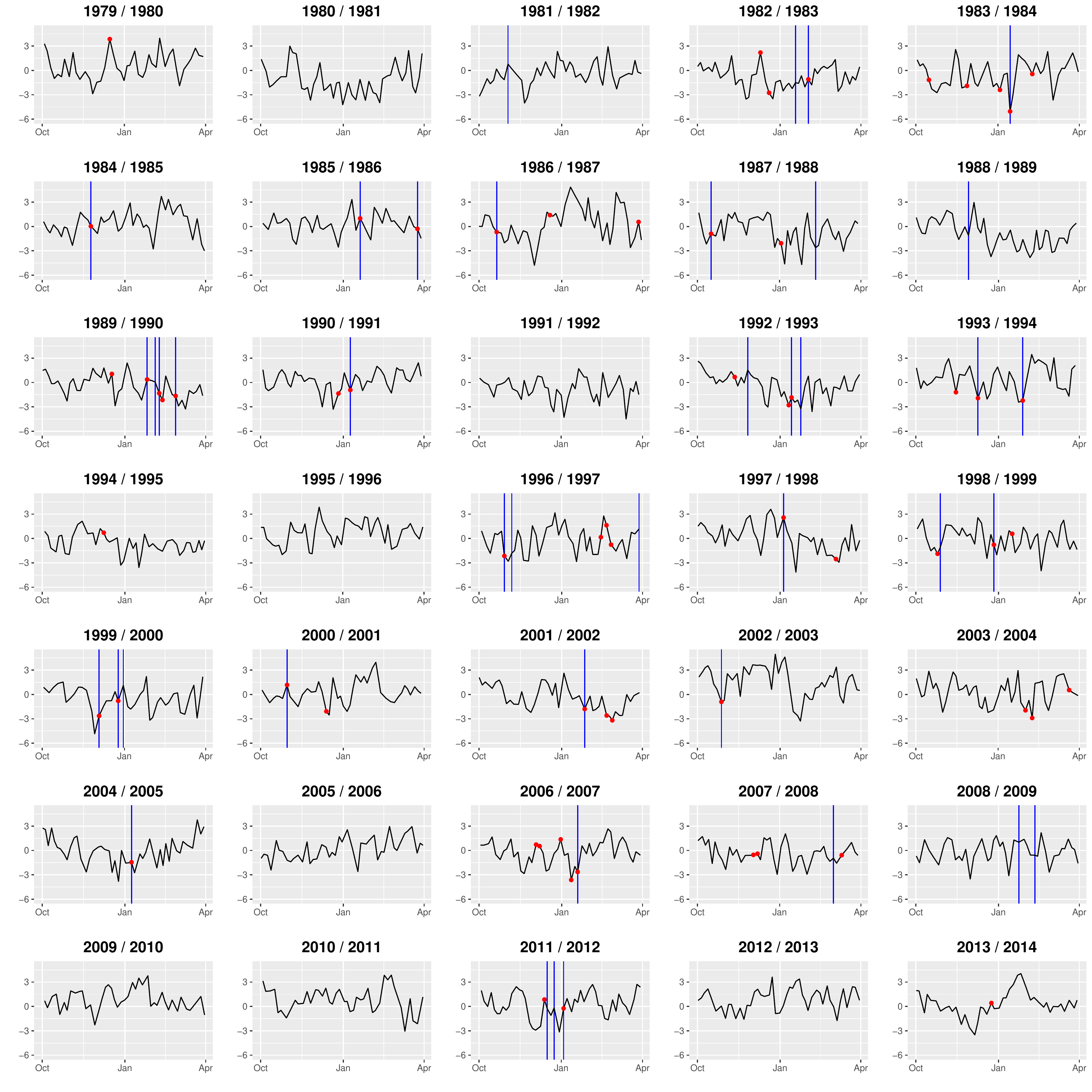}
\end{center}
\caption{{Three-hourly third eigenvalue} of the spatial EOF decomposition of the temperature anomaly computed on the ERA--Interim data set for each winter. $\R$-exceedances above the $0.96$ empirical quantile are represented by red dots and windstorms starting dates from XWS catalogue are represented by blue vertical lines.}
\label{fig: third}
\end{figure}

Here we give the plots for the logistic regression model for the distribution of the indicator $\indicatorFun\{\R(x) \geqslant u\}$ for storm occurrence in Europe.
The North Atlantic Oscillation (NAO) index and the first and third eigenvalues of the temperature anomaly, shown in Figures~\ref{fig: nao}, \ref{fig: first} and \ref{fig: third}, influence the occurrence of winter storms significantly at the $0.1\%$ level.

\section{Diagnostic plots for the windstorm probability exceedances}\label{app: sm winstorms}

\begin{figure}[!h]
\begin{center}
\includegraphics[width=\textwidth]{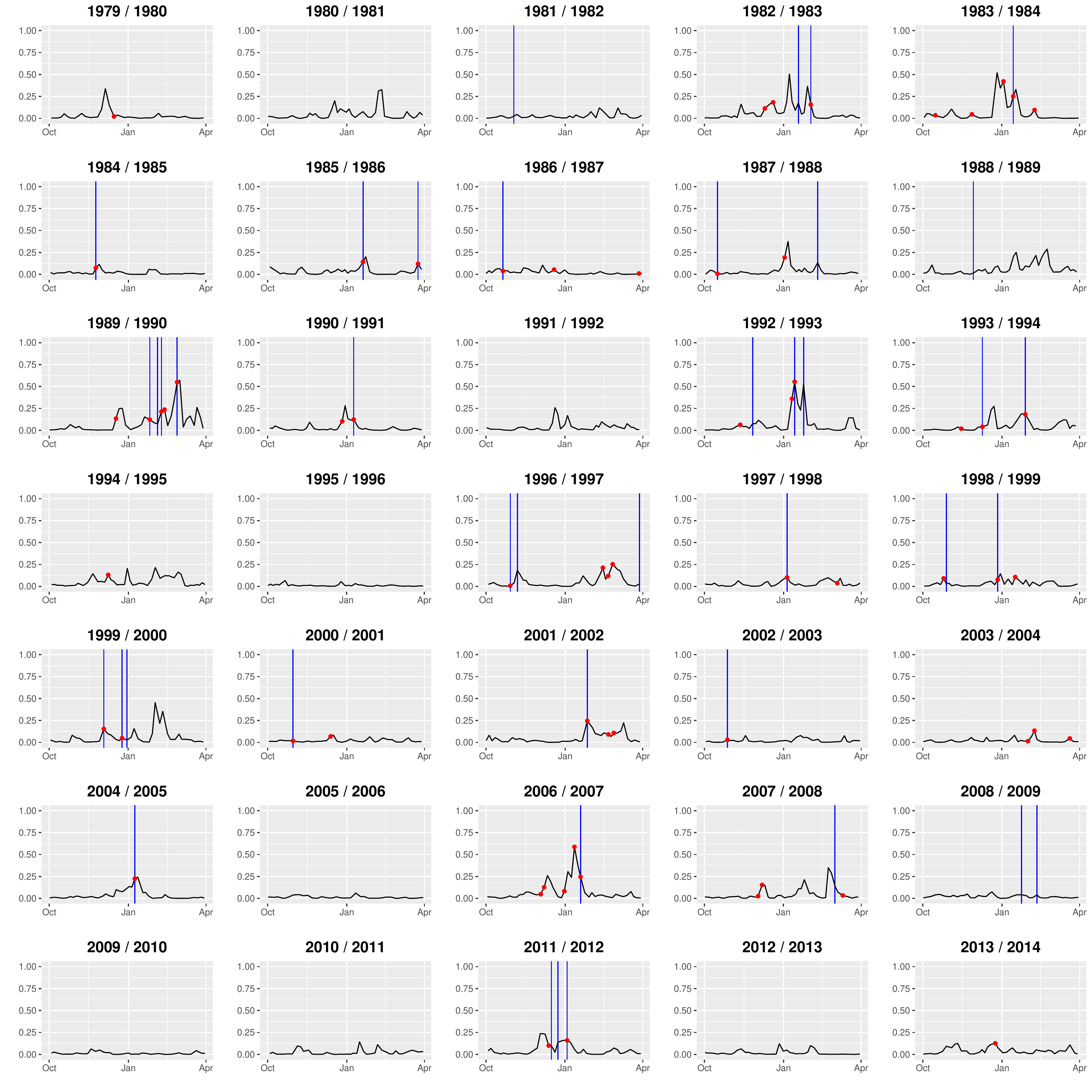}
\end{center}
\caption{Three-hourly probability of $\R$-exceedances using logistic regression model with the NAO index and the first and third temperature anomaly eigenvalues as covariates. Observed $\R$-exceedances are represented by red points and the blue vertical lines correspond to the storms starting dates from the XWS catalogue. }
\label{fig: log reg model}
\end{figure}

Figure~\ref{fig: log reg model} shows the fitted daily probabilities of $\R$-exceeedances for the European windstorms.  

\section{Rainfall model plots}

\begin{figure}
\begin{center}
\begin{tabular}{cc}
\includegraphics[width=0.4\textwidth]{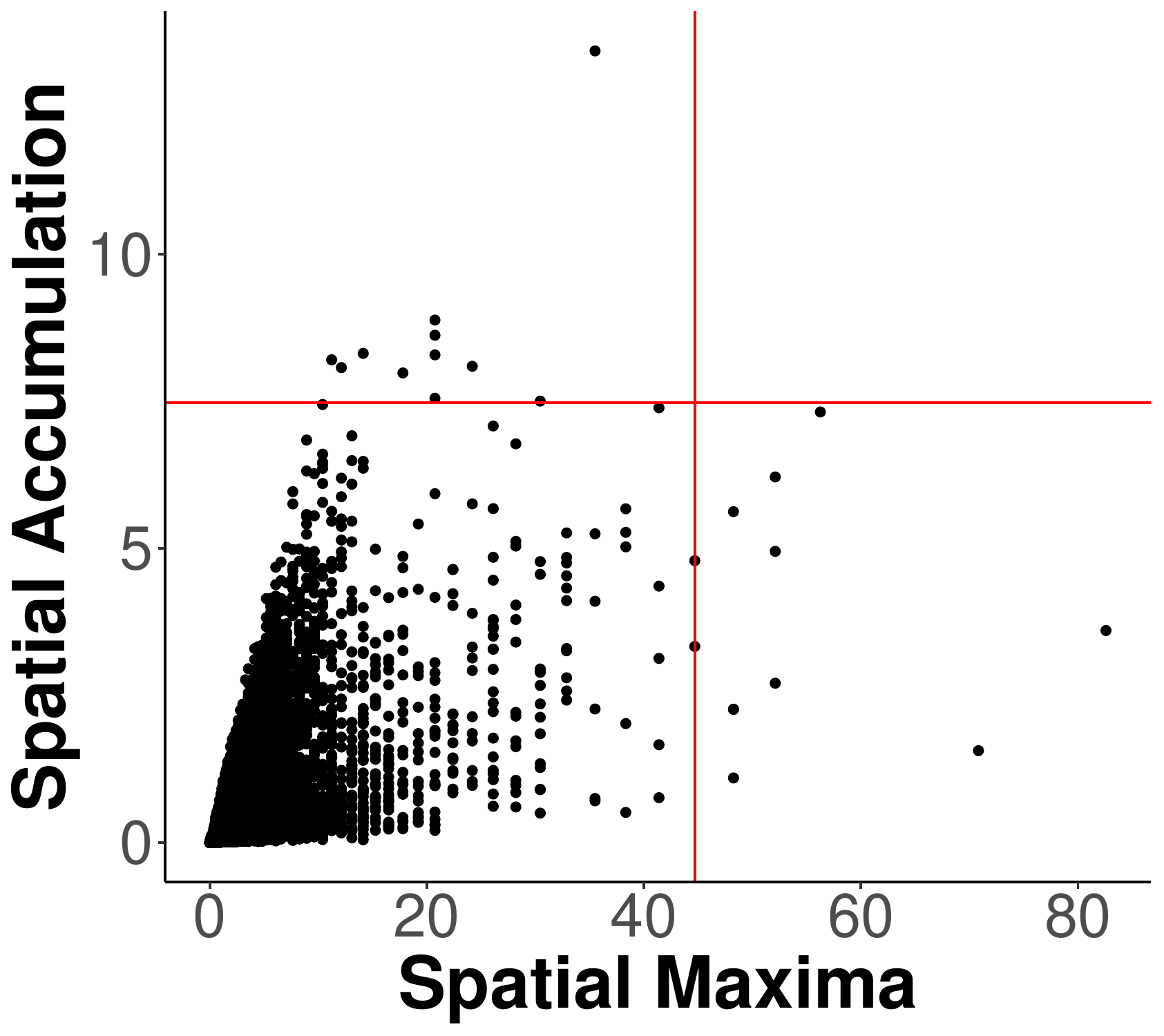}  &\includegraphics[width=0.4\textwidth]{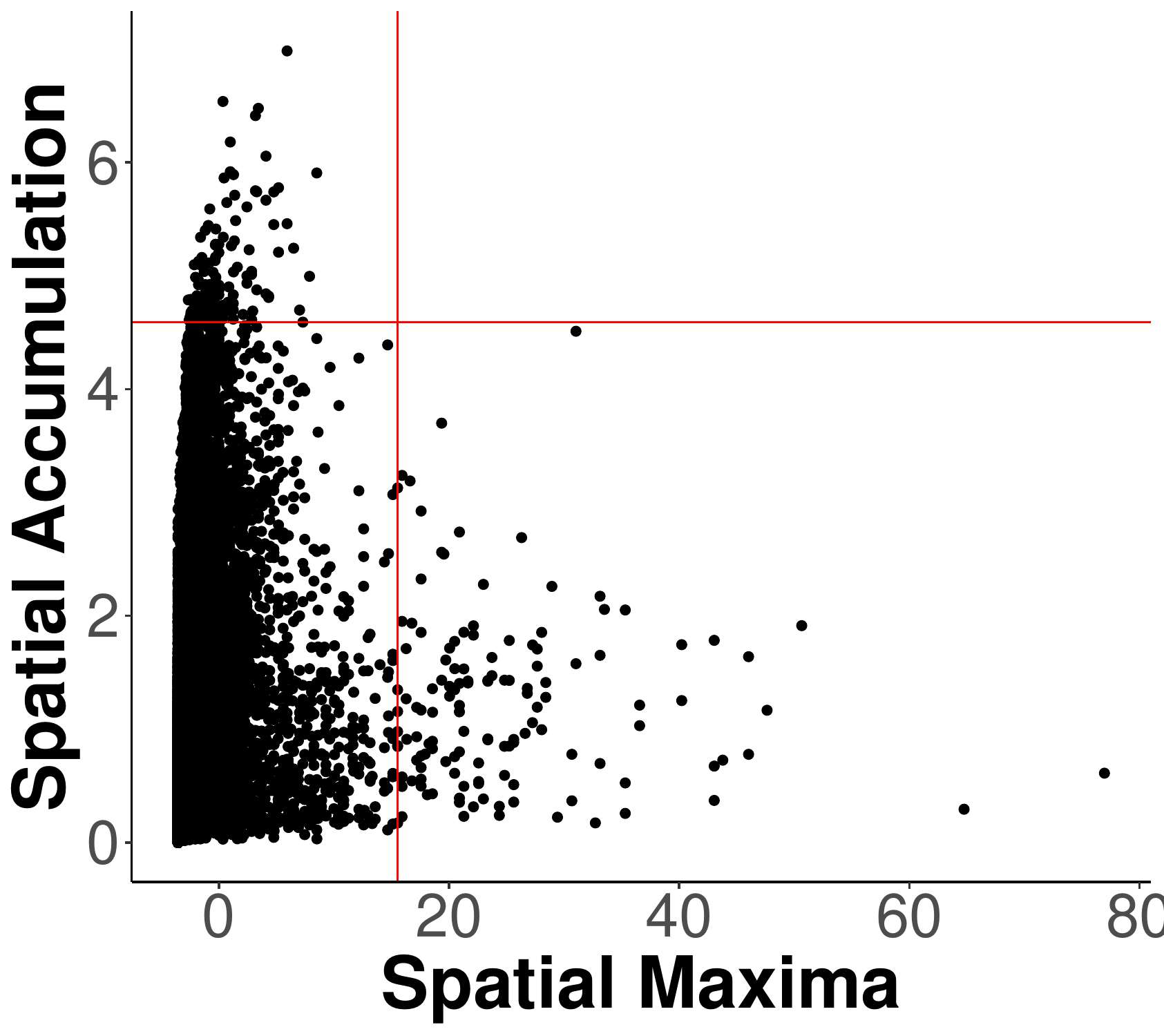} \\
\\ 
\includegraphics[width=0.45\textwidth]{images/chap4/sum_event.pdf} & \includegraphics[width=0.45\textwidth]{images/chap4/max_event.pdf}  \\
\end{tabular}
\end{center}
\caption{Extreme hourly rainfall events in the  Zurich region, 2013--2018, computed using radar rainfall measurements $X(s)$ (mm) on a grid $S$.  Top:  spatial averages $|S|^{-1}\int_S X(s)\, \D{s}$ and spatial maxima $\max_{s \in S} X(s)$, with red thresholds demarcating the largest $11$ events of each type. Top right:  likewise for modified spatial averages and spatial maxima and thresholds for the largest $132$ events of each type.
Bottom line: events corresponding to the largest spatial average (left) and the largest spatial maximum (right).}
\label{fig: rainfall observations app}
\end{figure}

Here we give detailed plots for the different risk functionals used in the rainfall model. Figure~\ref{fig: rainfall observations app} highlights the influence of the low-pass filter on the modified spatial average functional.

\end{document}